\documentclass[prd,aps,twocolumn,a4paper,floatfix,nofootinbib]{revtex4}

\usepackage{graphicx,psfrag}
\usepackage{mathrsfs,bbm}
\usepackage{amsmath,amsfonts,amssymb}
\usepackage{slashed}
\usepackage{comment}
\usepackage{hyperref}
\usepackage[normalem]{ulem}
\usepackage[utf8]{inputenc}
\usepackage{tikz}

\allowdisplaybreaks

\def\p{\partial}
\def\Lie{{\cal L}}
\def\ul{\underline}

\begin{document}

\title{Summation by Parts and Truncation Error Matching on
  Hyperboloidal Slices}

\author{Shalabh Gautam$^{1}$, Alex Va\~n\'o-Vi\~nuales$^{2}$, David
  Hilditch$^{2}$ and Sukanta Bose$^{1,3}$}

\affiliation{${}^1$Inter-University Centre for Astronomy and
  Astrophysics, Post Bag 4, Ganeshkhind, Pune 411007,
  India\\ ${}^2$CENTRA, Departamento de F\'isica, Instituto Superior
  T\'ecnico IST, Universidade de Lisboa UL, Avenida Rovisco Pais 1,
  1049-001 Lisboa, Portugal\\
  ${}^3$Department of Physics \& Astronomy,
  Washington State University, Pullman, WA 99164, USA}

\date{\today}

\begin{abstract}
We examine stability of summation by parts (SBP) numerical schemes
that use hyperboloidal slices to include future null infinity in the
computational domain. This inclusion serves to mitigate outer boundary
effects and, in the future, will help reduce systematic errors in
gravitational waveform extraction. We also study a setup with
truncation error matching. Our SBP-Stable scheme guarantees
energy-balance for a class of linear wave equations at the
semidiscrete level. We develop also specialized dissipation
operators. The whole construction is made at second order accuracy in
spherical symmetry, but could be straightforwardly generalized to
higher order or spectral accuracy without symmetry. In a practical
implementation we evolve first a scalar field obeying the linear wave
equation and observe, as expected, long term stability and norm
convergence. We obtain similar results with a potential term. To
examine the limitations of the approach we consider a massive field,
whose equations of motion do not regularize, and whose dynamics near
null infinity, which involve excited incoming pulses that can not be
resolved by the code, is very different to that in the massless
setting. We still observe excellent energy conservation, but
convergence is not satisfactory. Overall our results suggest that
compactified hyperboloidal slices are likely to be provably effective
whenever the asymptotic solution space is close to that of the wave
equation.
\end{abstract}

\maketitle

\section{Introduction}\label{Section:introduction}

A persistent problem in numerical relativity is the inclusion of
future null infinity~$\mathscr{I}^+$ in the computational domain. As
described by Penrose~\cite{Pen63} future null infinity is the set of
endpoints of future directed null geodesics. Ultimately this will
allow us to study the propagation of waves out to~$\mathscr{I}^+$. In
the modern era of gravitational wave astronomy, one landmark goal is
to compute waveforms from a binary merger in a completely satisfactory
manner. The present state of the art for extracting signals
at~$\mathscr{I}^+$ is to use
Cauchy-Characteristic-Extraction~\cite{BisGomLeh97a, BisGomLeh97,
  ZloGomHus03, HanSzi14, BarMoxSch19}. In this approach a standard
time evolution is performed, and data taken on a timelike world-tube
from that evolution serve as the {\it given data} for a tertiary
computation on outgoing characteristic slices compactified
to~$\mathscr{I}^+$. This approach suffers from the principle weakness
that data transfer is one-way, so eventually artificial outer boundary
conditions in the Cauchy domain corrupt the interior physically
correct data. Cauchy-Characteristic-Matching~\cite{Win12} proposes to
solve this shortcoming by evolving and coupling both domains
simultaneously. In practice interfacing two different formulations of
GR may not however result in a composite PDE problem that is
well-posed~\cite{GiaHilZil20}.

\begin{figure}[htbp]
\includegraphics[width=0.45\textwidth]{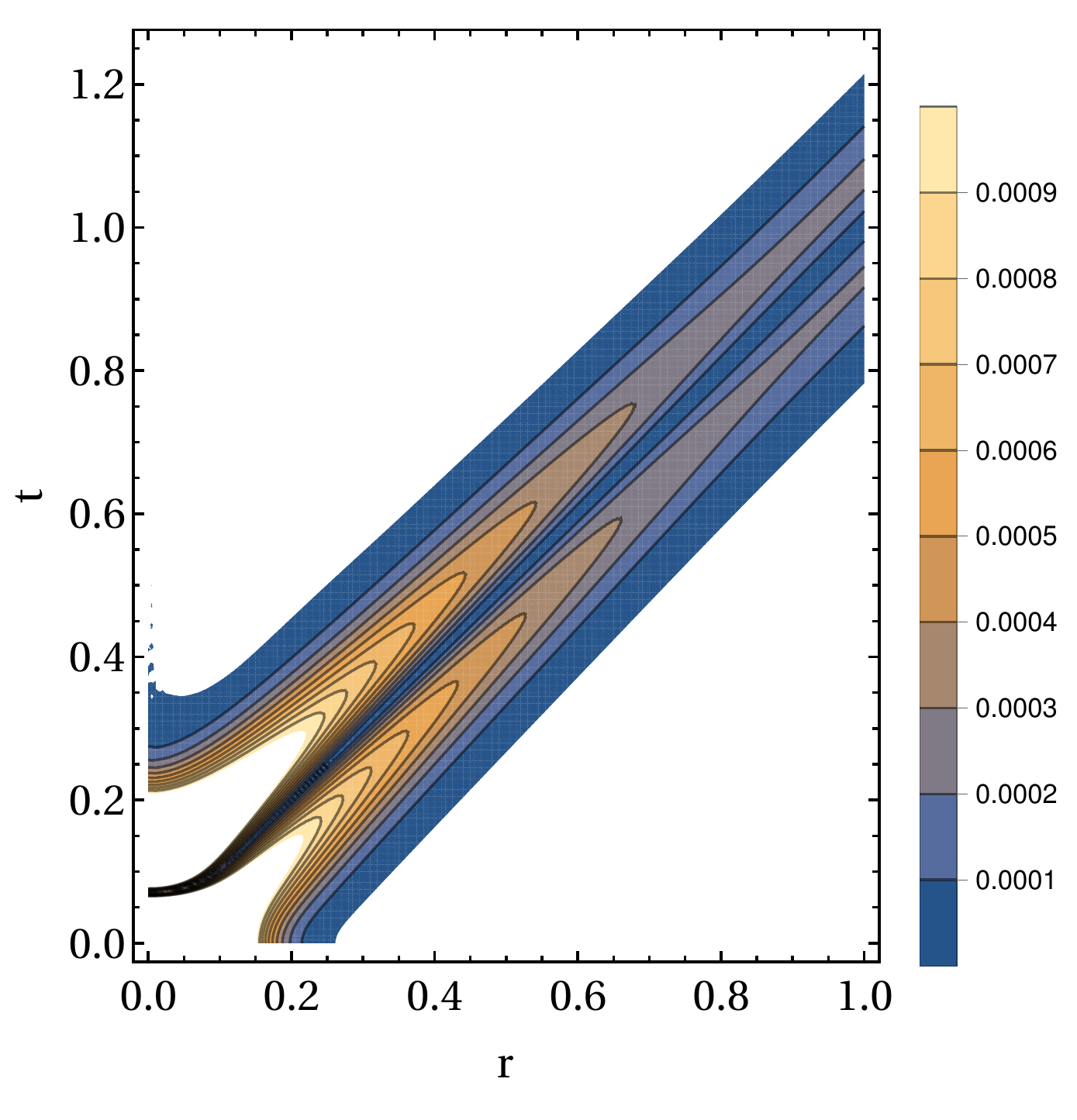}
\caption{A contour plot showing the propagation of a pulse of a scalar
  field~$\tilde{\psi}$, satisfying the wave equation, propagating
  to~$\mathscr{I}^+$, located here at~$r=1$. The solution is computed
  using our SBP-Stable scheme. The plot is cut-off at low and high
  amplitudes.\label{fig_LWE_Contour_plot}
}
\end{figure}

An alternative path, which we follow, is to use compactified
hyperboloidal slices, which are everywhere spacelike but which
terminate at~$\mathscr{I}^+$. Starting with the conformal field
equations~\cite{Fri81,Fri81a} hyperboloidal slices have been used with
several formulations of
GR~\cite{MonRin08,Zen08,Rin10,BarSarBuc11,VanHusHil14,Van15} all of
which have to render the field equations sufficiently regular for
numerical approximation in some way. The specific strategy we follow
was suggested in~\cite{HilHarBug16} and employs the Dual Foliation
(DF) formalism~\cite{Hil15}. The means to achieve regular equations
for regular unknowns is to use a carefully chosen tensor basis in
combination with hyperboloidal coordinates. Follow-ups on the
mathematical formalism~\cite{GasHil18} and numerical
implementation~\cite{GasGauHil19} have shown that it should be
possible to manage logarithmic divergences that appear in the
asymptotic solution space by a careful choice of variables.

Until now our analysis~\cite{HilHarBug16,GasHil18,GasGauHil19} has
always been performed at the continuum level, with verification of
convergence of numerical schemes being performed only
empirically. Thus the question arises whether a numerical scheme can
be given that provably converges to the continuum solution in the
limit of infinite resolution. This question is far too difficult to
tackle right away for GR. Even for systems used in the perturbative
studies~\cite{Zen10,Zen11} there is no rigorous numerical analysis. In
this paper, we therefore deal with the simplest case of a scalar field
obeying a linear wave equation with potential (LWEP). Special cases
occur when the potential vanishes (LWE) and for the massive
Klein-Gordon Equation (LMKGE).  We build two approximation schemes. In
some sense both use a summation by parts (SBP)
approach~\cite{Str94}. The first, which we call SBP-Stable, is
formally stable and captures at the semidiscrete level the energy
conservation properties of the continuum system. In the second scheme,
which we call SBP-TEM, we apply truncation error matching (TEM)
at~$\mathscr{I}^+$ (see for example~\cite{Pre04}) rather than
accepting the lower order accurate pointwise approximation that is
unfortunately necessary in the first approach. This helps minimize
unphysical reflections from the outer boundary. We work at second
order accuracy on a Minkowski background. In this work we restrict
ourselves to spherical symmetry. Technical difficulties arise because
of the coordinate singularity at the origin, but the key challenge we
face is in managing the asymptotics near~$\mathscr{I}^+$. We fully
expect a generalization of our scheme to hold in more general
scenarios. In Fig.~\ref{fig_LWE_Contour_plot} we present a contour
plot of a numerical solution for the wave equation, in which one can
see that the pulse leaves the domain in essentially two bursts, with
no visible numerical reflection.

It is intuitively clear that to reach~$\mathscr{I}^+$ some price must
be paid. By construction, our coordinates are well adapted to outgoing
radiation, but there is a key difficulty in resolving {\it incoming}
waves. To investigate this we perform tests with different potentials,
that result in a coupling between outgoing and incoming pulses. Of
particular interest is the LMKGE. It turns out that compactification
makes the mass term singular at~$\mathscr{I}^+$. But as described by
Winicour~\cite{Win88} solutions fall-off towards~$\mathscr{I}^+$
faster than any inverse power of areal radius~$R$, so the field decays
more rapidly towards~$\mathscr{I}^+$ than the rate at which the
coefficient of the mass term blows up. With this setup our scheme
guarantees perfect energy conservation, but we find that the
excitation of badly resolved incoming pulses prevents long-term
convergence. In practice this means that, at least for now, if one
wishes to use massive fields with hyperboloidal slices, we need to
keep the support of the fields away from the wavezone. Working with
less aggressive potentials we find that perfect long-term convergence
is, as expected, immediately recovered.

The paper is structured as follows. In Sec.~\ref{Intro_hyp_slice} we
begin with a summary of the specific hyperboloidal foliation that we
use. The foliation can be adjusted with only superficial changes to
the subsequent expressions. In Sec.~\ref{LWEP_hyp_slice} we present
our model equation and derive an appropriate energy-balance law on
hyperboloidal slices for the continuum equations. Building directly on
this, in Sec.~\ref{SBP_Scheme}, we construct our SBP
schemes. Afterwards in Sec.~\ref{Numerical_Evolution} numerical
evolutions are presented with a series of different potentials.
Finally we conclude in Sec.~\ref{Conclusions}. Geometric units are
used throughout.

\section{Hyperboloidal Slices Overview}
\label{Intro_hyp_slice}

We now briefly review geometric quantities describing a foliation of
spacetime, for which we use the standard notation, and evaluate them
under our choice of hyperboloidal slices. They will be used to obtain
an energy on such slices conserved up to boundary fluxes, which will
in turn underpin our numerical
scheme. Let~$x^{\ul{\mu}'}=(T,R,\theta^A)$ be the canonical spherical
polar coordinates on the Minkowski spacetime, so that the line-element
becomes
\begin{align}\label{line_element_shp_coord}
ds^2 = - dT^2 + dR^2 + R^2 d\Omega^2 \,,
\end{align}
where~$d\Omega^2$ is the line element on the unit round
two-sphere. Let~$x^\mu = (t,r,\theta^A)$ be the hyperboloidal
coordinates, defined by~$T = t + H(R)$ and~$R = R(r)$. Here,~$r \in
[0,r_{\mathscr{I}}]$ is a compactified radial coordinate
with~$r_{\mathscr{I}}$ a fixed positive number that denotes the value
of~$r$ at~$\mathscr{I}^+$. The angular coordinates~$\theta^A$ are held
fixed.

Let~$C^R_\pm=\pm 1$ denote the outgoing and incoming radial
lightspeeds in the original coordinates, and~$c^r_\pm$ those in
hyperboloidal coordinates. For the latter we get
\begin{align}
c^r_\pm = \pm \frac{1}{R'(1 \mp H')} \,,
\end{align}
with~$H' \equiv dH/dR$ and~$R'\equiv dR/dr$. Thus the
lightspeeds~$c^r_\pm$ are functions of~$r$. If we choose~$R$ and~$H$
carefully, we can restrict these functions to a desired form. Ideally,
we would have~$c^r_\pm = \pm 1$ so that both incoming and outgoing
pulses could be resolved. However, this is not completely compatible
with our wish to draw infinity to a finite coordinate distance by use
of a compactification~$R(r)$. In particular,
following~\cite{CalGunHil05}, we might take the simple
\begin{align}\label{Compactification}
  R(r) = \frac{r}{\Omega(r)^{\frac{1}{1-n}}} = r
  \left( 1 - \frac{r^2}{r_\mathscr{I}^2} \right)^{\frac{1}{1-n}}\,,
\end{align}
with~$1 < n \leq 2, r \in [0,r_{\mathscr{I}}]$. Later we adjust to a
slightly different compactification. With this choice~$R' \sim R^n$
as~$r \rightarrow r_\mathscr{I}$. Thus only the height function~$H$
remains to be chosen. To resolve outgoing pulses, of primary interest
in the asymptotically flat setting, we thus choose~$H$ so that~$c^r_+
= 1$. Throughout we choose~$R'(1-H') = 1$, or~$H' = 1 - 1/R'$, and
thus obtain~$c^r_+ = 1$ identically, and~$c^r_- = -1/(2R'-1)$. Note
that~$c^r_- = -1$ at the origin and decreases in magnitude
monotonically to~$0$ at~$r_\mathscr{I}$. The line element becomes
\begin{align}
  ds^2 = - dt^2 - 2 (R'-1) dt dr + (2R'-1) dr^2 + R^2 d\Omega^2 \,.
  \label{lowercase_metric}
\end{align}
The components of the spatial metric~$\gamma_{ij}$ here can be read
off from the spatial components. The lapse~$\alpha$ and only
non-trivial component of the shift~$\beta^i$ are given by
\begin{align}
\alpha = R'(2R'-1)^{-(1/2)}\,,\quad \beta^r=-\frac{R'-1}{2R'-1}\,.
\end{align}
Note that the shift is negative, but finite near and at~$\mathscr{I}$.
Finally, the extrinsic curvature~$K_{ab}$ can be computed
from~$\Lie_t\gamma_{ab}=0$, but is not explicitly needed in the
following.

\section{The Wave Equation with Potential on Hyperboloidal
  Slices}\label{LWEP_hyp_slice}

In this section, we formulate the LWE with a linear potential~$F$ on
our hyperboloidal slices and study its regularization. The case of the
LWE can be obtained simply by taking~$F=0$.

\subsection{The wave equation and regularity at the origin}

Consider a scalar field~$\psi$ satisfying a linear wave equation with
a potential~$F$
\begin{align}\label{LWEP}
(\Box - F)\psi = 0 \, ,
\end{align}
where~$\Box$ is the standard d'Alembertian. Imposing spherical
symmetry, we require that~$\psi = \psi(T,R)$, and also that the
potential be time-independent and non-negative, i.e.~$F=F(R)\geq
0$. Defining~$\pi \equiv - \p_T \psi$ and~$\phi_R \equiv \p_R \psi$,
we thus get a first order reduction of this equation in form of a
system of three first order equations in three variables,
\begin{align}
  \p_T \psi &= -\pi, \nonumber\\
  \p_T \phi_R &= - \p_R \pi + \gamma_2 (\p_R \psi - \phi_R),
  \nonumber\\
  \p_T \pi &= - \p_R \phi_R - \tfrac{2}{R} \phi_R + F \psi.
  \label{ppp-flat-coord}
\end{align}
The first equation comes directly from the definition of~$\pi$. The
second follows by equality of mixed partials~$\p_T$
and~$\p_R$. The~$\gamma_2$ coefficient serves to damp the reduction
constraints~\cite{GunGarCal05,LinSchKid05} associated with the
definition of~$\phi_R$. This constraint vanishes at the continuum
level in the original second order system, but should not be assumed
to necessarily vanish in the reduction or, in general, at the discrete
level. Later, we will choose~$\gamma_2 = 0$, and will study conditions
under which the constraint is satisfied, in some sense, even at the
discrete level. The third equation is obtained by substituting
for~$\p_T\psi$ and~$\p_R \psi$ within~\eqref{LWEP}.

Following the dual foliation~\cite{Hil15} strategy of our earlier
work~\cite{HilHarBug16,GasHil18,GasGauHil19} we rewrite this first
order reduction system in hyperboloidal coordinates while keeping the
reduction variables~$(\psi,\phi_R,\pi)$ unchanged. This approach has
the technical advantage that the same dynamical variables are evolved
in two coordinate systems. Changing to the coordinates~$x^\mu$
introduced in Section~\ref{Intro_hyp_slice},
Eqs.~\eqref{ppp-flat-coord} become
\begin{align}
\p_t \psi &=  - \pi \, , \nonumber\\
\p_t \phi_R &=  - \frac{R'}{2R'-1} \p_r \pi -
\frac{R'-1}{2R'-1} \left[ \p_r
  + \frac{2R'}{R} \right] \phi_R \nonumber\\
&\quad + \frac{\gamma_2 R'}{2R'-1} \left[ \p_r \psi
  + (R'-1) \pi - R' \phi_R \right]  \nonumber\\
&\quad + \frac{R'(R'-1)}{2R'-1} F \psi \, , \nonumber\\
\p_t \pi &=  - \frac{R'-1}{2R'-1} \p_r \pi
- \frac{R'}{2R'-1} \left[ \p_r + \frac{2R'}{R} \right]
\phi_R\nonumber\\
&\quad + \frac{\gamma_2 (R'-1)}{2R'-1}\left[ \p_r \psi  + (R'-1) \pi
  - R' \phi_R \right] \nonumber\\
&\quad + \frac{R'^2}{2R'-1} F \psi \, .\label{ppp-hyp-coord}
\end{align}
The origin of the~$\gamma_2$ terms appearing above is the single
constraint equation introduced in~\eqref{ppp-flat-coord}, expressed
here in lowercase coordinates.  Regularity at the origin is
well-understood, but as it will play an important role in our
numerical scheme we nevertheless provide a brief discussion about it.
Since we are working in spherical symmetry, we will substitute
the~$\theta$ and~$\phi$ coordinates by~$\theta^A=(\theta,\phi)$. The
radial coordinate goes from 0 to~$r_{\mathscr{I}}$. The origin is not
a physical boundary, but the artifact of the choice of spherical
coordinates.  The terms containing~$1/R$ become singular as~$R
\rightarrow 0$, but the~$1/R$ coefficient appears multiplying
only~$\phi_R$, which must vanish for~$r\rightarrow 0$. This is due to
regularity at the origin. If we were to extend all three
variables~$(\psi,\phi_R,\pi)$ from~$[0,r_{\mathscr{I}}]$
to~$[-r_{\mathscr{I}},r_{\mathscr{I}}]$ (equivalent to considering
positive~$r$ at~$\phi\to\phi+\pi$), parity of the fields requires
that~$\psi$ and~$\pi$ (scalars) be even functions of~$r$,
and~$\phi_R$, as the radial derivative of a scalar, be an odd
function. The result is that~$\p_t{\phi}_R = 0$ at the origin for all
times. Applying l'H\^opital's rule gives~$\phi_R/R \rightarrow
\phi_R'/R'$ as~$r \rightarrow 0$ and the equations at the origin
become
\begin{align}
\p_t \psi(t,0) = & - \pi(t,0) \, , \nonumber\\
0 = & -  \p_r \pi(t,0) \, , \nonumber\\
\p_t \pi(t,0) = & - 3 \phi_R'(t,0) + F \psi(t,0) \, ,
\end{align}
both in canonical spherical polars and in hyperboloidal coordinates.
The~$\gamma_2$ terms vanish because, due to the parity condition
above,~$\p_r \psi = 0$ and~$\phi_R = 0$ at the origin, while~$R'-1 =
0$ at the origin by definition. The second equation, therefore, gives
just an identity.

Equations~\eqref{ppp-hyp-coord}, which use hyperboloidal coordinates,
can be rewritten in terms of the incoming and outgoing characteristic
variables
\begin{align}\label{characteristic_variables}
  \sigma^+ \equiv - \pi + \phi_R, \qquad \sigma^- \equiv - \pi -
  \phi_R \,,
\end{align}
resulting in
\begin{align}
  \p_t \psi &=  \tfrac{1}{2}(\sigma^+ + \sigma^-) \, ,
  \nonumber\\
\p_t \sigma^+ &=  \frac{1}{2R'-1} \left[ \p_r \sigma^+
  + \frac{R'}{R} (\sigma^+ - \sigma^-) - R' F \psi \right]
  \nonumber\\
&\quad + \gamma_2 \left[ \frac{1}{2R'-1} \left( \p_r \psi
  + \frac{\sigma^-}{2} \right) - \frac{\sigma^+}{2} \right] \, ,
  \nonumber\\
\p_t \sigma^- &=  - \left[ \p_r \sigma^- + \frac{R'}{R}
  (\sigma^- - \sigma^+) + R' F \psi \right] \nonumber\\
&\quad - \gamma_2 \left[ \left( \p_r \psi + \frac{\sigma^-}{2} \right)
  - (2R'-1) \frac{\sigma^+}{2} \right] . 
\label{pss-hyp-coord}
\end{align}
Here again, the~$\gamma_2$ terms are proportional to the reduction
constraint. The equivalent
transformation for the flat equations~\eqref{ppp-flat-coord} can be
straightforwardly obtained from~\eqref{pss-hyp-coord} by
substituting~$R'\to 1$, $r\to R$ and~$t\to T$.

From \eqref{pss-hyp-coord} we see that
\begin{align}
  \phi_R(t,0) = 0 \Rightarrow \sigma^+(t,0) = \sigma^-(t,0) = -\pi(t,0)
\end{align}
and
\begin{align}
  \p_r \pi(t,0) = 0 \Rightarrow \p_r \sigma^+(t,0)
  = - \p_r \sigma^-(t,0) = \p_r \phi_R(t,0) , \end{align}
yielding
\begin{align}
  \p_t \psi (t,0) = & \sigma^+ (t,0) = \sigma^- (t,0) \, ,
  \nonumber\\
  \p_t \sigma^+(t,0) = & - 3 \p_r \sigma^-(t,0) - F \psi(t,0) \, ,
  \nonumber\\
  \p_t \sigma^-(t,0) = & - 3 \p_r \sigma^-(t,0) - F \psi(t,0)
= \p_t \sigma^+(t,0) \, ,\label{pss-hyp-coord_origin}
\end{align}
both in flat and hyperboloidal coordinates. The last equation makes
sense because~$\sigma^+(t,0) = \sigma^-(t,0)$ for all
times~$t$. 

\subsection{Regularization}

We now look at the behavior of the solutions near~$\mathscr{I}^+$, and
examine how it may be used to regularize terms appearing in the field
equations with~$R'$. For example, in
Eqs.~\eqref{pss-hyp-coord}~$(R'/R)\phi_R$ and potential terms appear,
with coefficients that become singular at~$r_{\mathscr{I}}$. But on
the other hand, we expect that the field variables~$(\psi,\phi_R,\pi)$
fall off as positive powers of~$1/R$ when~$r \rightarrow
r_{\mathscr{I}}$. Thus in order to regularize these terms, we seek a
suitable rescaling of the field variables. The expectation is that the
presence of a physically reasonable potential does not induce slower
decay towards~$\mathscr{I}^+$ than for a solution of the
LWE. Therefore we rescale our variables according to expected decay
rates for the LWE regardless of the form of~$F$, to be set later on.

\subsubsection{$(\psi,\phi_R,\pi)$ variables}

A scalar field~$\phi$ obeying the LWE falls off like~$1/R$
towards~$\mathscr{I}^+$. This suggests the rescaling
\begin{align}
  \tilde{\psi} \equiv \chi \psi \, , \quad \tilde{\phi}_R \equiv \chi
  \phi_R + \chi' \psi \, , \quad \tilde{\pi} \equiv \chi \pi \, ,
\end{align}
with~$\chi=\chi(R)\simeq R$ for large~$R$, and~$\chi\simeq 1$ near the
origin. These conditions ensure that the equations are unaffected at
the origin, but that the evolved variables become~$O(1)$
at~$\mathscr{I}^+$. As in~\cite{GasGauHil19} we take~$\chi \equiv
\sqrt{1+R^2}$. The system satisfies
\begin{align}
\p_t \tilde{\psi} &=  - \tilde{\pi} \, , \nonumber\\
\p_t \tilde{\phi}_R &=  - \frac{R'}{2R'-1} \p_r \tilde{\pi}
- \frac{R'-1}{2R'-1} \p_r \tilde{\phi}_R - \frac{2 R' (R'-1)}{(2R'-1)
  \chi^2 R} \tilde{\phi}_R \nonumber\\
&\quad + \frac{R'(R'-1)}{2R'-1} \left( \frac{3}{\chi^4} + F \right)
\tilde{\psi} + \frac{\gamma_2 R'}{2R'-1} \left[ \p_r \tilde{\psi} \right.
  \nonumber\\
  &\quad \left. + (R'-1) \tilde{\pi} - R' \tilde{\phi}_R  \right] \, ,
\nonumber \\
\p_t \tilde{\pi} &=  - \frac{R'-1}{2R'-1} \p_r \tilde{\pi}
- \frac{R'}{2R'-1} \p_r \tilde{\phi}_R
- \frac{2 R'^2}{(2R'-1)\chi^2 R} \tilde{\phi}_R \nonumber\\
&\quad + \frac{R'^2}{2R'-1} \left( \frac{3}{\chi^4} + F \right)
\tilde{\psi} + \frac{\gamma_2 (R'-1)}{2R'-1}
\left[ \p_r \tilde{\psi} \right. \nonumber\\
&\quad \left. + (R'-1) \tilde{\pi} - R' \tilde{\phi}_R \right] \,.
\label{ppp-hyp-coord_rescaled}
\end{align}
The potential term remains singular at~$r_\mathscr{I}$ if~$F$ does not
fall faster than~$1/R$. If at large~$R$ we have~$F \sim
1/R^{1+\epsilon}$, with~$\epsilon > 0$, we can choose~$n$
in~\eqref{Compactification} such that~$1 < n < 1 + \epsilon$, which
gives~$R' F \rightarrow 0$ as~$r \rightarrow r_\mathscr{I}$ and the
term becomes regular. The constraint damping terms are also tricky,
since, at least naively, they require~$\gamma_2$ to fall-off very
fast.

\subsubsection{$(\psi,\sigma^+,\sigma^-)$ variables}

A change of variables that captures more sharply the fall-off of
solutions, and that generalizes to nonlinear systems by the use of
asymptotic expansions~\cite{Hor97,LinRod04}, is offered by
\begin{align}\label{pss_rescaling}
  \tilde{\psi} \equiv \chi \psi \, , \quad \tilde{\sigma}^+
  \equiv \chi^2 \sigma^+ \, , \quad  \tilde{\sigma}^-
  \equiv \chi \sigma^- \, .
\end{align}
These variables satisfy the equations of motion
\begin{align}
  \p_t \tilde{\psi} &=  \frac{1}{2} \left[ \frac{\tilde{\sigma}^+}{\chi}
    + \tilde{\sigma}^- \right] \, , \nonumber\\
  \p_t \tilde{\sigma}^+ &=  \frac{1}{2R'-1} \left[ \p_r \tilde{\sigma}^+
    - \frac{2RR'}{\chi^2} \tilde{\sigma}^+ + \frac{R'}{R}
    \left( \tilde{\sigma}^+ - \chi \tilde{\sigma}^- \right) \right.
    \nonumber\\
    &\quad \left. - R' \chi F \tilde{\psi} \right] + \gamma_2
  \left[ \frac{1}{2R'-1}
    \left( \chi \p_r \tilde{\psi} - \frac{RR'}{\chi} \tilde{\psi}
    \right. \right. \nonumber\\
    &\quad \left. \left. + \chi \frac{\tilde{\sigma}^-}{2} \right)
    - \frac{\tilde{\sigma}^+}{2} \right] \, ,\nonumber\\
  \p_t \tilde{\sigma}^- &=  - \p_r \tilde{\sigma}^-
  + \left( \frac{RR'}{\chi^2} - \frac{R'}{R} \right) \tilde{\sigma}^-
  + \frac{R'}{R \chi} \tilde{\sigma}^+
  - R' F \tilde{\psi} \nonumber\\
  &\quad - \gamma_2 \left[ \p_r \tilde{\psi} - \frac{R R'}{\chi^2}
    \tilde{\psi} + \frac{\tilde{\sigma}^-}{2} - (2R'-1)
    \frac{\tilde{\sigma}^+}{2 \chi} \right] \, ,
  \label{pss-hyp-coord_rescaled}
\end{align}
in hyperboloidal coordinates. As before, the potential term can be
regularized only if~$F \sim 1/R^{1 + \epsilon}$, with~$\epsilon > 0$,
for large~$R$. Otherwise all terms, except the second on the right
hand side of the third equation are regular at~$r_\mathscr{I}$. A
closer inspection, however, shows that these two singular terms in
fact cancel each other at~$\mathscr{I}^+$, rendering the equations
regular. Regularity at the origin follows by the same considerations
as in the previous section. The regularization scheme for
the~$(\psi,\sigma^+,\sigma^-)$ system is sharper and simpler than that
for the~$(\psi,\phi_R,\pi)$ variables, and is thus preferred for
numerical work. Other advantages are that the~$\gamma_2$ constraint
terms are regular with no further thought, and, as we shall see, the
energy norm it provides is simpler than that of
the~$(\psi,\phi_R,\pi)$ system.

Inspired by~\cite{Eva84,CalNei04,NeiLehSar04}, for later application
in our numerical setup, we define the operator~$\tilde{\p}_r$ as:
\begin{align}\label{ptilde_r_1}
  \tilde{\p}_r\phi &\equiv \chi^2(\p_r + 2R'/R)(\phi/\chi^2) \, .
\end{align}
Using this operator, we can avoid the explicit appearance of the
term~$2R'/R$ which is singular at~$r_\mathscr{I}$ as~$\tilde{\p}_r$
corresponds to
\begin{align}\label{ptilde_r_2}
\tilde{\p}_r \phi =  \p_r \phi + \frac{2R'}{(1+R^2)R} \phi \, .
\end{align}
The motivation behind using this operator will become even more
apparent in the next section. With this definition we can write,
\begin{align}
  R'/R = \tfrac{1}{2}(\chi^{-2} \tilde{\p}_r \chi^{2} - \p_r) \, .
\end{align}
Substituting in~\eqref{pss-hyp-coord_rescaled} yields
\begin{align}
  \p_t \tilde{\psi} &= \frac{1}{2} \left[ \frac{\tilde{\sigma}^+}{\chi}
    + \tilde{\sigma}^- \right] \, , \nonumber\\
  \p_t \tilde{\sigma}^+ &=  \frac{1}{2R'-1}
  \left[ \left( \frac{\p_r + \tilde{\p}_r}{2} \right) \tilde{\sigma}^+
    + \chi \left( \frac{\p_r - \tilde{\p}_r}{2} \right)
    \tilde{\sigma}^- \right. \nonumber\\
  & \left. - \frac{RR'}{\chi^2} \tilde{\sigma}^+
    - \frac{RR'}{\chi} \tilde{\sigma}^-
    - R' \chi F \tilde{\psi} \right]
    + \gamma_2 \left[ \frac{1}{2R'-1} \right. \nonumber\\
  & \left. \left( \chi (\p_r \tilde{\psi})
    - \frac{R}{\chi} R' \tilde{\psi}
    + \chi \frac{\tilde{\sigma}^-}{2} \right)
    - \frac{\tilde{\sigma}^+}{2} \right] \, , \nonumber\\
  \p_t \tilde{\sigma}^- = & - \left[ \left(
    \frac{\p_r + \tilde{\p}_r}{2} \right)
    \tilde{\sigma}^- + \left( \frac{\p_r - \tilde{\p}_r}{2 \chi} \right)
    \tilde{\sigma}^+ \right. \nonumber\\
  & \left. - \frac{RR'}{\chi^3} \tilde{\sigma}^+
    + R' F \tilde{\psi} \right] - \gamma_2
    \left[ \p_r \tilde{\psi} - \frac{R}{\chi^2}
    R' \tilde{\psi} \right. \nonumber\\
  & \left. + \frac{\tilde{\sigma}^-}{2} - (2R'-1)
    \frac{\tilde{\sigma}^+}{2 \chi} \right] \, .
  \label{pss_rescaled_chi_chi_sq_cont}
\end{align}
At~$r_\mathscr{I}$, setting~$\gamma_2 \simeq 1/R$, the equations take
the form
\begin{align}
\p_t \tilde{\psi} = & \frac{\tilde{\sigma}^-}{2} \, , \nonumber\\
\p_t \tilde{\sigma}^+ = & - \frac{\tilde{\sigma}^-}{2}
- \frac{\chi F \tilde{\psi}}{2} \, , \nonumber\\
\p_t \tilde{\sigma}^- = & - \left[ \left(
  \frac{\p_r + \tilde{\p}_r}{2} \right) \tilde{\sigma}^-
  - \frac{R'}{\chi^2} \tilde{\sigma}^+
  + R' F \tilde{\psi} \right] \nonumber\\
& - \gamma_2 \left[ - \frac{R'}{\chi} \tilde{\psi}
  - (2R'-1) \frac{\tilde{\sigma}^+}{2 \chi} \right] \, .
\end{align}
Again, we see that the potential terms are singular
at~$r_\mathscr{I}$, unless~$F \sim 1/R^{1+\epsilon}$, with~$\epsilon >
0$, for large $R$.

\subsection{Conserved energy on hyperboloidal slices}
\label{Conserved_Energy}

As we saw in the equations of motion, if~$F$ falls off too slowly it
may result in singular equations near~$\mathscr{I}^+$, even when
working with the rescaled variables. A classical example is~$F=m^2$,
corresponding to the massive Klein-Gordon equation. As pointed out by
Winicour~\cite{Win88} with a conformal approach, whatever rescaling we
take, the mass term always remains singular at~$\mathscr{I}^+$. This
can lead to numerical problems, such as blow-up or a lack of
convergence. One way of trying to tackle such issues is to derive
special algorithms that respect physical restrictions, to make sure
that errors are well-behaved and that the code converges at the
desired accuracy. One constraint of physical interest is provided by
the energy conservation. If such a conserved energy is available, we
can utilize it as an additional constraint on the dynamics of the
field. In this subsection, we derive a conserved energy at the
continuum level. Later, in section~\ref{SBP_Scheme}, we will construct
an approximation to this norm in our discretization.

\subsubsection{Conserved energy with the original variables}

Consider the functional~$T_{\mu\nu}[\psi]$, with~$\psi$ satisfying the
LWEP, given by
\begin{align}\label{LMKGE-T-mu-nu}
  T_{\mu\nu} = \p_\mu \psi \p_\nu \psi - \tfrac{1}{2} g_{\mu\nu}
  ( \p^\alpha \psi \p_\alpha \psi + F \psi^2) \, ,
\end{align}
which we will refer to as the stress-energy tensor. The
time-independent potential~$F$ is non-negative for all~$R$, bounded,
and~$C^k$ for large~$k$. In our setup the stress-energy tensor is not
necessarily covariantly conserved but, for the purposes of this work,
its utility comes down to the fact that it provides coercive estimates
on solutions to the field equations. To see this, we follow the
standard steps of the vector-field method. A clear introduction to
this approach can be found in~\cite{Are13}. Consider a vector
field~$K^\mu$ and contract it with~$T_{\mu\nu}$. Taking the total
divergence and using the product rule, we get
\begin{align}
  \nabla^\mu (T_{\mu\nu} K^\nu) = (\nabla^\mu T_{\mu\nu}) K^\nu
  + T_{\mu\nu} \nabla^\mu K^\nu \, .
\end{align}
Using Eq.~\eqref{LWEP} in the first term on the right-hand side we
obtain
\begin{align}\label{Div-Tmunu-Vnu}
  \nabla^\mu (T_{\mu\nu} K^\nu) = - \tfrac{1}{2} K^\nu(\p_\nu F)
  \psi^2 + T_{\mu\nu} \nabla^\mu K^\nu \, .
\end{align}

\begin{figure}
\includegraphics[scale=0.5]{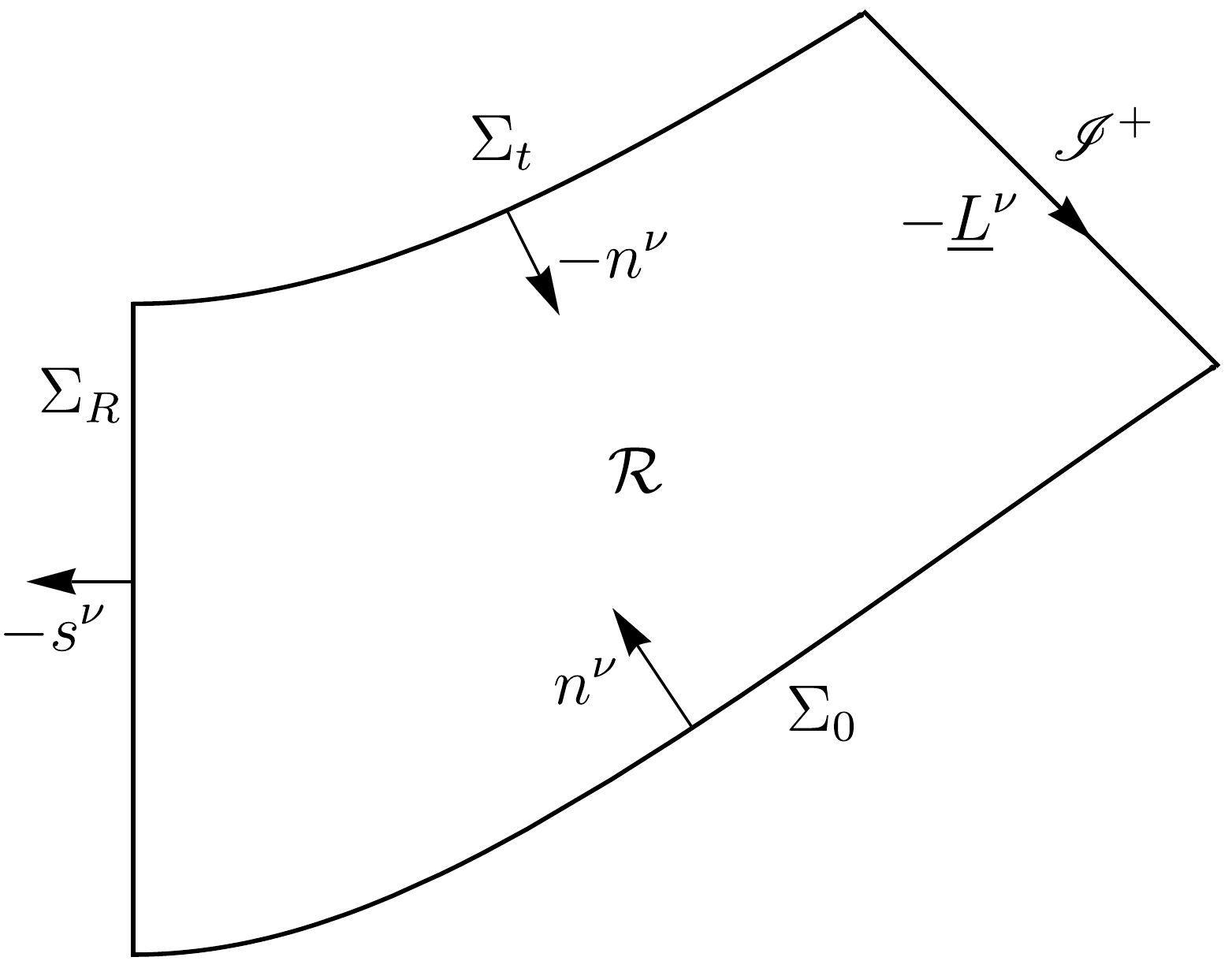}
\caption{Diagram depicting the spacetime region where Stoke's theorem
  is applied. The normal vectors follow the conventions of
  \cite{Are13}.}
\label{Stokes_diag}
\end{figure}

As shown in Fig.~\ref{Stokes_diag}, consider a region~$\mathcal{R}$
surrounded by a boundary~$\mathcal{\p R}$ consisting of the initial
hyperboloidal slice~$\Sigma_0$, some later hyperboloidal
slice~$\Sigma_t$, with~$t>0$, inner timelike constant radial
boundary~$\Sigma_R$ and future null
infinity~$\mathscr{I}^+$. Integrating~\eqref{Div-Tmunu-Vnu}
over~$\mathcal{R}$ and applying Stoke's theorem on the left side of
the equation yields
\begin{align}
  & \int_{\Sigma_0} T_{\mu\nu} K^\mu n^\nu
  + \int_{\Sigma_t} T_{\mu\nu} K^\mu (-n^\nu)
  + \int_{\Sigma_R} T_{\mu\nu} K^\mu (-s^\nu) \nonumber\\
  & + \int_{\mathscr{I}^+} T_{\mu\nu} K^\mu (-\underline{L}^\nu)
  = \int_\mathcal{R} [T_{\mu\nu} \nabla^\mu K^\nu
    - \tfrac{1}{2}  K^\nu(\p_\nu F) \psi^2] \, ,
\end{align}
where~$\underline{L}^\nu$ is the ingoing null vector
at~$\mathscr{I}^+$ as defined in Eq.~(5) in~\cite{GasGauHil19},
and~$s^\nu$ is the spatial normal vector to~$\Sigma_R$. The first term
in the bulk integral on the right vanishes if~$K^\mu$ is a Killing
vector. Taking furthermore~$K^\mu$ causal and recalling our
restriction on the sign of the potential, it follows that the
integrand on the two spatial slices~$\Sigma_0$ and~$\Sigma_t$ is sign
definite. If~$K^\mu = (\p_T)^\mu$, we get~$K^\nu(\p_\nu F)=\p_T F=0$,
resulting in a vanishing bulk integral. Henceforth we make this
choice. Taking~$R=0$ at the inner boundary makes the~$\Sigma_R$
integral vanish. What remains is
\begin{align}
  \int_{\Sigma_0} T_{\mu\nu} K^\mu n^\nu
  - \int_{\Sigma_t} T_{\mu\nu} K^\mu n^\nu
  - \int_{\mathscr{I}^+} T_{\mu\nu} K^\mu \underline{L}^\nu = 0 \, .
\label{Energy_Conservation}
\end{align}
From the line element~\eqref{lowercase_metric}, we can compute
\begin{align}
n^{\underline{\nu}'} &= \alpha(1,1-1/R',0,0),\qquad
\underline{L}^{\underline{\nu}} = (\p_T - \p_R)^{\underline{\nu}'}.
\end{align}
Substituting these into~\eqref{Energy_Conservation},
using~\eqref{LMKGE-T-mu-nu} and, moving now to work with our first
order reduction (using~$\pi$ and~$\phi_R$), we obtain
\begin{align}
  & \left( \int_{\Sigma_t} - \int_{\Sigma_0} \right) \frac{\alpha}{2}
  \left[\pi^2 - 2 \frac{R'-1}{R'} \pi \phi_R + \phi_R^2
    + F \psi^2 \right] \nonumber\nonumber\\
  &\qquad\qquad\qquad\quad = - \tfrac{1}{2}
  \int_{\mathscr{I}^+} \left[ (\sigma^-)^2
    + F \psi^2 \right] \leq 0 \, .\label{en_cons_int_form}
\end{align}
The reduction admits the natural analog of~\eqref{Energy_Conservation}
when~$\gamma_2=0$, which we assume henceforth. Because~$F \geq 0$
at~$\mathscr{I}^+$, the right side of~\eqref{en_cons_int_form} is
negative semi-definite. Thus the energy on our hyperboloidal slices,
given by the integrals on the left side, can leak out only
through~$\mathscr{I}^+$, and remains conserved only when the right
side is identically zero. The flux of radiation
through~$\mathscr{I}^+$ will depend on the form of~$F$, which plays a
crucial role in the dynamics of the field. In all cases, the energy on
the subsequent hyperboloidal slices is always upper bounded by the
initial energy. Therefore, integrating out the trivial dependence
on~$\theta^A$, the angular coordinates, we consider the energy norm
\begin{align}
  E(t) = \int_{\Sigma_t} \varepsilon dr \,.
  \label{cons_en_hyp_slice}
\end{align}
Depending on our choice of variables, we write either
\begin{align}
  \varepsilon &=  \frac{1}{2} \left[ \pi^2 - 2 \left( \frac{R' - 1}{R'}
    \right) \pi \phi_R + \phi_R^2 + F \psi^2 \right] R^2 R'\,,
\end{align}
or
\begin{align}
  \varepsilon
  &=  \frac{1}{2} \left[ \left( \frac{2 R' - 1}{2 R'} \right) (\sigma^+)^2
    + \left( \frac{1}{2 R'} \right) (\sigma^-)^2 + F \psi^2 \right]
  R^2 R' \,.\label{cons_en_density_hyp_slice}
\end{align}
Note the rather unusual convention, in which we are absorbing the
volume-form into the symmetrizer (the matrix that representing the
quadratic form). Some care is needed to obtain the flux at
infinity. Let us thus temporarily truncate the slices at an outer
radius~$r_o \leq r_{\mathscr{I}}$ and take~$r_o$ to be some function
of time~$t$. We obtain then~$E(t) = E(t,r_o(t))$ and
\begin{align}
  \frac{d}{dt}E(t,r_o) = \left. \p_t E(t,r_o)
  + \p_{r_o} E(t,r_o) \cdot \frac{dr_o}{dt}
  \right|_{r = r_o} \,. \label{time_deriv_cons_en}
\end{align}
Substitution from the equations of motion and integrating by parts
gives for the first term
\begin{align}
  \p_t E(t,r_o) = - R^2 \phi_R \pi|_{r = r_o} = \left.
  \tfrac{1}{4} R^2 [(\sigma^+)^2 - (\sigma^-)^2] \right|_{r = r_o} \, ,
  \label{pE_pt_continuous}
\end{align}
while from the definition of~$\varepsilon$, we get
\begin{align}
  \p_{r_o} E(t,r_o) = \lim_{\delta r_o \rightarrow 0}
  \frac{1}{\delta r_o}
  \left( \int_0^{r_o + \delta r_o} - \int_0^{r_o} \right)
  \varepsilon dr = \varepsilon(t,r_o) \, .
\end{align}
If the outer boundary of the system is a timelike constant radius
worldtube then~$dr_o/dt = 0$, the second term on the right
of~\eqref{time_deriv_cons_en} vanishes. If, instead, it is an incoming
null curve, so that~$\dot{r}_o = dr_o/dt = c^r_- = -1/(2R'-1)$, we
obtain
\begin{align}
\frac{d}{dt} E(t,r_o)
&=  - \left. \tfrac{1}{4} R^2 \frac{2R'}{2R'-1} \left[ F \psi^2
  + (\pi + \phi_R)^2 \right] \right|_{r = r_o} \nonumber\\
&=  - \tfrac{1}{4} R^2  \frac{2R'}{2R'-1} \left[ F \psi^2
  + (\sigma^-)^2 \right] \Bigg|_{r = r_o} \, .\label{dE_dt_cont}
\end{align}
In the limit~$r_o \rightarrow
r_{\mathscr{I}}$,~$2R'/(2R'-1) \rightarrow 1$,
we recover
\begin{align}
  \frac{d}{dt}E(t,r_{\mathscr{I}}) &=
  - \left. \tfrac{1}{4} R^2 \left[ F \psi^2
  + (\pi + \phi_R)^2 \right] \right|_{r = r_\mathscr{I}} \nonumber\\
  &=
  - \left. \tfrac{1}{4} R^2 \left[ F \psi^2
    + (\sigma^-)^2 \right] \right|_{r = r_\mathscr{I}} \, ,
  \label{en_cons_diff_form}
\end{align}
consistent with the right side of Eq.~\eqref{en_cons_int_form} - note
that a factor of~$1/2$ appeared above because~$\delta t = 2 \delta u$
there. In our case, we take~$\dot{r}_o = c^r_-$
because~$\mathscr{I}^+$ is incoming null. Note here that the right
hand side of this expression should still be understood in a limiting
sense (as~$r_o \to r_\mathscr{I}$). We will avoid this complication in
the following by the use of rescaled variables. In deriving
Eq.~\eqref{en_cons_diff_form}, the second term on the right-hand side
of~\eqref{time_deriv_cons_en} is important. Since the hyperboloidal
slices meet~$\mathscr{I}^+$, rather than~$i^0$, it appears naively
that if we had just taken~$\dot{r}_o = 0$ and let~$r_o \rightarrow
r_\mathscr{I}$, we would still get the correct expression
for~$dE/dt$. Contrarily, as the foregoing discussion shows, this is
not true.

\subsubsection{Conserved energy with rescaled variables}

In~\eqref{en_cons_diff_form}, as~$r_o \rightarrow r_{\mathscr{I}}$,~$R
\rightarrow \infty$ and it becomes difficult to express~$dE/dt$ in
closed form. But if we recall the rescaled
variables~\eqref{pss_rescaling},
\begin{align}
  \tilde{\psi} = \chi \psi \, , \quad \tilde{\sigma}^+ = \chi^2
  \sigma^+ \, , \quad \tilde{\sigma}^- = \chi \sigma^- \, ,
\end{align}
where~$\chi=\sqrt{1+R^2}$, all the coefficients in the above
expressions in fact regularize except possibly the potential
term. Examples that will be used later in the derivations are
\begin{align}
  \varepsilon = & \frac{1}{2} \left[F \tilde{\psi}^2 R'
    + \left( \frac{2 R' - 1}{2 \chi^2} \right) (\tilde{\sigma}^+)^2
    + \frac{(\tilde{\sigma}^-)^2}{2}  \right] \frac{R^2}{\chi^2}
  \label{cons_en_dens_hyp_slice_resc_pss}
\end{align}
for the energy density,
\begin{align}
  &\p_{r_o} E(t,r_o(t)) \cdot \frac{dr_o}{dt} =
  - \frac{1}{2} \frac{1}{2R'-1}
  \cdot\nonumber\\
  &\quad 
  \left[ F \tilde{\psi}^2 R' + \left( \frac{2 R' - 1}{2 \chi^2} \right)
    (\tilde{\sigma}^+)^2 + \frac{(\tilde{\sigma}^-)^2}{2} \right]
  \frac{R^2}{\chi^2}\Bigg|_{r = r_o}
\end{align}
for the moving-boundary term, and
\begin{align}\label{pE_pt_pss_resc}
  \p_t E(t,r_o) = \frac{1}{4}
  \left[ \frac{\tilde{\sigma}^{+2}}{\chi^2} - \tilde{\sigma}^{-2} \right]
  \frac{R^2}{\chi^2} \Bigg|_{r = r_o}
\end{align}
for the remaining boundary term. In the limit~$r_o = r_{\mathscr{I}}$,
we have~$R^2/\chi^2 \rightarrow 1$ and thus get
\begin{align}\label{en_cons_diff_form_resc}
  \frac{d}{dt}E(t) = - \frac{1}{4} \left( (\tilde{\sigma}^-)^2
  + F \tilde{\psi}^2 \right) \bigg|_{r = r_\mathscr{I}} \, .
\end{align}
The potential term in~$\varepsilon$ is thus still singular if~$F$ does
not fall off fast enough. In that case, we still can choose initial
data such that the product~$R' F \psi^2$ is finite
at~$r_\mathscr{I}$. Using inequality~\eqref{en_cons_int_form}, we can
make sure that this whole term remains regular at~$r_{\mathscr{I}}$
for all times~$t$.

\section{Summation by Parts Scheme on the Hyperboloidal Slices}
\label{SBP_Scheme}

Having laid out the continuum setup above, in this section we now
present our discrete approximation. This involves the evolution
equations, the conserved (up to boundary fluxes) energy, the use of
regularized variables and discrete operators satisfying SBP and TEM.

\subsection{SBP and TEM overview}\label{SBP_TEM_Overview}

The energy estimate~\eqref{en_cons_diff_form_resc} shows that for the
continuum equations, the size of the solution at any time is bounded
above by the size of the initial data plus an integral of the flux of
radiation that leaves the domain through~$\mathscr{I}^+$. The key idea
of a summation by parts scheme is to discretize such a system so that
the semidiscrete equations admit a similar estimate. Examples of the
use of SBP schemes in numerical relativity include~\cite{CalLehNei03,
  CalNei03,CalLehReu03,SeiSziPol08,TayKidTeu10}. See the
review~\cite{SarTig12} for a more thorough discussion. We now give a
brief summary of how that is achieved. Consider a first order linear
symmetric hyperbolic system for a state vector~$\mathbf{u}$ with~$k$
components, each a scalar quantity on spacetime. The equation of
motion is then written,
\begin{align}
  \p_t\mathbf{u}&= \mathbf{A}^p(x^\mu) \p_p\mathbf{u} +
  \mathbf{S}(x^\mu)\,, \label{eqn:FO_SH_system}
\end{align}
where~$x^\nu$ are the spacetime coordinates. Here~$p$ is summed over
all spatial index values, and the principal part
matrices~$\mathbf{A}^p$ and source terms~$\mathbf{S}(x^\mu)$ have the
obvious dimensionality. We use boldface symbols to represent objects
and operators with the dimensionality of the state vector. Symmetric
hyperbolicity means that there exists a symmetrizer, a symmetric
positive definite matrix~$\mathbf{H}$, with the
product~$\mathbf{H}\mathbf{A}^p$ symmetric for each~$p$. Suppose that
we solve the initial boundary value problem for this system on a
compact spatial domain~$V(t)$ with boundary~$\p V(t)$. Then we have
the energy, 
\begin{align}
  E(t) =  \int_{V(t)}\tfrac{1}{2} \mathbf{u}^T\mathbf{H}\mathbf{u}\,,
  \label{eqn:FO_SH_energy}
\end{align}
where the superscript~`$T$' denotes the matrix transpose. This energy
norm satisfies
\begin{align}
  \frac{d}{dt} E(t)&=\int_{\p V(t)}
  \tfrac{1}{2}\mathbf{u}^T\mathbf{H}(\mathbf{A}^ps_p+v^ps_p\mathbf{1})
  \mathbf{u}+\textrm{Bulk
    Term}\,,\label{eqn:FO_SH_energy_balance}
\end{align}
where the bulk term, an integral over~$V(t)$, can in general be seen
to be bounded using a combination of the Gr\"onwall and Cauchy-Schwarz
inequalities,~$v^p=\p_tx^p(t)$ denotes the local velocity of the outer
boundary, and~$s_p$ denotes the outward pointing unit normal to the
domain at the boundary. We assume, as in our hyperboloidal setup, that
the bulk term vanishes, and that the boundary integral is
non-positive. Now discretize the equations first by introducing the
spatial grid~$x^p_I$, with index~$I$ labeling the grid points. We
replace the continuum state vector~$\mathbf{u}$ by a discrete
analog~$\mathbf{U}$ that lives on the grid, and thus has
components~$U^{\alpha}_I$ for each continuum field, with~$\alpha$ here
labeling the different continuum fields. We need an approximation to
the spatial derivative~$\p_p$, which we denote here as~$D_i$. In our
specific setup, this last step is more subtle because,
following~\cite{Eva84,CalNei04,NeiLehSar04}, the use of
shell-coordinates (spherical polars) requires us to introduce two
different approximations. But to illustrate the general approach we
sweep non-essential complications under the carpet at this stage. We
work with the semidiscrete approximation, writing the large collection
of ODEs for the components of~$\mathbf{U}$ as,
\begin{align}
  \frac{d}{dt}\mathbf{U}
  &= \mathbf{A}^p(t,x_I)D_p\mathbf{U}+\mathbf{S}(t,x_I)\,.
\end{align}
At this point different options are available, and we will choose the
simplest. See~\cite{CalLehReu03} for a discussion of the alternatives.
Consider now the discrete approximation to~\eqref{eqn:FO_SH_system},
given by the sum over grid points
\begin{align}
  E(t)=\tfrac{1}{2} (\mathbf{U},\mathbf{U})_{\mathbf{H}} \equiv
  \tfrac{1}{2} \sum_I \mathbf{U}_I^T \Upsilon_I\mathbf{H}_I
  \mathbf{U}_I\,,
\end{align}
where~$\mathbf{H}_I=\mathbf{H}(t,x_I)$
and~$\Upsilon_I\equiv\Upsilon_I\mathbf{1}$, which we call the
quadrature or quadrature matrix ($\mathbf{1}$ here is the~$k\times k$
identity matrix associated with the state space) encodes information
about the local gridspacing at point~$I$. For simplicity the norm is
taken to be diagonal over grid points. Computing the time derivative we
get,
\begin{align}
  \frac{d}{dt}(2E(t))&=
  (\mathbf{A}^pD_p\mathbf{U},\mathbf{U})_{\mathbf{H}}
  +(\mathbf{U},\mathbf{A}^pD_p\mathbf{U})_{\mathbf{H}}\nonumber\\
  &\quad+((\ln \Upsilon)\dot{}\,\mathbf{U},\mathbf{U})_{\mathbf{H}}
  +\textrm{Bulk Term}\,,
\end{align}
with the shorthand~$((\ln \Upsilon)\dot{}\,\mathbf{U})_I= (d/dt(\ln
\Upsilon_I))\,\mathbf{U}_I$. Now observe that the discretization can
be carefully chosen so that
\begin{align}
  &(\mathbf{A}^pD_p\mathbf{U},\mathbf{U})_{\mathbf{H}}
  +(\mathbf{U},\mathbf{A}^pD_p\mathbf{U})_{\mathbf{H}}
  +((\ln \Upsilon)\dot{}\,\mathbf{U},\mathbf{U})_{\mathbf{H}}
  \nonumber\\
  &=(\mathbf{U},(\mathbf{A}^ps_p+\mathbf{1}v^ps_p)
  \mathbf{U})_{\mathbf{H}, \p V}
  +\textrm{Bulk Term}\,,\label{eqn:SBP_defn}
\end{align}
where it must be possible to bound the bulk term by~$E(t)$ multiplied
by a constant that is independent of resolution, encoded in our
notation by~$\Upsilon_I$, times~$E(t)$. Here we have defined a
boundary inner product and associated norm,
\begin{align}
  (\mathbf{U},\mathbf{V})_{\mathbf{H},\p V}&=
  \sum_B \mathbf{U}_B^T \Upsilon_B\mathbf{H}_B \mathbf{V}_B\,,
\end{align}
with the sum here taken over the set of boundary points, denoted
throughout by an index~$B$. We may then conclude that the continuum
energy conservation equation~\eqref{eqn:FO_SH_energy_balance} has the
semidiscrete analog
\begin{align}
  \frac{d}{dt}E(t)&=(\mathbf{U},(\mathbf{A}^ps_p+\mathbf{1}v^ps_p)
  \mathbf{U})_{\mathbf{H}, \p V}
    +\textrm{Bulk Term}\,.
\end{align}
If a condition like~\eqref{eqn:SBP_defn} is satisfied, the method is
called a summation by parts (SBP) scheme. Such a scheme has the
advantage that, under mild assumptions on the coefficient matrices and
source terms, and the use of suitable boundary conditions, solutions
of the approximation are guaranteed to converge to solutions of the
continuum system in the limit of infinite
resolution~\cite{Tho98c}. The specific rate of convergence is
determined by the choice of approximation to the spatial
derivative. In numerical relativity the two most popular choices are
to use a spectral approximation or finite differences.

Despite its strengths, naively applying the SBP approach may result in
a scheme with undesirable features.  For example, insisting that the
semidiscrete and continuum energies match exactly, say by careful
adjustment of the derivative operators near the boundary, can result
in the creation of numerical noise that propagates into the domain.
In the hyperboloidal setting we might already suspect that such noise
would be problematic, since everything about our coordinates is
engineered with the resolution of outgoing rather than incoming waves
in mind. One might counter that since the method would be guaranteed
to converge in some norm, we could just increase resolution to
suppress the noise, but several points stand against this
perspective. The final aim of our research program is to provide
gravitational waveforms at null-infinity. Since these waveforms will
be used for modeling they should be as clean as possible {\it
  pointwise} even at finite resolution. In other words the natural
mathematical measure of error provided by the problem may not
correspond with the notion of error required of the data in
applications. Second, the norm that the equations naturally
provide~\eqref{cons_en_dens_hyp_slice_resc_pss} in fact degenerates in
the incoming characteristic variable~$\tilde{\sigma}^+$, in that the
coefficient multiplying that variable goes to zero
near~$\mathscr{I}^+$, if the compactification parameter~$n<2$. In some
of the models in this paper we could choose~$n=2$. However, owing to
the presence of log-terms in the natural expansion
near~$\mathscr{I}^+$ in our gauge, no formulation is presently
available for GR in which~$n=2$ is permissible under our
approach. Observe, in passing that this degeneracy appears also on
null-slices even for the wave equation, and so is not surprising. We
aim therefore here to develop a method that satisfies semidiscrete
estimates like~\eqref{cons_en_dens_hyp_slice_resc_pss}, but which
minimizes dangerous reflections from the outer boundary.

To that end let us illustrate, as summarized nicely in~\cite{Pre04},
the utility of truncation-error-matching (TEM) by considering two
finite difference approximations to the first derivative. Suppose that
in the bulk domain we have a one-dimensional uniform grid with
spacing~$h$, and the approximation
\begin{align}
DF_I&= \tfrac{1}{2h}(F_{I+1}-F_{I-1}) \,,
\end{align}
to the first derivative using the arbitrary grid function~$F$, which
should not be confused with the potential, but that at the outer
boundary~$I=N$ we take one of
\begin{align}
  DF_N &=\tfrac{1}{2h}(3F_N-4F_{N-1}+F_{N-2})\,,\nonumber\\
  DF_N &= \tfrac{1}{2h}(4 F_N - 7 F_{N-1} + 4 F_{N-2}-F_{N-3}) \, .
\end{align}
Either choice results in a second order accurate approximation to the
first derivative. Assuming we are approximating with~$F$ a~$C^3$
continuum function~$f$ we can Taylor expand, and find, using the
standard little-oh notation, that the error coefficient takes the form
\begin{align}
DF_I &= f'(x_I) + \tfrac{1}{6}h^2f'''(x_I)+o(h^2)\,, &\quad I<N\,,
 \nonumber\\
DF_N&=  f'(x_I) - \tfrac{1}{3}h^2f'''(x_I)+o(h^2)\,,&\quad I=N\,.
\end{align}
in the first case and 
\begin{align}
DF_I &= f'(x_I) + \tfrac{1}{6}h^2f'''(x_I)+o(h^2)\,,
\end{align}
everywhere in the second. In the first case the coefficient in front
of the~$h^2$ error term is different, which will induce (convergent)
high-frequency noise, whereas in the second the approximation was
carefully chosen at the boundary so that the errors match up. In what
follows, we exploit this, the basic idea of TEM, to minimize
high-frequency reflections from~$\mathscr{I}^+$.

\subsection{Discretization}\label{Discretization}

All our derivations will be done for a non-staggered grid, which
includes a grid point at the boundaries. Let the radial coordinate~$r$
take discrete values~$\{r_0,\ldots,r_N\}$. We take a uniform grid with
step size~$h$, which gives
\begin{align}
r_I = Ih, \quad \mbox{with} \quad I = 0,\ldots,N \, ,
\end{align}
where~$N$ is a positive integer. Here,~$N$ corresponds to the grid
index at the outer boundary. Although the origin is not a physical
boundary point, it is convenient to treat it as a boundary while
defining the grid on the closed interval~$[0,r_o]$ and give boundary
conditions in terms of the parity of the various fields. As
before,~$r_o$ corresponds to the compactified radial coordinate at the
outer boundary.

We will define our discretization using a single grid variable~$\psi$
instead of the whole state vector~$\mathbf{u}$, since the basic idea
remains the same. Define~$\Psi_I(t) = \psi (t,r_I)$, assuming
that~$\psi(t,r)$ is a sufficiently smooth function of~$r$, and for
convenience, we drop the argument~$t$. Let~$\Psi$ to be the column
vector with~$\Psi_I$ as its~$I$th element. We express every linear
operator, e.g. the finite difference operator~$D$, acting on~$\Psi$ as
an~$(N+1) \times (N+1)$ matrix. While writing the discrete form of the
continuum equations, we express various coefficients that appear,
which are in general functions of~$r$, as~$(N+1) \times (N+1)$
matrices. This step will become clearer in the next subsection. In the
next paragraph, we describe the general notation and properties of
these multiplication operators.

We denote the operators in the approximation corresponding to various
coefficient functions of~$r$ in the equations of motion by writing
their continuum names in square brackets. For example, we denote
by~$[f(r)]$ the operator, or~$(N+1) \times (N+1)$ matrix,
corresponding to the function~$f(r)$ in the continuum limit. For
simplicity, we take these matrices to be diagonal with diagonal
entries~$[f(r)]_{II} = f(r_I)$. Being diagonal, all these operators
satisfy the same basic algebraic properties, such as commutativity, as
the corresponding continuum functions.

We define all our discrete norms using a centered grid, in which each
interval, of size~$h$, in the bulk is taken symmetrically about its
respective grid point. Therefore the boundary points are left with the
intervals of size~$h/2$ which lie only towards the bulk, so that the
sum of intervals remains~$Nh$. If the state vector~$\mathbf{U}$
contains only a single variable~$U^1=\Psi$, the quadrature reduces to
a~$1 \times 1$ matrix in the state space,~$\mathbf{\Upsilon} =
[\Upsilon]$. Here~$\Upsilon$ is a scalar in the state space but
an~$(N+1) \times (N+1)$ matrix in the grid space. For simplicity, we
take it to be a diagonal matrix
\begin{align}\label{Static_Quadrature}
\Upsilon = \textrm{diag}(h/2,h,\ldots,h,h/2)\,.
\end{align}
The same arguments apply to the symmetrizer as well, which takes the
form~$\mathbf{H} = [W]$, where~$W$ is just a scalar in the state space
and an~$(N+1) \times (N+1)$ diagonal matrix in the grid space
\begin{align}\label{W_diag}
W = \textrm{diag}(w_0,\ldots,w_N) \,.
\end{align}
These conventions lead to the following definition for the norm of
a single grid-function~$\Psi$:
\begin{align}\label{disc_norm}
||\Psi||_{\mathbf{H}} = \Psi^T \Upsilon W \Psi \,.
\end{align}
Assuming~$W$ is time independent, the only time dependence appears
in~$\Psi$.

In our system, we will face the situation in which the coordinate
position of the outer boundary is a continuous function of time. To
realize this in our numerics, we keep our grid uniform in the bulk,
with width~$=h$, but make the position of the last grid point a
continuous function of time, so that it moves with the outer
boundary. However, we impose that the maximum value the last grid
width can take is~$h$. If the outer boundary moves further, we create
a new,~$(N+1)$th grid point, at~$r_N$ at that instant which moves with
the outer boundary. If, let us say at time~$t_1$, the outer boundary
reaches a distance~$h$ away from~$r_N$ and still keeps moving outward,
this~$(N+1)$th grid point gets fixed there and a newer,~$(N+2)$th grid
point is created at~$r_{N+1} = (N+1)h$ at that instant, and so on. We
can model the reverse situation, in which the outer boundary moves
inwards, in exactly the reverse way. That is, when the last grid point
merges with the penultimate one, it vanishes and the penultimate one
becomes the last grid point, and so on.

The next problem is to incorporate the moving outer boundary in the
definition of the norm. As before, we keep the elements of~$W$ time
independent, but make its dimensionality a function of time. The
latter condition also applies to the quadrature~$\Upsilon$, but we
make its last entry a function of time by redefining it as
\begin{align}\label{Dynamic_Quadrature}
  \Upsilon = \textrm{diag}(h_0,\ldots,h_N) \, ,
\end{align}
with~$h_0 = h/2$ and~$h_I = h$ for~$I = 1,\ldots,N-1$. Here,~$h_N$
can only take values in~$0 < h_N \leq h$. Its relationship with the
creation or annihilation of the last grid point is `out of phase' as
follows. Whenever~$(r_N - r_{N-1}) > h/2$,~$\Upsilon$ is given
by~\eqref{Dynamic_Quadrature} with~$h_N$ given by
\begin{align}\label{h_N}
h_N = r_N - r_{N-1} - \frac{h}{2}
\end{align}
and the norm is given by~\eqref{disc_norm}. However, when~$(r_N -
r_{N-1}) \leq h/2$, we do not consider the contribution of the last
grid point to the norm, which is the same as removing the last row
of~$\Psi$,~$W$ and~$\Upsilon$ and the last column of~$W$
and~$\Upsilon$ in~\eqref{disc_norm}, with~$\Upsilon$ as given
by~\eqref{Dynamic_Quadrature}. In this case, the effect of the moving
boundary is captured by~$h_{N-1}$ and its value is given by
\begin{align}\label{h_N-1}
  h_{N-1} = r_N - r_{N-1} + \frac{h}{2} \quad
  \textrm{for } 0 \leq r_N - r_{N-1} \leq \frac{h}{2} \, .
\end{align}

In summary, for the case of a moving outer boundary, we define the
discrete norm by~\eqref{disc_norm}, taking~$\Psi =
(\Psi_0,\ldots,\Psi_M)^T$,~$W = \textrm{diag}(w_0,\ldots,w_M)$
and~$\Upsilon = \textrm{diag}(h_0,\ldots,h_M)$, with~$h_0 = h/2$,~$h_i
= h$ for~$i = 1,\ldots,M-1$. We take~$M = N$ whenever~$(r_N - r_{N-1})
> h/2$, in which case~$h_N$ is given by~\eqref{h_N}, and~$M = N-1$
whenever~$(r_N - r_{N-1}) \leq h/2$, with~$h_{N-1}$ given
by~\eqref{h_N-1}. On the initial slice, we set~$h_N = h/2$.

Therefore, the total time derivative of the norm becomes
\begin{align}\label{dNorm_dt}
  \frac{d}{dt} ||\Psi||_\mathbf{H} = 2 \, \Psi^T \Upsilon W \dot{\Psi}
  + \Psi^T \dot{\Upsilon} W \Psi \, ,
\end{align}
where~$\dot{\Psi} = d \Psi / dt$ and~$\dot{\Upsilon} =
d\Upsilon/dt$. The second term appears solely because of the moving
outer boundary. Using the chain rule, we get
\begin{align}
\dot{\Upsilon} = \frac{\p \Upsilon}{\p r_N} \dot{r}_N \, .
\end{align}
Substituting the definitions given in the previous paragraph along
with~\eqref{h_N} and~\eqref{h_N-1}, we obtain
\begin{align}\label{ups_dot}
\Psi^T \dot{\Upsilon} W \Psi = w_M \Psi_M^2 \, \dot{r}_N \, ,
\end{align}
with~$M = N$ for~$(r_N - r_{N-1}) > h/2$ and~$M = N-1$ for~$(r_N -
r_{N-1}) \leq h/2$. When the trajectory of~$r_N$ is of an incoming
radial null ray, we get
\begin{align}\label{r_N_dot}
\dot{r}_N = c^r_- \Big|_{r = r_N} = - \frac{1}{2R_N'-1} \, ,
\end{align}
where~$R_N' = R'(r_N)$. Therefore, when the outer boundary is
at~$r_\mathscr{I}$, this gives~$\Psi^T \dot{\Upsilon} W \Psi = 0$.

All these computations are easily generalized to a state
vector~$\mathbf{U}$ belonging to the higher dimensional state space by
working in a basis which diagonalize~$\mathbf{H}$
and~$\mathbf{\Upsilon}$ in that space. Since the
quadrature~$\mathbf{\Upsilon}$ depends only on the grid spacing and
not on the dynamical variables, it should be a scalar multiple of the
identity matrix acting on the state space. In that case, we can write
\begin{align}
\mathbf{H} = \textrm{diag}(W_1,\cdots,W_k)
\end{align}
where, as introduced in~\eqref{eqn:FO_SH_system}, $k$ is the dimension
of the state space.

\subsection{$(\psi,\sigma^+,\sigma^-)$ system}

In the next two subsections we now discretize
the~$(\psi,\sigma^+,\sigma^-)$ system of
equations~\eqref{pss-hyp-coord} and define a semidiscrete
energy. Demanding conservation of this discrete energy up to a
boundary term in the usual way we obtain our SBP scheme. Whenever
working with the semidiscrete setting we set the constraint damping
parameter~$\gamma_2=0$. This has the advantage of simplifying the
energy estimates by rendering the bulk term trivial and, at least in
the linear setting we shall see has no negative consequences for
constraint violation. The latter may need revisiting when we tackle
nonlinear problems like GR, but since we are developing a scheme with
the linear-dominated wavezone in mind, seems reasonable. Following the
conventions of section~\ref{Discretization}, we define~$\Psi_I(t) :=
\psi(t,r_I)$,~$\Sigma^+_I(t) := \sigma^+(t,r_I)$ and~$\Sigma^-_I(t) :=
\sigma^-(t,r_I)$ and suppress the~$t$ dependence. Define the column
vectors~$\Psi$,~$\Sigma^+$ and~$\Sigma^-$ with~$I$th entries
as~$\Psi_I$,~$\Sigma^+_I$ and~$\Sigma^-_I$ respectively. The state
vector for our system is then~$\mathbf{U} =
(\Psi,\Sigma^+,\Sigma^-)^T$.

Let~$D$ and~$\bar{D}$ denote the finite difference operators
represented by~$(N+1) \times (N+1)$ matrices such that~$\Upsilon^{-1}
D$ and~$\Upsilon^{-1} \bar{D}$ approximate~$\p_r$ and~$\p_r + 2R'/R$,
respectively, at the discrete level. Here, $\Upsilon$ is the same
quadrature matrix defined in~\eqref{Dynamic_Quadrature}. Therefore,
motivated from~\eqref{pss-hyp-coord}, we define our finite difference
scheme as
\begin{align}
\dot{\Psi} = & \frac{\Sigma^+ + \Sigma^-}{2} \, , \nonumber\\
\dot{\Sigma}^+ = & \left[ \frac{1}{2R'-1} \right] \Upsilon^{-1}
\left( \frac{(D + \bar{D})}{2} \Sigma^+
  + \frac{(D - \bar{D})}{2} \Sigma^- \right) \nonumber\\
& - \left[ \frac{R'}{2R'-1} \right] [F] \Psi \, , \nonumber\\
\dot{\Sigma}^- = & - \Upsilon^{-1}
\left( \frac{(D + \bar{D})}{2} \Sigma^-
  + \frac{(D - \bar{D})}{2} \Sigma^+ \right) - [R'] [F] \Psi \, .
\label{pss_disc_eqs}
\end{align}
As introduced in the previous section, the quantities in square
brackets denote the discrete operators corresponding to the continuum
functions written inside them. This makes sense because~$R'$ and~$F$
are functions of $r$.

Motivated by~\eqref{cons_en_hyp_slice}
and~\eqref{cons_en_density_hyp_slice}, we define our discrete energy
norm as
\begin{align}
  \hat{E} = \tfrac{1}{2}(\Psi^T [F] \Upsilon W \Psi + (\Sigma^+)^T
  \Upsilon W^+\Sigma^+ + (\Sigma^-)^T \Upsilon W^- \Sigma^-)
  \,.\label{semi_energy_physical}
\end{align}
Here, the various~$W$'s are the~$(N+1) \times (N+1)$ weight matrices
just like~$W$ in the last subsection. Therefore, the symmetrizer
matrix here is the diagonal matrix with blocks~$\mathbf{H} =
\textrm{diag}(W,W^+,W^-)$. So far, we only demand that the
matrices~$W$,~$W^+$ and~$W^-$ are positive and diagonal.

The discrete energy defined above is a function of time and the outer
boundary~$r_N$, which again is a function of time, i.e.~$\hat{E} =
\hat{E}(t,r_N(t))$. The contribution to change in energy solely from
the evolved variables is
\begin{align}
  \p_t \hat{E} = [\Psi^T [F] \Upsilon W \dot{\Psi} + (\Sigma^+)^T \Upsilon W^+
    \dot{\Sigma}^+ + (\Sigma^-)^T \Upsilon W^- \dot{\Sigma}^-].
\end{align}
Substituting the evolution equations~\eqref{pss_disc_eqs} and using
various algebraic relations and symmetry properties, we obtain
\begin{align}\label{dEdt_presub}
  \p_t \hat{E} = & \left[ (\Sigma^+)^T \frac{\Upsilon}{2}
    \left( W - W^+ \left[ \frac{2R'}{2R'-1} \right] \right)
    + (\Sigma^-)^T \frac{\Upsilon}{2} \right. \nonumber \\
    & \bigg( W - W^- \left[2R' \right] \bigg) \bigg] [F] \Psi
  + \left[ (\Sigma^+)^T W^+ \left[ \frac{1}{2R'-1} \right] \right. \nonumber \\
    & \left. \left( \frac{D + \bar{D}}{2} \right) \Sigma^+
    - (\Sigma^-)^T W^- \left( \frac{D + \bar{D}}{2} \right)
    \Sigma^- \right] \nonumber \\
    & + \left[ (\Sigma^+)^T W^+ \left[ \frac{1}{2R'-1} \right]
    \left( \frac{D - \bar{D}}{2} \right) \Sigma^- \right. \nonumber \\
    & \left. - (\Sigma^-)^T W^- \left( \frac{D - \bar{D}}{2} \right)
    \Sigma^+ \right].
\end{align}
In order to derive an SBP scheme, this energy is required to be
conserved up to the boundary term, which was not imposed up to this
point. Hence, motivated by the continuum
expression~\eqref{pE_pt_continuous}, we demand
\begin{align}\label{pEhat_pt_pss}
\p_t \hat{E} = (\Sigma^+)^T B \Sigma^+ - (\Sigma^-)^T B \Sigma^- \, ,
\end{align}
which gives first
\begin{align}\label{Us_pss}
W = W^+ \left[ \frac{2R'}{2R'-1} \right] = W^- [2R'] \, ,
\end{align}
then
\begin{align}
W^- \left( \frac{D + \bar{D}}{2} \right) = B \, ,
\end{align}
and finally
\begin{align}
  W^- \left( \frac{D - \bar{D}}{2} \right)
  - \left( W^- \left( \frac{D - \bar{D}}{2} \right) \right)^T = 0 \, .
\end{align}
This gives our SBP scheme 
\begin{align}\label{sbp_pss_1}
W^- \bar{D} + D^T W^- = B + B^T \, ,
\end{align}
or, isolating instead~$\bar{D}$,
\begin{align}\label{sbp_pss_2}
\bar{D} = - (W^-)^{-1} D^T W^- + (W^-)^{-1} (B + B^T) \, .
\end{align}
Here,~$B$ is called the boundary matrix, or boundary operator. As the
name suggests, this matrix is expected to be nonzero only at (or near)
the outer boundary. These relations are analogous to those given in
the continuum energy norm~\eqref{cons_en_dens_hyp_slice_resc_pss},
which is already promising.

We take the outer boundary to be an incoming null ray at a finite
coordinate radius~$r_N$. Using the same argument as
in~\eqref{dNorm_dt} and~\eqref{ups_dot}, and using~\eqref{r_N_dot} the
effect of the moving outer boundary to the change in energy is
\begin{align}\label{pE_p_rN}
  \left( \p_{r_N} \hat{E} \right) \dot{r}_N &=
  - \frac{1}{2} \left[ F_N w_N \Psi_N^2 + w^+_N (\Sigma^+_N)^2
    + w^-_N (\Sigma^-_N)^2 \right] \nonumber \\
& \cdot \frac{1}{2R'_N-1} \nonumber \\
  &=  - \frac{1}{2} w^-_N \bigg[ F_N \frac{2R'_N}{2R'_N-1} \Psi_N^2
    + (\Sigma^+_N)^2 \nonumber \\
& + \frac{1}{2R'_N-1} (\Sigma^-_N)^2 \bigg] \, ,
\end{align}
where we use the obvious generalization of the notation~\eqref{W_diag}
for~$W^\pm$ and the relations~\eqref{Us_pss}. Unlike in the last
subsection, we do not need to use the index~$M$ instead of~$N$ here
because, as~$r_N \rightarrow r_\mathscr{I}$,~$\dot{r}_N \rightarrow
0$. Therefore, the total change in energy becomes
\begin{align}\label{Ehat_dot}
  \dot{\hat{E}} = & \frac{d}{dt} \hat{E} = \p_t \hat{E}
  + \left( \p_{r_N} \hat{E} \right) \dot{r}_N \nonumber \\
  &=  - \frac{1}{2} w^-_N F_N \frac{2R'_N}{2R'_N-1} \Psi_N^2
  - \frac{1}{2} w^-_N (\Sigma^+_N)^2 + (\Sigma^+)^T B \Sigma^+ \nonumber \\
  & - \frac{1}{2} w^-_N \frac{1}{2R'_N-1} (\Sigma^-_N)^2
  - (\Sigma^-)^T B \Sigma^- \, .
\end{align}
The SBP relation~\eqref{sbp_pss_1}, or equivalently~\eqref{sbp_pss_2},
dictates the way in which the four operators~$D$,~$\bar{D}$,~$U^-$
and~$B$ should be related. Therefore, given three of them, it can be
used to derive the fourth one. We will choose~$D$ and~$B$ by hand and
describe a method to choose~$U^-$, and hence derive~$\bar{D}$. We
define our methods with a second order accurate operator~$D$ given by
\begin{align}\label{D_bulk}
(D \Psi)_I = \frac{\Psi_{I+1} - \Psi_{I-1}}{2} \,
\end{align}
in the bulk. A similar method can be applied for any higher order
accurate operator~$D$ as well. Choosing
\begin{align}
B = \textrm{diag}(0,\ldots,0,B_N) \, ,
\end{align}
the SBP relation~\eqref{sbp_pss_2} gives
\begin{align}\label{Dtili}
  (\bar{D}\Psi)_I = -(W^{-1} D^T W \Psi)_I =
  \frac{w_{I+1} \Psi_{I+1} - w_{I-1} \Psi_{I-1}}{2 w_I}
\end{align}
in the bulk. We use two methods described in~\cite{GunGarGar10}. One
method is by Evans, given in~\cite{Eva84} and described as
follows. The continuum identity
\begin{align}
\p_r \psi + \frac{2R'}{R} \psi = 3R' \frac{d(R^2 \psi)/dr}{d(R^3)/dr}
\end{align}
suggests one form for~$\bar{D}$. To keep it consistent
with~\eqref{Dtili}, we define
\begin{align}
  (\bar{D} \Psi)_I = \frac{\frac{(R_{I+2}^3 - R_I^3)}
    {12 h R'_{I+1}}\Psi_{I+1} - \frac{(R_I^3 - R_{I-2}^3)}
    {12 h R'_{I-1}}\Psi_{I-1}}{2 \frac{(R_{I+1}^3 - R_{I-1}^3)}
    {12 h R'_I}}
\end{align}
in the bulk. This suggests to us the choice~$w^-_I = (R_{I+1}^3 -
R_{I-1}^3)/12 h R'_I$ for all~$I$, which reduces to $R^2/2$ in the
continuum limit. This extra half factor makes the discrete energy norm
compatible with the continuum one.

The other method is described in~\cite{CalNei03,NeiLehSar04}, and uses
the identity
\begin{align}
\p_r \psi + \frac{2R'}{R} \psi = \frac{\p_r(R^2 \psi)}{R^2} \, .
\end{align}
This suggests~$w^-_I = R_I^2/2$ for all~$I$. Following the terminology
of~\cite{GunGarGar10} we refer to this as the Sarbach method. Note
that it is simpler to define~$w^-_I$ using this method at larger radii
than Evan's method, but Evan's method is more convenient near~$r = 0$,
since it avoids the singular form of~$1/R$ at the origin.

\subsection{Regularization Scheme}\label{Regularization Scheme}

We know from the previous section that~$R_N \rightarrow \infty$
as~$r_N \rightarrow r_\mathscr{I}$, making~$B$,~$w^-_N$ and,
hence,~$\bar{D}$ singular at~$r_\mathscr{I}$. Although this singular
nature of~$\bar{D}$ is expected due to the singular nature of~$R'/R$
at~$\mathscr{I}^+$, it becomes impossible to define it
at~$r_\mathscr{I}$. To avoid this, we need to regularize~$\bar{D}$ as
well. In this section, we therefore study the regularization scheme
of~$\bar{D}$ obtained with Sarbach's method, which is simple and
naturally allows us to set the outer boundary at~$\mathscr{I}^+$. The
regularization is not only `nice', but also necessary to
straightforwardly apply results, such as the Lax equivalence theorem,
from numerical analysis; see for example~\cite{Tho98c} for
details. This is because the formal definition of numerical stability
requires arbitrary given data with finite norm to be admissible, which
will not be the case if there are singular coefficients in the
problem. Similar issues arise when treating the origin in spherical
polar coordinates, but unfortunately at infinity we can not rely on
parity to help.

\subsubsection{Regularized~$\bar{D}$}

We define a new operator~$\tilde{D}$ by
\begin{align}\label{Dtilde_P}
\tilde{D} = P^{-2} \bar{D} P^2
\end{align}
in such a way that all the entries of the matrix~$\tilde{D}$
are~$O(1)$ in the interval~$[0,r_{\mathscr{I}}]$. Since, using
Sarbach's method, the coefficients in~$\bar{D}$ blow up like~$R^2$
at~$\mathscr{I}^+$, the entries of~$P$ should fall off
like~$R^{-1}$. We take $P$ to be an~$(N+1) \times (N+1)$ diagonal
matrix with diagonal elements
\begin{align}
  P_{II} = \frac{1}{\sqrt{1+R^2_{II}}}
  = \frac{1}{\chi_I} \quad \textrm{for } I = 0,\ldots,N \, .
\end{align}
We choose~$1+R^2$, instead of~$R^2$, simply to avoid singularities at
the origin. With this definition, the singular part of~$\bar{D}$
at~$\mathscr{I}^+$ is absorbed by the matrices~$P^2$ and~$P^{-2}$.

Since, in the continuum limit,~$P(r) = 1/\chi$, where~$\chi^2 = 1+R^2$
as defined previously, one can check that this choice of~$\tilde{D}$
corresponds to the operator defined in~\eqref{ptilde_r_1},
equivalently~\eqref{ptilde_r_2}, as
\begin{align}\label{Dtilde_P_continuum}
  \tilde{D} \rightarrow \chi^2 \left( \p_r
  + \frac{2R'}{R} \right) \chi^{-2}
  = \p_r + \frac{2R'}{(1+R^2)R} = \tilde{\p}_r \,,
\end{align}
thus justifying the definition.

\subsubsection{Regularized variables and operators}

Naively, one might expect that writing a discrete version
of~\eqref{pss_rescaled_chi_chi_sq_cont} and defining~$\tilde{W}^- =
P^T W^- P$ and~$\tilde{B} = P^T B P$, one obtains the SBP scheme with
\begin{align}\label{sbp_pss_resc_1}
  \tilde{D} = - (\tilde{W}^-)^{-1} D^T \tilde{W}^-
  + (\tilde{W}^-)^{-1}  (\tilde{B} + \tilde{B}^T)
\end{align}
for the regular equations. But this turns out not to be the case,
because the additional rescaling of~$\sigma^+$ by a factor of~$\chi$,
compared to the other variables, does not play any role in the
definition of~$\tilde{D}$. This is also evident
from~\eqref{pE_pt_pss_resc}, as we want the same boundary matrix
acting on all of the dynamical variables. Therefore, to derive an SBP
scheme for the regular equations with the most aggressively rescaled
variables, we first rescale all the dynamical variables by a single
power of~$\chi$, derive equations of motion, and {\it then}
replace~$\bar{\sigma}^+ (\equiv \chi \sigma^+)$ by~$(\tilde{\sigma}^+
/ \chi)$. This reduces~\eqref{pss_rescaled_chi_chi_sq_cont} to the
following form
\begin{align}\label{pss_rescaled_chi_cont}
  \p_t \tilde{\psi} &=  \frac{1}{2} \left[ \frac{\tilde{\sigma}^+}{\chi}
    + \tilde{\sigma}^- \right] \, , \nonumber \\
  \p_t \tilde{\sigma}^+ &=  \frac{1}{2R'-1} \left[ \chi
    \left( \frac{\p_r + \tilde{\p}_r}{2} \right)
    \left( \frac{\tilde{\sigma}^+}{\chi} \right)
    + \chi \left( \frac{\p_r - \tilde{\p}_r}{2} \right)
    \tilde{\sigma}^- \right. \nonumber \\
    &\quad \left. - \frac{RR'}{\chi} \tilde{\sigma}^-
    - R' \chi F \tilde{\psi} \right] + \gamma_2
  \left[ \frac{1}{2R'-1} \left( \chi (\p_r \tilde{\psi})
    \right. \right. \nonumber \\
    &\quad \left. \left. - \frac{R}{\chi} R' \tilde{\psi}
    + \chi \frac{\tilde{\sigma}^-}{2} \right)
    - \frac{\tilde{\sigma}^+}{2} \right] \, , \nonumber \\
  \p_t \tilde{\sigma}^- = & - \left[ \left( \frac{\p_r
      + \tilde{\p}_r}{2} \right) \tilde{\sigma}^-
    + \left( \frac{\p_r - \tilde{\p}_r}{2} \right)
    \left( \frac{\tilde{\sigma}^+}{\chi} \right)
    \right. \nonumber \\
    &\quad \left. - \frac{RR'}{\chi^3} \tilde{\sigma}^+
    + R' F \tilde{\psi} \right] - \gamma_2
  \left[ \p_r \tilde{\psi} - \frac{R}{\chi^2} R'
    \tilde{\psi} \right. \nonumber \\
    &\quad \left. + \frac{\tilde{\sigma}^-}{2}
    - (2R'-1) \frac{\tilde{\sigma}^+}{2 \chi} \right] \, .
\end{align}
The semidiscrete form, with~$\gamma_2 = 0$, is
\begin{align}
  \dot{\tilde{\Psi}} &=  \frac{1}{2} \left[ P \tilde{\Sigma}^+
    + \tilde{\Sigma}^- \right] \, , \nonumber\\
  \dot{\tilde{\Sigma}}^+ &=   \left[ \frac{1}{2R'-1} \right]
    \left[ P^{-1} \Upsilon^{-1} \left( \frac{D + \tilde{D}}{2} \right) P
    \tilde{\Sigma}^+ + P^{-1} \Upsilon^{-1} \right. \nonumber\\
  &\quad \left. \left( \frac{D - \tilde{D}}{2} \right) \tilde{\Sigma}^-
    - P[RR'] \tilde{\Sigma}^- - P^{-1} [R'] [F]
    \tilde{\Psi} \right] \, , \nonumber\\
  \dot{\tilde{\Sigma}}^- &=  - \left[ \Upsilon^{-1}
    \left( \frac{D + \tilde{D}}{2} \right) \tilde{\Sigma}^-
    + \Upsilon^{-1} \left( \frac{D - \tilde{D}}{2} \right)
    P \tilde{\Sigma}^+ \right. \nonumber\\
  &\quad \left. - P^2 [RR'] P \tilde{\Sigma}^+
    + [R'] [F] \tilde{\Psi} \right] \, .
\label{pss_rescaled_chi_disc}
\end{align}
Defining the discrete energy as
\begin{align}\label{Ehat_resc}
\hat{E} = & \tfrac{1}{2} [\tilde{\Psi}^T [F] \Upsilon \tilde{W} \tilde{\Psi} +
  (\tilde{\Sigma}^+)^T \Upsilon \tilde{W}^+ \tilde{\Sigma}^+
  + (\tilde{\Sigma}^-)^T \Upsilon \tilde{W}^- \tilde{\Sigma}^-]
\end{align}
and following the same procedure as above, we demand
\begin{align}\label{pEhat_pt_pss_resc}
\p_t \hat{E} = (\tilde{\Sigma}^+)^T P^T \tilde{B} P
  \tilde{\Sigma}^+ - (\tilde{\Sigma}^-)^T \tilde{B} \tilde{\Sigma}^- \, ,
\end{align}
which gives
\begin{align}\label{Us_pss_resc}
  \tilde{W} = P^{-2} \tilde{W}^+ \left[ \frac{2R'}{2R'-1} \right]
  = \tilde{W}^- [2R'] \, ,
\end{align}
\begin{align}\label{bound}
\tilde{W}^- \left( \frac{D + \tilde{D}}{2} \right) = \tilde{B} \, ,
\end{align}
and
\begin{align}\label{sbp_pss_resc_2}
\tilde{W}^- \left( \frac{D - \tilde{D}}{2} \right) - \left(
\tilde{W}^- \left( \frac{D - \tilde{D}}{2} \right) \right)^T = 0 \, .
\end{align}
This leads to the SBP scheme given
by~\eqref{sbp_pss_resc_1}. Equations~\eqref{pss_rescaled_chi_disc} are
the ones used in the code. Provided the potential function~$F$ falls
off fast enough they are formally regular and satisfy the SBP
property.

\subsubsection{Constraints}\label{Subsection:Constraints}

The reduction constraint~$C=\p_R \psi - \phi_R$, written in terms of
the rescaled fields~$(\tilde{\psi},\tilde{\sigma}^+,\tilde{\sigma}^-)$
and using a suitable rescaling, takes the form
\begin{align}
\mathcal{C}&=\frac{R'\chi }{2R'-1}C\nonumber\\
  &= \frac{1}{2R'-1} \left( \p_r \tilde{\psi} - \frac{R R'}{\chi^2}
  \tilde{\psi} + \frac{\tilde{\sigma}^-}{2} \right)
  - \frac{\tilde{\sigma}^+}{2 \chi}\,.\label{Constraint_cont}
\end{align}
It also appears as the coefficient of~$\gamma_2\chi$ in the second
equation of~\eqref{pss-hyp-coord_rescaled}. In the continuum case, if
the constraint is satisfied by the initial data, it will remain
satisfied in the time development. However, this might not be the case
at the discrete level. Defining the discrete form
of~\eqref{Constraint_cont} as
\begin{align}
\hat{\mathcal{C}}&= \left[ \frac{1}{2R'-1} \right] \left( \Upsilon^{-1} D
\tilde{\Psi} - P^2 [R R'] \tilde{\Psi} + \frac{\tilde{\Sigma}^-}{2}
\right)- \frac{P}{2} \tilde{\Sigma}^+ \,,\label{Constraint_disc}
\end{align}
taking the time derivative and substituting the equations of
motion~\eqref{pss_rescaled_chi_disc}, we
obtain~$\dot{\hat{\mathcal{C}}}=0$. Therefore, in our discretization
scheme as well, if the constraint is satisfied on the initial data, it
will remain satisfied forever. Thus, taking~$\gamma_2 = 0$ in our
discretization scheme is perfectly justified. But, in general, this
will not necessarily be the case for a system of nonlinear equations.

\subsection{Truncation Error Matching}\label{TEM}

Taking~$\Upsilon = \textrm{diag}(h/2,h,\ldots,h,h/2)$, the operator~$D$
defined with second order accuracy by~\eqref{D_bulk} in the bulk has
the Taylor expansion
\begin{align}\label{D_ser_expan}
  (Df)_I = h \left[ f'_I + \frac{h^2}{6} f_I''' + \cdots \right],
  \quad I = 0,\ldots,N-1  \, .
\end{align}
Incorporating the TEM property at the last grid point,~$D$ is defined
there as
\begin{align}
(D f)_N = \frac{-f_{N-3} + 4 f_{N-2} - 7 f_{N-1} + 4 f_N}{4} \, .
\end{align}
An extra half factor appears in this definition, compared with the one
given in~\cite{Pre02}, because of the half factor
in~$\Upsilon_{NN}$. Taylor expanding~$f_I$ for~$I = N-3, \ldots, N$ at
the~$N$th grid point and substituting all these expansions in the
previous equation above gives us a series expansion of~$(Df)_N$ with
terms up to~$h^2$ the same as given in~\eqref{D_ser_expan},
with~$I=N$. As we will see shortly, our dissipation operator vanishes
like~$h^3$ as~$h\rightarrow 0$. Also, the~$h^3$ term in the Taylor
expansion of~$(Df)_I$ is zero for~$I = 0,\ldots,N-1$ but nonzero
for~$I=N$. This residual term at the last grid point interferes with
the dissipation operator and may cause the code to blow up from the
outer boundary. Therefore, we redefine the operator~$D$ at the last
grid point so that the~$h^3$ term in~$(Df)_N$ vanishes
identically. This leads to the following definition of~$D$ at the last
grid point:
\begin{align}\label{D_out_bound}
(D f)_N = \frac{f_{N-4} - 5 f_{N-3} + 10 f_{N-2} - 11 f_{N-1} + 5 f_N}{4} \, .
\end{align}
Here also, an extra half factor is introduced because of the half
factor in~$\Upsilon_{NN}$. Thus, the matrix form of~$D$ near the outer
boundary is
\begin{align}\label{matrix_D_2nd_ord_accu}
D = \left( \begin{array}{ccccccc}
\cdot & \cdot & \cdot & \cdot & \cdot & \cdot & \cdot \\
\cdot & 0 & 1/2 & 0 & 0 & 0 & 0 \\
\cdot & -1/2 & 0 & 1/2 & 0 & 0 & 0 \\
\cdot & 0 & -1/2 & 0 & 1/2 & 0 & 0 \\
\cdot & 0 & 0 & -1/2 & 0 & 1/2 & 0 \\
\cdot & 0 & 0 & 0 & -1/2 & 0 & 1/2 \\
\cdot & 0 & 1/4 & -5/4 & 5/2 & -11/4 & 5/4 \\
\end{array} \right) \, .
\end{align}
Now, using this definition of~$D$, and taking~$\tilde{W}^- =
\textrm{diag}(\tilde{w}^-_0,\ldots,\tilde{w}^-_N)$ and~$\tilde{B} =
\textrm{diag}(0,\ldots,1/4)$, we define~$\tilde{D}$
using~\eqref{sbp_pss_resc_1}. However, applying this operator to some
smooth function~$f$, one can see that the Taylor expansion
of~$(\tilde{D}f)_I$ not only violates the TEM property at~$I =
N-4,\ldots,N$, but also does not even give the leading
term~$h(\tilde{w}^- f)'/\tilde{w}^-$, which we will expect from the
Sarbach's method along with~\eqref{Dtilde_P}. Instead, looking at the
corresponding Taylor expansions, their leading terms are~$\sim
f/\tilde{w}^-$. Therefore, when divided by~$h$, these terms will blow
up at~$\mathscr{I}^+$ with increasing resolution. To make the system
consistent, we define the operator~$\tilde{D}$ by
hand. Using~\eqref{Dtili} in~\eqref{Dtilde_P}, we notice that~$D$
and~$\tilde{D}$ are related as~$\tilde{D} = (\tilde{W}^-)^{-1} D
\tilde{W}^-$ in the bulk. Inspired from this, define~$\tilde{D}$ on
the whole grid as
\begin{align}\label{Dtilde_P_TEM}
\tilde{D} \equiv (\tilde{W}^-)^{-1} D \tilde{W}^- \, .
\end{align}
This gives
\begin{align}\label{DtildeP_bulk}
(\tilde{D} f)_I = \frac{(\tilde{w}^- f)_{I+1} - (\tilde{w}^-
    f)_{I-1}}{2 \tilde{w}^-_I} = \frac{D(\tilde{W}^- f)_I}{\tilde{w}^-_I}
\end{align}
for~$I = 0,\ldots,N-1$ and
\begin{align}
  (\tilde{D} f)_N = & \frac{1}{4 \tilde{w}^-_N}
  \bigg[(\tilde{w}^- f)_{N-4} - 5 (\tilde{w}^- f)_{N-3}
    + 10 (\tilde{w}^- f)_{N-2} \nonumber\\
    & - 11 (\tilde{w}^- f)_{N-1}
    + 5 (\tilde{w}^- f)_N \bigg] \,,\label{DtildeP_out_bound}
\end{align}
for the term~$I = N$. This definition of~$\tilde{D}$ not only
approximates~\eqref{Dtilde_P_continuum} at second order accuracy but
also satisfies the TEM property at all grid points.

Since we are defining~$D$ and~$\tilde{D}$ by hand and
choosing~$\tilde{W}^-$ by Sarbach's method, in order to incorporate
the SBP property, we calculate the boundary matrix~$\tilde{B}$
using~\eqref{bound}. We do not need to worry about the
relation~\eqref{sbp_pss_resc_2} as it is automatically satisfied by
this new choice of~$D$ and~$\tilde{D}$. Unlike what was assumed so
far, this new~$\tilde{B}$ has nonzero entries in the bulk as well as
at the outer boundary, i.e. we also have~$\tilde{B}_{IJ} \neq 0$ for
both~$I,J < N$. However, we realize from~\eqref{pEhat_pt_pss_resc}
that only the symmetric part of~$\tilde{B}$ contributes to the energy
flux. Interestingly, it turns out that the symmetric part of the
new~$\tilde{B}$ has nonzero entries only at~$(I,N)$ and~$(N,I)$
positions, with~$I = (N-4),\ldots,N$.

In summary, we initially chose the operator~$D$, the weight
matrix~$\tilde{W}^-$ and the boundary matrix~$\tilde{B}$ by hand and
derived~$\tilde{D}$ using the SBP
relation~\eqref{sbp_pss_resc_1}. Doing this, we lost all control over
the properties of~$\tilde{D}$ near the outer boundary. As a result, we
obtained a form of~$\tilde{D}$ which is inconsistent
with~\eqref{Dtilde_P_continuum} near the outer boundary. In order to
resolve this issue, we adopted the reverse strategy. We first
chose~$D$,~$\tilde{D}$ satisfying the~TEM property everywhere
and~$\tilde{W}^-$ from the Sarbach's method, which we preferred over
the Evan's one. Using these operators, we then used the~SBP property
to calculate the boundary matrix~$\tilde{B}$. Since~$\tilde{B}$ merely
gives the energy flux at the outer boundary, the price we pay in order
to incorporate both~SBP and~TEM properties is that we lose control
over the boundary flux.

Another method for incorporating TEM could be adapted
from~\cite{GunGarGar10}, which employs the outer boundary condition in
the SBP scheme in~$1+1$ dimensions. Demanding specific relations
between the weight and boundary matrices between~$1+1$
and~$(j+1)$-dimensions they derive the operator~$\bar{D}$ in~$j+1$
dimensions. In our setup we instead saw how~$D$ and~$\tilde{D}$ are
related without invoking a~$1+1$ dimensional system. Given the above,
we expect that our method generalizes for any spatial dimension.

\subsection{The SBP-TEM and SBP-Stable Methods}
\label{SBP_vs_TEM}

In this section, we give two numerical schemes, obtained by
approximating the continuum equations at the outer boundary in two
different ways.  We will compare both in our numerical experiments,
and see that, empirically, give satisfactory norm convergence but have
slightly different pointwise convergence. It is observed empirically
in many cases that the first scheme, which we call the~SBP-TEM
discretization, gives perfect pointwise convergence but is not
formally stable. The second scheme, the SBP-Stable scheme, is provably
stable but has a lower order pointwise errors near the outer boundary.

\subsubsection{SBP-TEM} \label{SBP_TEM}

As before, the total change in energy is given by
\begin{align}\label{Ehat_dot_resc}
  \dot{\hat{E}} &=  - \frac{1}{2} \tilde{w}^-_N F_N
  \frac{2R'_N}{2R'_N - 1} \tilde{\Psi}_N^2
  + (\tilde{\Sigma}^+)^T P \tilde{B} P \tilde{\Sigma}^+ \nonumber \\
  &\quad - \frac{1}{2} \tilde{w}^-_N P_N^2 (\tilde{\Sigma}^+_N)^2
  - (\tilde{\Sigma}^-)^T \tilde{B} \tilde{\Sigma}^- \nonumber\\
  &\quad - \frac{1}{2} \frac{1}{2R'_N - 1}
  \tilde{w}^-_N (\tilde{\Sigma}^-_N)^2 \, ,
\end{align}
where the first, third and fifth terms arise because of the moving
outer boundary. The boundary matrix~$\tilde{B}$ here is the one
obtained by using the~SBP and~TEM properties.

Defining~$\tilde{B}_s := (\tilde{B} + \tilde{B}^T)/2$, we observe that
only the last row and the last column of~$\tilde{B}_s$ are
nonzero. As~$r_o \rightarrow r_\mathscr{I}$, $P_N = 1/\chi_N
\rightarrow 0$ and~$1/(2R_N'-1) \to 0$. Thus, the second, third and
fifth terms in~\eqref{Ehat_dot_resc} vanish, as all other factors in
these terms are~$O(1)$. The second term vanishes because~$\tilde{B}_s$
has nonzero elements only in its last row and last column. Since~$P$
is diagonal, multiplying~$P$ on left of~$\tilde{B}_s$, gives the
matrix~$P \tilde{B}_s$ with all elements zero in its last row and
nonzero elements only in its last column. Multiplying~$P \tilde{B}_s$
on the right by~$P$ gives all elements zero in the last column of the
resulting matrix. This gives~$P \tilde{B}_s P = 0$. Therefore, the
total change in energy reduces to
\begin{align}
  \label{Ehat_dot_resc_2}
  \dot{\hat{E}} &=  - \frac{1}{2} \tilde{w}^-_N F_N
  \frac{2R'_N}{2R'_N - 1} \tilde{\Psi}_N^2
  - (\tilde{\Sigma}^-)^T \tilde{B}_s \tilde{\Sigma}^- \, .
\end{align}
which is analogous to~\eqref{Ehat_dot_resc_2} in the continuum
problem. Here there is, however, an important subtlety. The first term
does indeed directly map to the potential term on the right-hand-side
of~\eqref{Ehat_dot_resc_2}. But the second contains cross-terms
between points at the boundary and points in the interior. In this
sense, one might argue that the SBP-TEM scheme is not truly an SBP
discretization, but we nevertheless keep the name to indicate the
origin of the method. This shortcoming means that there is a deviation
of the discrete energy flux at~$\mathscr{I}^+$ from the continuum one.
To understand this deviation we ignore the potential
term. Expanding~$\tilde{B}_s$ then, we get
\begin{align}\label{Ehat_dot_SBP-TEM}
  \dot{\hat{E}}(t) &=  - \frac{5}{4} \tilde{w}^-_N (\tilde{\Sigma}^-_N)^2
  + \bigg[ \frac{9}{8} (\tilde{w}^-_{N-1} + \tilde{w}^-_N)
    \tilde{\Sigma}^-_{N-1} \nonumber \\
    &\quad - \frac{5}{4} (\tilde{w}^-_{N-2} + \tilde{w}^-_N)
    \tilde{\Sigma}^-_{N-2} + \frac{5}{8}
    (\tilde{w}^-_{N-3} + \tilde{w}^-_N)
    \tilde{\Sigma}^-_{N-3} \nonumber \\
    &\quad - \frac{1}{8} (\tilde{w}^-_{N-4} + \tilde{w}^-_N)
    \tilde{\Sigma}^-_{N-4} \bigg] \tilde{\Sigma}^-_N \, .
\end{align}
We can furthermore rewrite this expression by separating the continuum
part out from this expression to obtain
\begin{align}
  \dot{\hat{E}}(t) = & - \frac{1}{2} \tilde{w}^-_N (\tilde{\Sigma}^-_N)^2
  - \tilde{\Sigma}^- \tilde{W}^-
  (\Delta^2 + \tilde{\Delta}^2) \tilde{\Sigma}^-
\end{align}
where~$\Delta^2 \tilde{\Sigma}^-_I = 0$ for~$I = 0,\ldots,N-1$ and
\begin{align}
  \Delta^2 \tilde{\Sigma}^-_N = \frac{\tilde{\Sigma}^-_{N-4}
    - 5 \tilde{\Sigma}^-_{N-3} + 10 \tilde{\Sigma}^-_{N-2}
    - 9 \tilde{\Sigma}^-_{N-1} + 3 \tilde{\Sigma}^-_N }{8} 
\end{align}
and finally
\begin{align}
\tilde{\Delta}^2 = (\tilde{W}^-)^{-1} \tilde{\Delta}^2 \tilde{W}^-\,.
\end{align}
At the last grid point, the operator~$\Delta^2$ corresponds to the
continuum operator
\begin{align}
\Delta^2 f = \frac{h^2}{8} (f'' + \frac{h^4}{12} f^{(4)} + \cdots) \, .
\end{align}
{\it Assuming} convergence, this gives
\begin{align}
  \dot{\hat{E}}(t) = - \frac{1}{2} \tilde{w}^-_N (\tilde{\Sigma}^-_N)^2
  - h^2(\cdots) \, ,
\end{align}
as resolution increases, so that the deviation diminishes like~$h^2$,
consistent with the~TEM property.

Therefore, convergence of the SBP-TEM scheme is the only remaining
aspect to prove. A standard way to do so is to first prove stability
and then use the Lax Equivalence Theorem~\cite{Tho98c,GusKreOli95} to
ensure convergence. Unfortunately the quadratic form
in~$\tilde{\Sigma}^-$ on the right-hand-side
of~\eqref{Ehat_dot_SBP-TEM} is not sign definite, and so formal
stability does not follow. This implies that the energy at any later
hyperboloidal time slice is not (in general) upper bounded by that on
the initial slice. It is important to realize that this shortcoming
does not mean that the method will not converge for any given initial
data. Rather it means that there is no {\it guarantee} of
convergence. It would be interesting to know the specific class of
data that does converge. To find examples of `bad' data we need to
look at the eigenvectors of the boundary matrix associated with
positive eigenvalues. Instead of going in to more detail along these
lines, in section~\ref{Numerical_Evolution} we study empirically
convergence of the scheme for various choices of initial data.

\subsubsection{SBP-Stable} \label{Stable_SBP}

We now present an alternative discretization which gives a provably
stable numerical scheme, but requires a drop in the pointwise
convergence order at the outer boundary. This scheme is obtained by
adding~$A\tilde{\Sigma}^-$ to the right hand side
of~$\dot{\tilde{\Sigma}}^-$ in~\eqref{pss_rescaled_chi_disc}, with
\begin{align}\label{Artificial_Boundary}
A = \Upsilon (\Upsilon^{-2} \Delta^2 + \Upsilon^{-2} \tilde{\Delta}^2) \, .
\end{align}
This adds a new term in~\eqref{Ehat_dot_resc_2}, which is
\begin{align}\label{Artificial_Boundary_2}
(\Sigma^-)^T \Upsilon \tilde{W}^- A \tilde{\Sigma}^- \, .
\end{align}
Since only the symmetric part of~$\Upsilon \tilde{W}^- A$ contributes
to~$\dot{\hat{E}}$, when added to~$\tilde{B}_s$, it gives
\begin{align}
- \tilde{B}_s + \frac{\Upsilon \tilde{W}^- A + (\Upsilon \tilde{W}^- A)^T}{2
} = \textrm{diag}(0,\ldots,-\tilde{w}^-_N/2)\,,\label{stability}
\end{align}
and so, for this adjusted scheme, we get~$\dot{\hat{E}} = \dot{E}$,
the continuum energy decay rate, which is negative semidefinite, and
the resulting semidiscrete scheme is stable. Choosing a suitable time
integrator, we can make the whole discrete scheme stable. Therefore,
by the Lax Equivalence Theorem, the resulting scheme is
convergent. However,~\eqref{Artificial_Boundary} shows that
the~$A\tilde{\Sigma}^-$ term vanishes like~$h$ rather than~$h^2$ with
increasing resolution. Thus, it decreases the convergence order of the
numerical scheme. We call~$A \tilde{\Sigma}^-$ an ``artificial
boundary" term, as it vanishes in the continuum limit.

When the outer boundary is not at~$\mathscr{I}^+$, we need to add more
such artificial boundary terms to the equations. Interestingly, it
turns out that adding these terms to the discrete equations of motion
is equivalent to rather change the definition of~$D$ at the outer
boundary:
\begin{align}\label{D_stable}
(Df)_N = \frac{f_N - f_{N-1}}{2} \, .
\end{align}
The operator~$\tilde{D}$ is automatically redefined
from~\eqref{Dtilde_P_TEM}, when~$D$ is defined by~\eqref{D_bulk}
for~$I = 0,\ldots,N-1$ and~\eqref{D_stable}. This clarifies how these
artificial boundary terms are decreasing the accuracy of the numerical
scheme at the outer boundary, effectively by decreasing the accuracy
of~$D$ and~$\tilde{D}$ at the last grid point. Therefore, just to keep
the generality, we will drop the accuracy of~$D$ and~$\tilde{D}$
instead of using the artificial boundary terms. This result is unique
because the choice of artificial boundary terms depends uniquely on
the definition of~$D$ and~$\tilde{D}$ at the last grid point and
demanding that~\eqref{stability} is satisfied.

Interestingly, dropping the accuracy of~$D$ and~$\tilde{D}$ does not
affect the norm convergence. As we saw above, this is equivalent to
using~$D$ and~$\tilde{D}$ satisfying the~TEM property and adding
suitable artificial boundary terms. Since these artificial boundary
terms in the equations vanish like~$h$ with increasing resolution, we
can infer from~\eqref{Artificial_Boundary}
and~\eqref{Artificial_Boundary_2} that their contribution
to~$\dot{\hat{E}}$ vanishes like~$h^2$. Therefore, the norm of errors
should still converge at second order accuracy. On the other hand the
artificial boundary terms do run the risk of badly damaging pointwise
convergence, as they may reflect a lot of noise into the bulk.

\subsection{Origin} \label{Origin}

In order to calculate various derivatives at the origin, which we
treat as an interior point, using centered finite difference stencils,
we introduce ghost points to the left of the origin in our numerical
grid, see Fig.~\ref{fig_stencil}. We fill these ghost zones using the
suitable parity conditions
\begin{align}
  \tilde{\psi}_{-I} = \tilde{\psi}_I \, , \quad \tilde{\pi}_{-I}
  = \tilde{\pi}_I \, \mbox{ and } \, (\tilde{\phi}_R)_{-I}
  = - (\tilde{\phi}_R)_I 
\end{align}
or, equivalently,
\begin{align}\label{Parity}
  \tilde{\psi}_{-I} = \tilde{\psi}_I \, ,
  \quad \tilde{\sigma}^+_{-I}
  = \tilde{\sigma}^-_I  \, \mbox{ and }
  \,  \tilde{\sigma}^-_{-I}
  = \tilde{\sigma}^+_I  \, ,
\end{align}
where~$\chi_I = \sqrt{1 + R_I^2}$. These parity conditions are
obtained by using the rescaling~\eqref{pss_rescaling} for~$r \geq 0$
and~$\tilde{\psi} \equiv \chi \psi$,~$\tilde{\sigma}^+ \equiv \chi
\sigma^+$ and~$\tilde{\sigma}^- \equiv \chi^2 \sigma^-$ for~$r<0$. The
latter rescaling gives all the rescaled variables~$O(1)$ for all~$r<0$
as~$\sigma^+$ becomes the outgoing characteristic variable for~$r<0$,
and hence falls like~$1/R$ and~$\sigma^-$ becomes the incoming one,
and hence falls like~$1/R^2$. Note that this extension renders the
evolved fields non-smooth at the origin, a shortcoming that could be
easily overcome by adjusting the rescaling slightly. This could be
done, for example, by choosing~$\chi$ to be~$1$ identically in a
neighborhood of the origin. Since we are concerned primarily with the
behavior of the approximation near infinity we do not do so, and will
instead rely on artificial dissipation to suppress any noise produced.
The above parity conditions are appropriate if and only if~$R$ is
taken to be an odd function of~$r$ and~$H$ an even function
of~$R$. For~$r \geq 0$,~$H'(R(r)) = 1 - 1/R'(r)$ gives~$H(r) \equiv
H(R(r)) = R(r) - r$. To impose evenness, we must define~$H$ by
\begin{align}
H(r) = \bigg\lbrace \begin{array}{cc}
R(r) - r \, , & \quad \textrm{for } r \geq 0 \\
r - R(r) \, , & \quad \textrm{for } r < 0
\end{array} \, .
\end{align}
Moreover, for~$r<0$, we must take~$c_- = -1$ and~$c_+ = 1/(2R'-1)$,
as~$c_+ $ and~$c_-$ switch roles as incoming and outgoing coordinate
lightspeeds, respectively. Taking~$R$ defined
by~\eqref{Compactification}, we see that~$H(r)$ is only~$C^1$ at the
origin. This is problematic, because due to this we can never expect a
smooth evolution of the fields at the origin. To overcome this
problem, we redefine~$\Omega(r)$ as given in~\eqref{Compactification}
by
\begin{align}\label{Compactification_regular}
  \Omega(r) = 1 - \frac{1}{2} \frac{r^2}{r_\mathscr{I}^2}
  \left[ \tanh \left\lbrace \tan
  \left( \pi \left( \frac{r}{r_\mathscr{I}} - \frac{1}{2} \right) \right)
    \right\rbrace + 1 \right] \, .
\end{align}
This choice of~$\Omega(r)$ not only has similar asymptotics to the
compactification function as the one defined in
Eq.~\eqref{Compactification} but also gives~$R^{(m)}(0) = 0$ for every
integer~$m > 1$, as~$\Omega^{(m)}(0) = 0$ for~$m \geq 1$. Therefore,
the height function~$H(r)$ so obtained is~$C^\infty$ at the origin
with~$H^{(m)}(0) = 0$ for all~$m \geq 0$.

Since there is a~$1/R$ singularity at the origin, there are two
methods to tackle it. One is using l'H\^opital's rule and the other is
using Evan's method, as described before. Using l'H\^opital's rule, we
completely get rid of the operator~$\tilde{D}$ at the origin, whereas,
using Evan's method, we get the value of~$\tilde{w}^-_0$ given by
\begin{align}
  \tilde{w}^-_0 = \frac{(D R^3)_0}{6 h R'_0 (1 + R_0^2)} =
  \frac{R_1^3 - R_{-1}^3}{12h R'_0 (1 + R_0^2)} = \frac{R_1^3}{6 h} \, .
\end{align}
This form of~$\tilde{w}^-_0$ has the following series expansion
\begin{align}
  \tilde{w}^-_0 = \frac{1}{6 R'_0 (1 + R_0^2)} \left[ (R_0^3)'
    + \frac{h^2}{6} (R_0^3)''' + \cdots \right] = \frac{h^2}{6} \, .
\end{align}
The series terminates because~$R_0 = 0$,~$R'_0 = 1$ and~$R^{(m)}(0) =
0$ for every integer~$m > 1$. Therefore, defining~$\tilde{w}^-_I$
using Sarbach's for every~$I \neq 0$ and Evan's method to
define~$\tilde{w}^-_0$, we get
\begin{align}
  (\tilde{D} f)_0 &=  \,\, \frac{6}{h^2} \left[ (\tilde{w}^- f)_0'
    + \frac{h^2}{6} (\tilde{w}^- f)_0''' + \frac{h^4}{120}
    (\tilde{w}^- f)_0^{(5)} + \cdots \right] \nonumber \\
  &=  \,\, 3 \left[ f_0' + \frac{h^2}{6} f_0''' - h^2 f_0'
    + \cdots \right] \, .
\end{align}
To calculate all these derivatives, we used the continuum
values~$\tilde{w}^- = R^2/[2(1+R^2)]$ and~$R^{(m)}(0) = 0$ for every
integer~$m > 1$. It is therefore clear that using Evan's method at the
origin is effectively the same as using l'H\^opital's rule there.

But now, we encounter a problem. If we use Sarbach's method to
define~$\tilde{D}$ for all~$I \neq 0$ and at some instant~$\Psi_I
\approx R_I$ near the origin, then, from~\eqref{D_ser_expan}
and~\eqref{Dtilde_P_TEM}, the associated error near the origin goes
like~$h^2/R^2 \approx h^2/r^2 = h^2/(I^2 h^2) = 1/I^2$, which does not
converge. Our strategy to overcome this problem is to use dissipation
(as outlined in the following section). At the origin, we therefore
simply use l'H\^opital's rule.

The choice~$n=2$ of the compactification parameter
in~\eqref{Compactification} gives a nonzero weight to
the~$(\tilde{\sigma}^+)^2$ term at~$r_\mathscr{I}$ in the energy
defined by~\eqref{cons_en_hyp_slice}
and~\eqref{cons_en_dens_hyp_slice_resc_pss}, and hence to its discrete
version, and makes the discrete energy a norm, so that the discrete
energy has a positive weight at all grid points, with a possible
exception at the origin. The origin has a positive weight whenever we
use Evan's method to define~$\tilde{D}$ there, and has a zero weight
whenever we rewrite the equations there using the l'H\^opital's rule
instead. In the latter case, we do not include the origin in our
definition of discrete energy and define all the operators as~$N
\times N$ matrices over the space of the grid functions defined on the
grid points~$I = 1,\ldots,N$. Thus, the choice~$n=2$ still makes the
discrete energy a norm.

\subsection{Fixing up the Energy}\label{Energy_Adjustment}

To this point our SBP-Stable scheme has been built for optimality in
the energy given by~\eqref{Ehat_resc}. As it is built directly on the
physical energy, this has the advantage that the resulting method
satisfies a precise energy balance relation with~$\dot{\hat{E}}\leq0$.
Unfortunately, however, in the massless case this physical energy is
degenerate, in that the rescaled field~$\tilde{\Psi}$ is completely
absent. A similar degeneracy happens near~$\mathscr{I}^+$ whenever the
potential~$F$ falls off fast enough. Fortunately we can easily adjust
the energy, taking instead
\begin{align}
  \tilde{E}&= \hat{E}+\tfrac{1}{2}\Psi^T\Upsilon [r^2] \Psi\nonumber\\
  &=\tfrac{1}{2} [\tilde{\Psi}^T
    \Upsilon( [r^2] + F\tilde{W} )\tilde{\Psi} +
  (\tilde{\Sigma}^+)^T \Upsilon \tilde{W}^+ \tilde{\Sigma}^+\nonumber\\
    &\quad+ (\tilde{\Sigma}^-)^T \Upsilon \tilde{W}^- \tilde{\Sigma}^-]\,,
  \label{adjusted_energy}
\end{align}
but keeping the exact same discretization as before. Using the Gr\"onwall
inequality we easily obtain the estimate
\begin{align}
\tilde{E}(t)\leq C(t_{\textrm{max}}) \tilde{E}(0)\,,
\end{align}
for all~$0\leq t\leq t_{\textrm{max}}$ with~$C(t_{\textrm{max}})>0$ a
constant independent of initial data for any~$t_{\textrm{max}}$. In
other words, by sacrificing strict stability (working with this
adjusted energy) we gain non-degenerate estimates and, because we have
not actually changed the discretization, we still have
strict-stability in the degenerate physical energy.

\subsection{Dissipation Operator} \label{Dissipation}

In this subsection we first give a brief discussion of standard
dissipation operators before showing, in the second part, how these
operators can be naturally included within our framework, both at the
origin in spherical polar coordinates and near null-infinity.

\subsubsection{For 1D and in the Trivial~$L^2$ Norm}
\label{Diss_L2_Norm}

We start by considering the fourth order Kreiss–Oliger dissipation
operator~\cite{KreOli73,KreLor89,GusKreOli95},
\begin{align}
  Q_{\textrm{KO}} = - \epsilon h^3 h^{-4} (D_+ D_-)^2,
  \label{Diss_Bulk_TEM}
\end{align}
where~$\epsilon$ is the dissipation
parameter whose value is set in our numerical evolutions. 
We will assume that we have an operator~$Q_d$ which agrees with this
in the bulk of the grid, and taking an alternative form to be fixed
just at a small number of grid points near the boundaries. Ultimately
we will `thread' the weights present in our norms through this
operator to render it suitable for use with the second order
accurate~$D$ and~$\tilde{D}$ operators. Here,
\begin{align}
(D_\pm f)_I = \pm \frac{f_{I \pm 1} - f_I}{2} \, ,
\end{align}
are the forward and backward finite difference operators, denoted by
plus and minus signs respectively. This
dissipation operator, which is centered, is only defined in the bulk
and corresponds to the fourth order derivative of a dynamical variable
at second order accuracy suppressed by a power of the grid spacing,
\begin{align}
  [h^{-4} (D_+ D_-)^2 f]_I &=
  \frac{f_{I-2} - 4 f_{I-1} + 6 f_I - 4 f_{I+1} + f_{I+2}}{h^4} \nonumber\\
  &= \left[ f^{(4)}_I + \frac{h^2}{6} f^{(6)}_I + \cdots \right] \, .
\label{Diss_bulk_error}
\end{align}
Ideally, we wish to define~$Q_d$ at the outer boundary in such a way
that the following desirable properties are satisfied:
\begin{enumerate}
\item It satisfies the dissipative property~(DP), as detailed
  momentarily.
\item It should be~$h^3$ times a discrete approximation of the fourth
  order derivative, as in~\eqref{Diss_bulk_error}, of the dynamical
  variable on which it acts.
\end{enumerate}

The DP, as described in~\cite{CalLehReu03}, is the requirement that in
the inner product that induces the norm used to establish
stability,~$Q_d$ satisfies the inequality
\begin{align}\label{Dissipative_Property}
(\Psi, Q_d \Psi) \leq 0 \, ,
\end{align}
for any state vector~$\Psi$. In this subsubsection, for simplicity, we
assume that the state vector consists of a single gridfunction~$\Psi$
and work with the norm
\begin{align}\label{S_norm}
  (\Psi,\Psi)_\Upsilon
  = \Psi^T \Upsilon \Psi \,.
\end{align}
The second desirable property assures that this operator vanishes
like~$h^3$ in the continuum limit. In other words, it assures that the
dissipation term in each equation acts like a higher order error
associated with a finite differencing scheme, which we want to match
for every grid point.

We next consider the form that each of these properties alone gives
to~$Q_d$ at the outer boundary. We denote the dissipation operator
obtained by demanding the first property alone by~$Q_{d1}$ and that
obtained from the second property alone by~$Q_{d2}$. Presently, we do
not know how, or if, both can be imposed simultaneously. We will
ignore the coefficient~$\epsilon$ in our calculations, as it plays no
role there.

Substituting~\eqref{Diss_Bulk_TEM} in the norm on the left
of~\eqref{Dissipative_Property}, with the norm defined
by~\eqref{S_norm}, gives
\begin{align}\label{DP_with_S_norm}
  \Psi^T \Upsilon Q_d \Psi = - \| D^{(2)} \Psi \|^2
  + (\textrm{boundary terms}) \, ,
\end{align}
where
\begin{align}
\| D^{(2)} \Psi \|^2 = (D^{(2)} \Psi)^T (D^{(2)}\Psi) \, ,
\end{align}
is the trivial~$l^2$-norm. Here,~$D^{(2)}$ is a centered finite
difference operator which approximates the second order derivative of
a smooth function~$f$ projected on the grid at second order accuracy
and is defined as
\begin{align}
  (\Upsilon^{-2} D^{(2)} f)_I = \frac{f_{I-1} - 2 f_I + f_{I+1}}{h^2} =
  f_I'' + \frac{h^2}{12} f_I'''' + \cdots
  \label{D2}
\end{align}
in the bulk, and~$\Upsilon$ is defined by~\eqref{Dynamic_Quadrature},
with~$h_N = h/2$. The form of the first term on the right
of~\eqref{DP_with_S_norm} is not surprising because, in the continuum
setting, we have
\begin{align}
(f,f^{(4)}) \equiv & \int_0^{r_\mathscr{I}} f f'''' dr \nonumber \\
= & \int_0^{r_\mathscr{I}} (f'')^2 dr + (f f''' - f' f'') \Big|_0^{r_\mathscr{I}} \, .
\end{align}
The form of the boundary terms above will depend on the definition
of~$Q_d$ at the boundary. In order to satisfy the DP, as defined
by~\eqref{Dissipative_Property}, one possibility is to force the
boundary terms to be identically zero. This leads to defining~$Q_{d1}$
from the equation~$(\Psi,Q_{d1} \Psi) = - \| D^{(2)} \Psi \|^2$, to
  get
\begin{align}\label{Qd_DP}
Q_{d1} = - \Upsilon^{-1} (D^{(2)})^T D^{(2)} \, .
\end{align}
Here,~$D^{(2)}$ is defined by~\eqref{D2} for~$I = 0,\ldots,N-1$. At
the last grid point, we define~$D^{(2)}$ as
\begin{align}
  (\Upsilon^{-2} D^{(2)} f)_N &=  \frac{f_{N-2} - 2 f_{N-1} + f_N}{h^2}
  = f_N'' + h f_N''' + \cdots \, .
\end{align}
This gives
\begin{align}
(D^{(2)} f)_N = \frac{f_{N-2} - 2 f_{N-1} + f_N}{4} \, .
\end{align}
The resulting~$Q_{d1}$ is the same as~\eqref{Diss_Bulk_TEM} in the
bulk and takes the following form at the outer boundary:
\begin{align}\label{Qd_DP_matrix}
\left(
\begin{array}{ccccccc}
\cdot & \cdot & \cdot & \cdot & \cdot & \cdot & \cdot \\
\cdot & -\frac{6}{h} & \frac{4}{h} & -\frac{1}{h} & 0 & 0 & 0 \\
\cdot & \frac{4}{h} & -\frac{6}{h} & \frac{4}{h} & -\frac{1}{h} & 0 & 0 \\
\cdot & -\frac{1}{h} & \frac{4}{h} & -\frac{6}{h} & \frac{4}{h} & -\frac{1}{h} & 0 \\
\cdot & 0 & -\frac{1}{h} & \frac{4}{h} & -\frac{97}{16 h} & \frac{33}{8 h} & -\frac{17}{16 h} \\
\cdot & 0 & 0 & -\frac{1}{h} & \frac{33}{8 h} & -\frac{21}{4 h} & \frac{17}{8 h} \\
\cdot & 0 & 0 & 0 & -\frac{17}{8 h} & \frac{17}{4 h} & -\frac{17}{8 h} \\
\end{array}
\right) \, .
\end{align}
With this definition,~$Q_{d1} f \approx h^3 f^{(4)}$ in the
bulk and is~$\approx h f^{(2)}$ at the last three
grid points. Therefore, it is expected that~$Q_{d1}$ affects the
pointwise convergence at the last three grid points, at least in the
TEM scheme, as it dominates the truncation error there, which
is~$\approx h^2 f^{(3)}$ for that scheme. There is no sense in
incorporating the~TEM property in the definition of~$D^{(2)}$ in the
construction of~$Q_{d1}$ because doing so does not avoid these
lower-order terms in the final operator.

If we instead prioritize the second desirable property when
defining~$Q_d$ near the outer boundary, we need to redefine the
operator only at the last two grid points. This property assures that
the dissipation operator does not affect the pointwise accuracy of the
numerical scheme at any grid point. We do not need to incorporate TEM
to define this operator at the last two grid points, as it is
already~$O(h^3)$ and we ignored all the terms in our TEM
discretization of order higher than~$h^3$. Remember that we matched
all the~$h^2$ and~$h^3$ coefficients in the finite difference
approximation~$D$ to the partial derivative~$\p_r$ at all grid points
to derive our~TEM scheme. Demanding only that the dissipation operator
should correspond to~$h^3 f^{(4)}$ at its lowest order and ignoring
the associated errors, we need only a five point stencil to define it
at the last two grid points. From this, we obtain
\begin{align}\label{Diss_N-1}
  (Q_{d2} f)_{N-1} &=  \epsilon h^{-1}
  (-f_{N-4} + 4 f_{N-3} - 6 f_{N-2} + 4 f_{N-1} \nonumber\\
  & \quad - f_N) \nonumber\\
  &=  - \epsilon h^3 [h^{-4} D_-^3 D_+ f]_{N-1} \, ,
\end{align}
and
\begin{align}\label{Diss_N}
  (Q_{d2} f)_N &=  \epsilon h^{-1}
  (-f_{N-4} + 4 f_{N-3} - 6 f_{N-2} + 4 f_{N-1} \nonumber\\
  & \quad - f_N) \nonumber\\
  &=  - \epsilon h^3 [h^{-4} D_-^4 f]_N \, .
\end{align}
In this case, the operator~$Q_{d2}$ satisfies~\eqref{DP_with_S_norm},
where the boundary terms are merely obtained from the difference
between~$Q_{d2}$ and~$Q_{d1}$, and we get
\begin{align}
  (\Psi,Q_{d2} \Psi)_\Upsilon
  & = - \| D^{(2)} \Psi \|^2
  + \bigg( - \Psi_{N-4} \Psi_{N-1}
  - \frac{\Psi_{N-4} \Psi_N}{2} \nonumber \\
  & + 5 \Psi_{N-3} \Psi_{N-1}
  + 2 \Psi_{N-3} \Psi_N
  + \frac{\Psi_{N-2}^2}{16} \nonumber \\
  & - \frac{41 \Psi_{N-2} \Psi_{N-1}}{4}
  - \frac{15 \Psi_{N-2} \Psi_N}{8}
  + \frac{37 \Psi_{N-1}^2}{4} \nonumber \\
  & - \frac{13 \Psi_{N-1} \Psi_N}{4}
  + \frac{9 \Psi_N^2}{16} \bigg) \, .
\end{align}
It is not immediately clear if~$Q_{d2}$ satisfies the DP. Assuming
that we are treating the initial data for which the TEM scheme is
convergent, we can Taylor expand all~$\Psi_I$'s in the boundary term
at the last grid point to obtain
\begin{align}
  (\Psi,Q_{d2} \Psi)_\Upsilon
  = - \| D^{(2)} \Psi \|^2 + O(h^3) \, .
\end{align}
Therefore, at sufficient resolution, we can make the~$h^3$ term
smaller such that only the bulk term, which is negative definite,
dominates. In this weak sense~$Q_{d2}$ is still dissipative, even if
it does not satisfy the DP.

\subsubsection{In 3D, Spherical Polar Coordinates and Energy Norm}
\label{Subsec:Dissipation_Threading}

As we will be using the energy norm to perform our norm convergence
tests, the next step is to construct a dissipation operator which
satisfies the~DP directly in our energy norm, and in spherical polar
coordinates. Since the weights of~$\tilde{\Psi}$,~$\tilde{\Sigma}^+$
and~$\tilde{\Sigma}^-$ in our energy norm differ, we need to define
these operators differently for each gridfunction. This needs to be
done in such a way that a non-trivial dissipative effect is maintained
on the solution at the origin itself. Our basic strategy is to take an
operator~$Q_d$ known to satisfy the DP for a single gridfunction in
the~$(\cdot,\cdot)_{\Upsilon}$ norm used in the last section, and then
`thread' our weights into it. Schematically this looks like
like~$(W^{-\frac{1}{2}}) Q_d (W^{\frac{1}{2}})$ away from the
origin. Recalling that each~${W}\sim r^2$ near the origin we use
l'H\^opital's rule to regularize the operator there. The remaining
subtlety to overcome is the parity of our evolved variables, which are
a combination of even and odd quantities that makes the application of
l'H\^opital's rule delicate for general fields. To see this, note for
example that the second order differential operator~$\Delta \psi\equiv
r^{-1}\p^2_r (r\psi)$ is defined only on even functions, so a vector
Laplace operator (or some such) is required.

We now outline the complete construction. We start by taking the DP
operator~$Q_{d1}$ from before, now replacing the~$\epsilon$
parameter. From this we define two auxiliary operators
\begin{align}
\hat{Q}_1 = Q_{d1} \,,
\end{align}
which is well-defined on odd gridfunctions, and
\begin{align}\label{Diss_odd}
  \hat{Q}_2 = [r]^T \, Q_{d1} \, [r] \, ,
\end{align}
which is well-defined on even gridfunctions. Both satisfy the DP
using~$(\cdot,\cdot)_{\Upsilon}$. The next question is,
given~$\hat{Q}_1$ and~$\hat{Q}_2$, how to use them with our equations
of motion. Considering our evolution system we know that~$\Psi$ is an
even function. Using the parity conditions~\eqref{Parity} we can also
separate~$\tilde{\Sigma}^+$ and~$\tilde{\Sigma}^+$ into their even and
odd parts with,
\begin{align}
  \tilde{\Sigma}^+ &= \frac{\tilde{\Sigma}^+ + \tilde{\Sigma}^-}{2}
  + \frac{\tilde{\Sigma}^+ - \tilde{\Sigma}^-}{2} \, ,\nonumber\\
  \tilde{\Sigma}^- &= \frac{\tilde{\Sigma}^- + \tilde{\Sigma}^+}{2}
  + \frac{\tilde{\Sigma}^- - \tilde{\Sigma}^+}{2} \,.
\end{align}
The first terms on the right are the even parts of~$\tilde{\Sigma}^+$
and~$\tilde{\Sigma}^-$ respectively, and the second their odd
parts. Defining
\begin{align}
  \tilde{\Sigma}^e := \frac{\tilde{\Sigma}^+
    + \tilde{\Sigma}^-}{2}\,, \qquad \,
  \tilde{\Sigma}^o := \frac{\tilde{\Sigma}^+
    - \tilde{\Sigma}^-}{2} \, ,
\end{align}
we get
\begin{align}
  \tilde{\Sigma}^+ = \tilde{\Sigma}^e + \tilde{\Sigma}^o \,,\qquad
  \, \tilde{\Sigma}^-
  = \tilde{\Sigma}^e - \tilde{\Sigma}^o \, .
\end{align}
Observe that the state vector~$\mathbf{U}$ can be written
as~$\mathbf{U} = (\tilde{\Psi}, \tilde{\Sigma}^+, \tilde{\Sigma}^-)^T$
or as~$\mathbf{V} := (\tilde{\Psi}, \tilde{\Sigma}^e,
\tilde{\Sigma}^o)^T$. These two representations are related
as~$\mathbf{U} = \mathbf{T} \mathbf{V}$, with
\begin{align}
\mathbf{T} = \left( \begin{array}{ccc}
1 & 0 & 0 \\
0 & 1 & 1 \\
0 & 1 & - 1
\end{array} \right)\,.
\end{align}
Observe that~$\mathbf{T}=\mathbf{\hat{T}}\mathbf{\Lambda}$
with~$\mathbf{\Lambda}=\textrm{diag}(1,\sqrt{2},\sqrt{2})$
and~$\mathbf{\hat{T}}$ a symmetric, orthogonal matrix. The weight
matrices in our energy norm satisfy the parity conditions
\begin{align}
  \tilde{W}(-r) = \tilde{W}(r) \, \textrm{ and }
  \, \tilde{W}^\pm(-r) = \tilde{W}^\mp(r) \, .
\end{align}
Now, away from the origin, we can define the dissipation operator as,
\begin{align}\label{Diss_DP_en_norm}
\mathbf{Q = H^{-\frac{1}{2}} \hat{T} Q_d \hat{T} H^\frac{1}{2}} \, ,
\end{align}
with 
\begin{align}\label{Diss_oe}
  \mathbf{Q_d}
  = \left( \begin{array}{ccc}
    \hat{Q}_1 & 0 & 0 \\
    0 & \hat{Q}_1 & 0 \\
    0 & 0 & \hat{Q}_2
  \end{array} \right) \,.
\end{align}
At the origin we simply apply l'H\^opital's rule which, as mentioned
above, results in a regular operator. Crucial here is
that~$\mathbf{Q_d}$ satisfies the DP in the~$(\cdot,\cdot)_\Upsilon$
norm. This definition guarantees that the dissipation operators
respect the parity of the fields to which they are applied, because
\begin{align}
  \mathbf{T^{-1}QU} =
  (\mathbf{T}^T\mathbf{H}^{\frac{1}{2}}\mathbf{T})^{-1}\mathbf{Q_d}
  (\mathbf{T}^T\mathbf{H}^{\frac{1}{2}}\mathbf{T}) \mathbf{V}\,,
\end{align}
where both~$\mathbf{Q_d}$, the matrix given in parentheses on the
right, and its inverse, respect parity. To verify that this choice
satisfies the DP in our energy norm we compute directly obtaining
\begin{align}
  \mathbf{U}^T\Upsilon\mathbf{H}\mathbf{Q}\mathbf{U}
  = (\mathbf{\hat{T}}\mathbf{H}^{\frac{1}{2}}\mathbf{U})^T
  \Upsilon\mathbf{Q_d}
  (\mathbf{\hat{T}}\mathbf{H}^\frac{1}{2}\mathbf{U})\leq 0\,,
\end{align}
as desired. This requires the fact~$\mathbf{H}^{\frac{1}{2}}$
and~$\mathbf{\hat{T}}$ commute with~$\Upsilon$, along with the other
properties noted above. In our discretization we {\it use} the
operator by choosing
\begin{align}
 \frac{d}{dt}\mathbf{U}&=\dots+\mathbf{Q}\mathbf{U}\,.
\end{align}
where the ellipses denote right-hand-sides obtained solely from the
earlier scheme. More explicitly, we can write this as
\begin{align}
  \dot{\tilde{\Psi}} = & \cdots + ([r^2] + [F] \tilde{W})^{-\frac{1}{2}}
  \hat{Q}_1  ([r^2] + [F] \tilde{W})^\frac{1}{2} \tilde{\Psi} \,, \nonumber \\
  \dot{\tilde{\Sigma}}^+ = & \cdots
  + \frac{1}{4} \Big[(\tilde{W}^+)^{-\frac{1}{2}}
    (\hat{Q}_1 + \hat{Q}_2) (\tilde{W}^+)^\frac{1}{2}
    \tilde{\Sigma}^+ \nonumber \\
    & + (\tilde{W}^+)^{-\frac{1}{2}} (\hat{Q}_1 - \hat{Q}_2)
    (\tilde{W}^-)^\frac{1}{2} \tilde{\Sigma}^- \Big] \, , \nonumber\\
  \dot{\tilde{\Sigma}}^- = & \cdots
  + \frac{1}{4} \Big[ (\tilde{W}^-)^{-\frac{1}{2}}
    (\hat{Q}_1 - \hat{Q}_2) (\tilde{W}^+)^\frac{1}{2}
    \tilde{\Sigma}^+ \nonumber \\
    & + (\tilde{W}^-)^{-\frac{1}{2}} (\hat{Q}_1 + \hat{Q}_2)
    (\tilde{W}^-)^\frac{1}{2} \tilde{\Sigma}^- \Big] \,,
\end{align}
with suitable application of l'H\^opital's rule understood at the
origin. To derive this we use the adjusted energy
norm~\eqref{adjusted_energy}. To see that the dissipation effectively
removes energy from the system we need only compute the time
derivative of the energy norm, obtaining,
\begin{align}
  \dot{\tilde{E}} = \cdots + \tfrac{1}{2} \mathbf{U}^T \mathbf{H}
  \Upsilon \mathbf{Q} \mathbf{U} \, ,
\end{align}
as desired. We close with the observation that the dissipation
operator is not defined at all grid points for which~$F R'$ becomes
unbounded, as is the case with~LMKGE at~$\mathscr{I}^+$. (Although in
that case, neither the SBP-Stable or the SBP-TEM scheme are defined
anyway).

\section{Numerical Evolution}\label{Numerical_Evolution}

\subsection{Code Description}

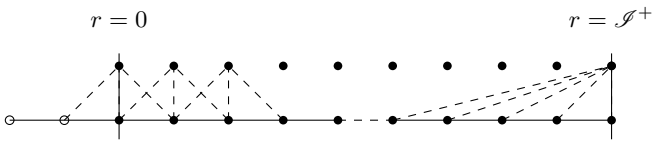
\begin{figure}[tp]
\center
\begin{tikzpicture}[scale=0.36]
  \draw (-2cm, 0cm) -- (10cm, 0cm);
  \draw[dashed] (10cm, 0cm) -- (12cm, 0cm);
  \draw (12cm, 0cm) -- (20cm, 0cm);
  \draw (2cm, 70pt) -- (2cm, - 20pt);
  \draw (2cm, 0cm) node[above=32pt] {$r=0$};
  \draw (20cm, 70pt) -- (20cm, - 20pt);
  \draw (20cm, 0cm) node[above=32pt] {$r=\mathscr{I}^+$};
  \foreach \x in {-2, 0} \draw (\x cm, 0cm) circle (4.5pt);
  \foreach \x in {2, 4, 6, 8, 10} \fill (\x cm, 0cm) circle (4.5pt);
  \foreach \x in {12, 14, 16, 18, 20} \fill (\x cm, 0cm) circle (4.5pt);
  \fill (2cm, 2cm) circle (4.5pt); \foreach \x in {0,2,4}
  \draw[dashed] (\x cm,0cm) -- (2cm, 2cm);
  \fill (20cm, 2cm) circle (4.5pt); \foreach \x in {12,14,16,18,20}
  \draw[dashed] (\x cm,0cm) -- (20cm, 2cm);
  \foreach \x in {4, 6, 8, 10, 12, 14, 16, 18, 20}
  \fill (\x cm, 2cm) circle (4.5pt);
  \foreach \x in {2, 4, 6} \draw[dashed] (\x cm,0cm) -- (4cm, 2cm);
  \foreach \x in {4, 6, 8} \draw[dashed] (\x cm,0cm) -- (6cm, 2cm);
\end{tikzpicture}
\caption{A schematic diagram showing the non-staggered grid and second
  order finite-difference stencils (dashed line segments) that we use
  in the numerical implementation. All the grid points are uniformly
  spaced. The stencil is centered everywhere except on the boundary,
  which is at~$\mathscr{I}^+$, where it leans left and uses five grid
  points in the~SBP-TEM scheme and two for the SBP-Stable one. The
  values of the variables on the black-filled points are evolved using
  the equations of motion. The empty circles on the left denote the
  ghost points, which are required to calculate derivatives at the
  origin and are filled using the parity conditions described in
  subsection~\ref{Origin}.}
\label{fig_stencil}
\end{figure}

We employ a 1-dimensional code, written for spherically symmetric
systems in spherical polar coordinates on hyperboloidal slices, using
the same infrastructure as that of the work
in~\cite{VanHusHil14,VanHus17,Van15}. We use a compactified radial
coordinate and hyperboloidal time as explained in
Sec.~\ref{Intro_hyp_slice}. The implementation uses the method of
lines with a fourth order Runge-Kutta for time integration. We work
with second order accurate finite difference operators~$D$
and~$\tilde{D}$ to approximate the spatial derivatives derived from
the~SBP-TEM scheme, as given
by~\eqref{D_bulk},~\eqref{D_out_bound},~\eqref{DtildeP_bulk}
and~\eqref{DtildeP_out_bound}, and the SBP-Stable scheme, as explained
in Sec.~\ref{Stable_SBP}. Our spatial grid has grid points at the
origin and at~$\mathscr{I}^+$, as shown in
Fig.~\ref{fig_stencil}. Regarding dissipation, we use a fourth order
Kreiss–Oliger-like dissipation operator~$Q_{d2}$ satisfying the~TEM
property, given by~\eqref{Diss_Bulk_TEM},~\eqref{Diss_N-1}
and~\eqref{Diss_N}, with the~SBP-TEM scheme and by the
operator~$\mathbf{Q}$ acting on the whole state vector and satisfying
the~DP, as constructed in Sec.~\ref{Subsec:Dissipation_Threading},
with the SBP-Stable scheme. We treat the origin as an inner grid
point, for which we introduce ghost points on its left with the same
grid spacing as on the physical grid, and populate them using the
parity conditions~\eqref{Parity}. Then all the finite difference
operators at the origin are defined in the same way as on a typical
interior grid point, using a centered stencil,
cf. Fig.~\ref{fig_stencil}. It suffices to have a single ghost point
in order to define~$D$ and~$\tilde{D}$ at the origin, but we need two
such ghost points to define the dissipation operators there. In
contrast, the outer boundary is a true boundary which is placed
at~$\mathscr{I}^+$. All the operators defined there are completely
left sided.

\subsection{Implementation}

We experimented with various different values of the compactification
parameter~$n$, defined in~\eqref{Compactification}, obtaining
qualitatively similar results. For brevity, in our presentation we
choose the compactification function~$\Omega(r)$ given
by~\eqref{Compactification_regular} with~$n=2$ and~$r_\mathscr{I} =
1$. We observe that using Evan's method at the origin gives both the
pointwise and norm convergence plots visually indistinguishable from
those obtained by rewriting the equations there using the
l'H\^opital's rule. This is exactly what we expect from the
explanations given in Sec.~\ref{Origin}. In our implementation, we use
the l'H\^opital's rule as it has an advantage that the energy norm
becomes independent of the resolution. We set the height function~$H$
such that~$H'(R(r)) = 1 - 1/R'(r)$. We use~$\chi = \sqrt{1 + R^2}$ as
a rescaling function and~$\gamma_2 = 0$ for all our purposes, as
justified in subsection~\ref{Subsection:Constraints}. We take~$N=200$
as our base resolution and increase this number by a factor of~$2$
whenever performing convergence tests. This gives~$h = r_\mathscr{I}/N
= 0.005$ at the original resolution. The~Courant-Friedrichs-Lewy
factor, defined as the ratio between the timestep and grid
spacing~$\delta t/h$, is taken to be~$0.5$ unless stated otherwise. We
work with the~$(\tilde{\Psi}, \tilde{\Sigma}^+, \tilde{\Sigma}^-)$
system for all three choices of~$F$ considered here. We have tested
several families of initial data, but, in our presentation, we take,
\begin{align}\label{inidatagauss}
\psi(0,R) = a e^{-\lambda R^2} \, \textrm{, and, } \, \pi(0,R) = 0 \, ,
\end{align}
with~$a = 0.01$ and $\lambda=1$, unless stated otherwise, and compute
the initial data for the~$(\tilde{\psi}, \tilde{\sigma}^+,
\tilde{\sigma}^-)$ variables according to the transformation
rules~\eqref{characteristic_variables} and~\eqref{pss_rescaling}.

\subsection{Results and Interpretation}

\subsubsection{Linear Wave Equation,~$F=0$}

Without adding dissipation, the evolved variables look quite noisy at
the origin, for both SBP-TEM and SBP-Stable schemes. The reason for
the noise is most likely the non-smoothness, mentioned in
section~\ref{Origin}, that arises from our choice of~$\chi$ in
combination with our parity conditions.  Since our primary interest is
in the regularization at~$\mathscr{I}^+$, and in the future we will
employ a multipatch method that avoids the coordinate singularity at
the origin, we have not invested a huge effort in improving the
treatment there. Instead we use a small amount of dissipation to
suppress the noise. Interestingly, setting for the dissipation
parameter~$\epsilon = 0.002$ suffices to damp almost all of this noise
by~$t=2$; with this level of dissipation the amplitude of the solution
at our base resolution is down to~$\sim 10^{-8}$ by~$t=10$.

Each of our schemes is naturally associated with a different
dissipation operator, SBP-Stable with the dissipation
operator~$\mathbf{Q}$ which acts on the whole state vector and
satisfies the~DP, and SBP-TEM with~$Q_{d2}$ which acts
variable-by-variable and has clean pointwise properties. If we use
instead~$\mathbf{Q}$ with SBP-TEM we see that pointwise convergence is
damaged, whereas if we use~$Q_{d2}$ with SBP-Stable we see at
particular times a small, though convergent, growth in the energy of
the solution. Matching the dissipation operators with their natural
discretization plays to the strengths of each of the two methods and
works well.

Returning to Fig.~\ref{fig_LWE_Contour_plot}, we see the basic
behavior of the massless scalar field satisfying the LWE in our
simulations. The initial narrow pulse at the origin, chosen to be
Gaussian-like as in~\eqref{inidatagauss} with~$a=0.01$
and~$\lambda=100$, propagates to~$\mathscr{I}^+$ with speed equal to
unity, as expected from our construction in
Sec.~\ref{Intro_hyp_slice}. Here, we plot the absolute value of the
rescaled field~$|\tilde{\psi}|$. The plot shows two bursts of the
pulse because of the time symmetry in our initial data obtained by
taking~$\pi(T=0,R)=0$. Most of the region looks white because for
clarity we only show the values
for~$10^{-6}<|\tilde{\psi}|<10^{-3}$. This plot was generated using
the SBP-Stable scheme with a little ($\epsilon=0.002$)
dissipation. This plot also shows a small amount of noise at the
origin which gets damped with time because of the dissipation.

\begin{figure}
\includegraphics[scale=0.45]{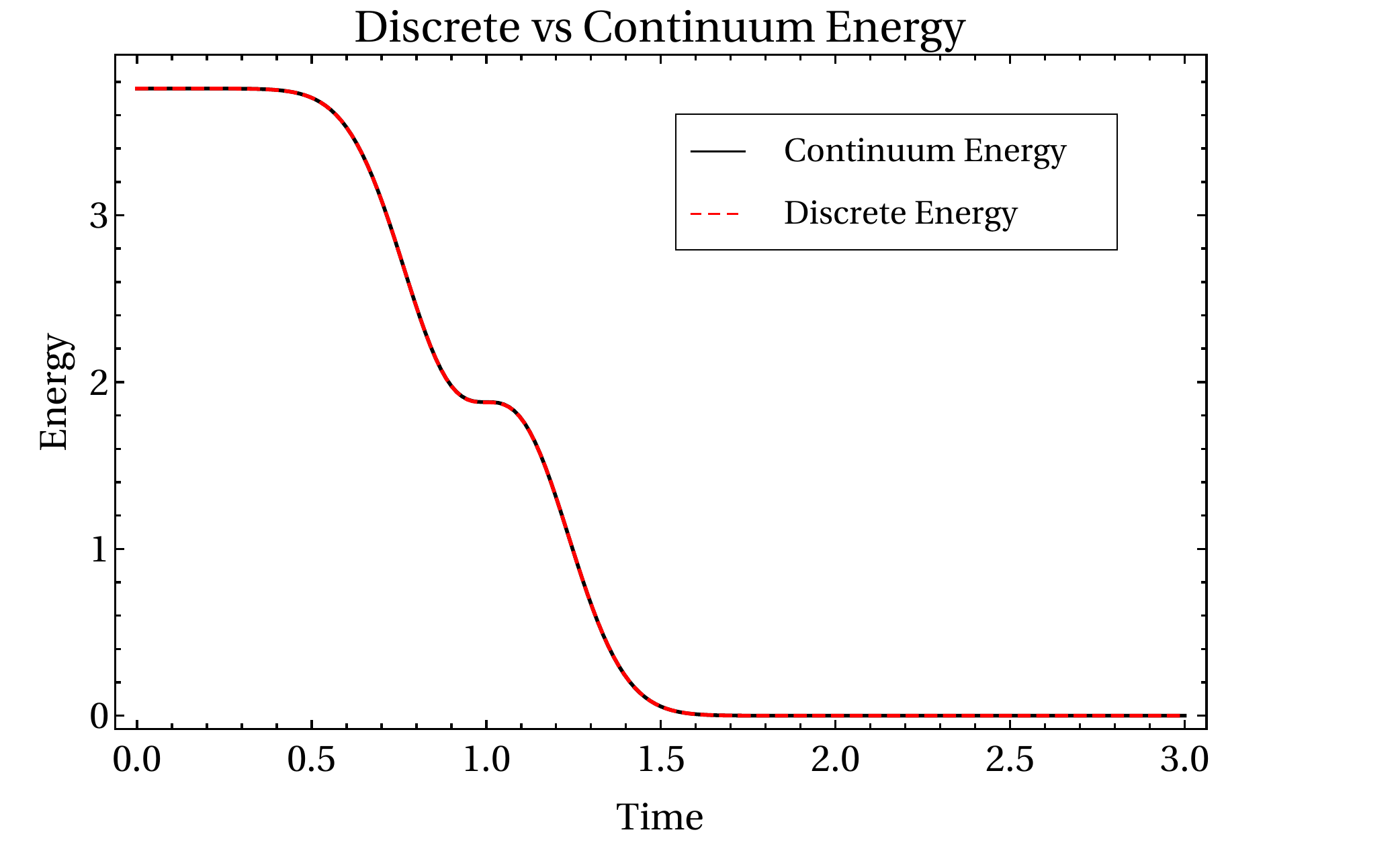}
\caption{Comparison of the continuum and discrete energies as a
  function of time for the initial data specified in the main text.
  This is the same data as plotted in Fig.~\ref{fig_LWE_Contour_plot}, 
  and so it makes sense that as each of the two pulses hit the outer
  boundary the energy drops rapidly.
  \label{fig_LWE_Disc-Cont_En}}
\end{figure}

In order to test the correctness of the implementation, we compare the
decay rate over time of our approximation to the physical
energy~\eqref{Ehat_resc} with that of the analytical one. Complete
agreement between the two is demonstrated in
Fig.~\ref{fig_LWE_Disc-Cont_En} for the SBP-Stable scheme. To generate
these curves, we consider the general solution of the LWE in spherical
symmetry,
\begin{align}
\psi(T,R) = \frac{f(T+R) - f(T-R)}{R} \, ,
\end{align}
and then rewrite it in terms of hyperboloidal coordinates and
choose~$f(x) = e^{-x^2}$. With this~$f$ we build the initial data for
the corresponding numerical setup. The numerical solution plotted is
constructed at our lowest resolution,~$N=200$.

\begin{figure}
\includegraphics[scale=0.45]{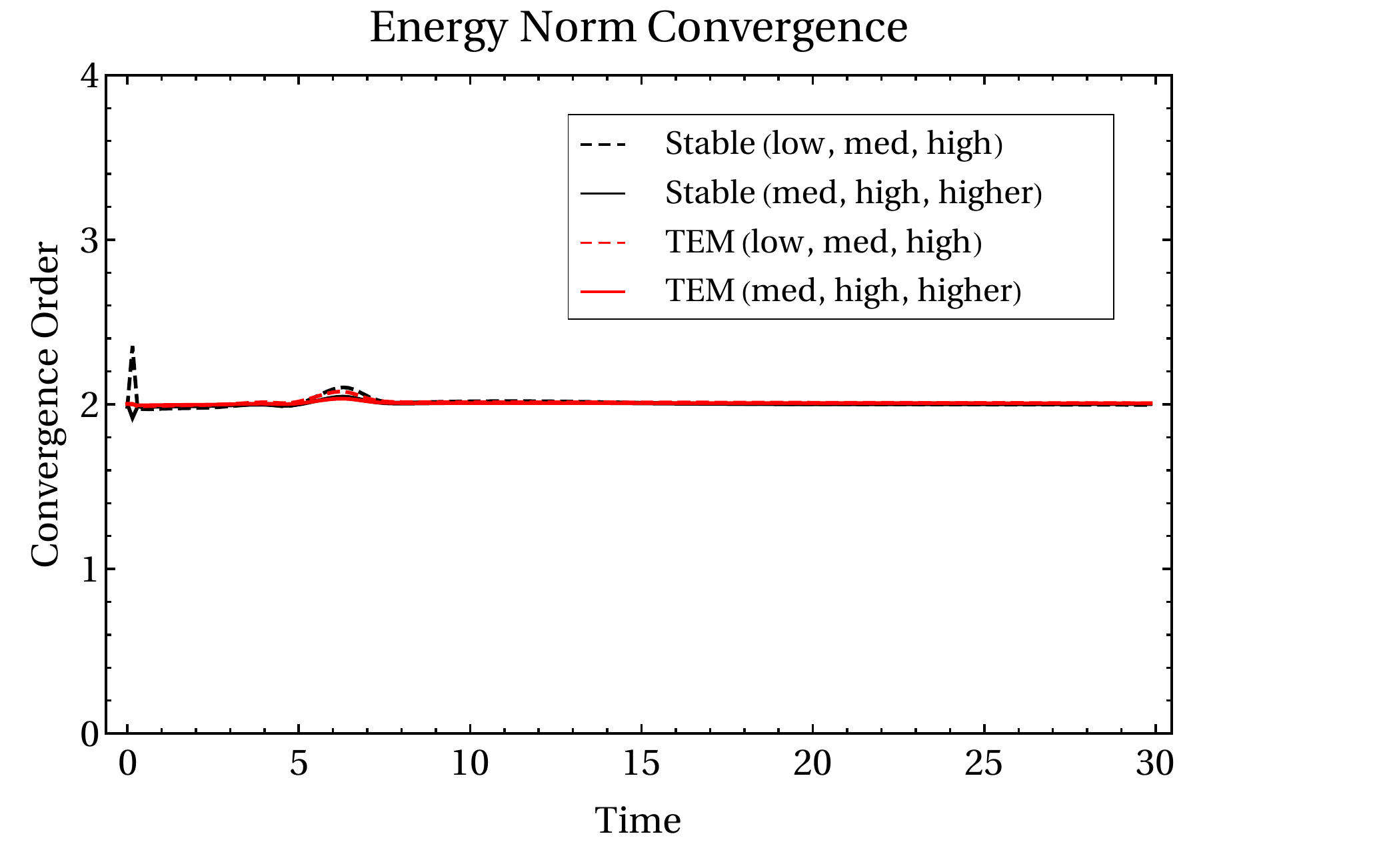}
\caption{Convergence order of our scalar field obeying LWE in the
  adjusted energy norm given by~\eqref{adjusted_energy}. The red
  curves show the same for the~SBP-TEM discretization and the black
  curves for the SBP-Stable one. Both are near-perfect, although it is
  true that the SBP-Stable plot would not be as clean if we focused
  on the physical energy~\eqref{Ehat_resc} instead, because,
  as can be understood from Fig.~\ref{fig_LWE_Disc-Cont_En} the energy
  present after~$t\sim 3$ is negligible, and the effects of
  dissipation start to dominate the error in~$\tilde{\Sigma}^\pm$
  after around~$t\sim 7$.}
\label{fig_conv_order_en_norm}
\end{figure}

We now compare the~SBP-TEM and SBP-Stable schemes through the norm and
pointwise convergence curves with a specific focus
on~$\mathscr{I}^+$. In Fig.~\ref{fig_conv_order_en_norm}, we see the
norm convergence plots in the adjusted norm~\eqref{adjusted_energy}
for the two schemes, plotted in different colors, and for different
resolutions plotted in solid and dashed curves. At late times a small,
smooth, stationary, though convergent feature remains in~$\Psi$ (not
shown here). We interpret this as the constraint violation induced by
the dissipation. This violation dominates the other errors by about
three orders of magnitude towards the end of the
evolution. Fig.~\ref{fig_conv_order_en_norm} shows almost perfect
second order convergence for all times in both schemes, as expected.

If we construct a similar plot using the physical
energy~\eqref{Ehat_resc} that is, without adding the~$\tilde{\Psi}^2$
term, the stationary error is completely eliminated and the remaining
errors start dominating. In the~SBP-TEM scheme, all these remaining
errors still converge at second order and we again observe a perfect
second order norm convergence with only small wiggles in some time
intervals. These wiggles are observed to be completely dependent on
the dissipation, as increasing the dissipation parameter~$\epsilon$
increases their amplitude. Since these errors converge faster than
those produced by the~SBP-TEM scheme, these wiggles diminish rapidly
by increasing the resolution. On the other hand, in the SBP-Stable
scheme, this convergence order starts drifting to~$\sim 3$ at late
times. This is because, at late times, errors introduced by the
dissipation operator start dominating. Although these errors converge
like~$h$ pointwise at the last three grid points,
cf. Sec~\ref{Diss_L2_Norm}, they can be easily seen to converge
like~$h^3$ in the norm. This appears to be the price for guaranteed 
stability. We do not observe this behavior in
the~SBP-TEM scheme because in this scheme the energy flux
through~$\mathscr{I}^+$ depends on the resolution. 

\begin{figure}
\includegraphics[scale=0.425]{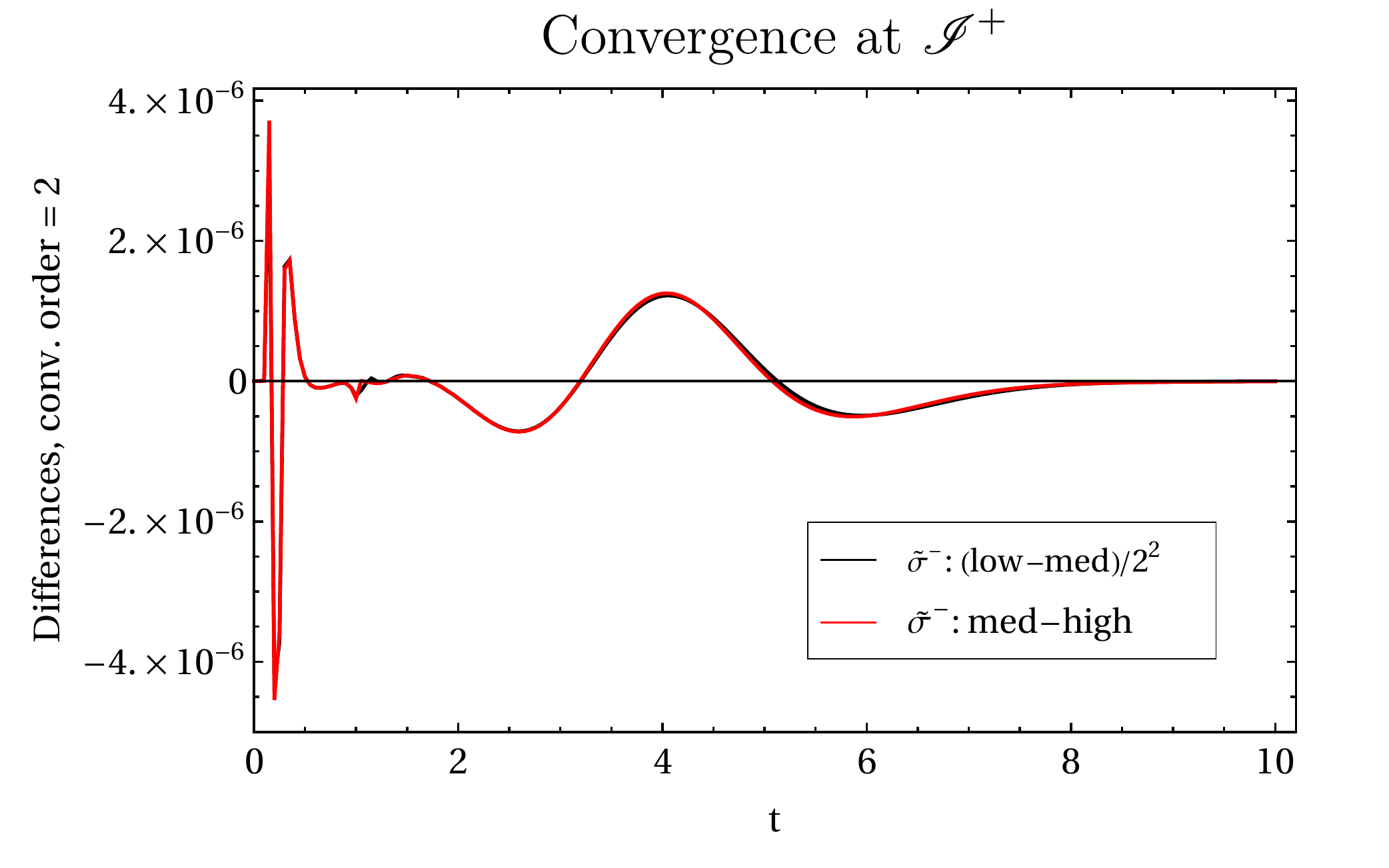}
\caption{Convergence of~$\tilde{\sigma}^-$ at~$\mathscr{I}^+$ for the
  scalar field obeying the LWE in the SBP-Stable discretization.}
\label{fig_conv_scri_stable}
\end{figure}

We now consider pointwise convergence. Since the SBP-TEM scheme is
designed to converge at second order at all grid points for suitable
initial data, we expect all the errors to converge pointwise
like~$h^2$ even at~$\mathscr{I}^+$, at least for a large class of
initial data. On the other hand, the SBP-Stable scheme uses various
finite difference operators at the last grid point, some of which are
only~$O(h)$, so we might expect a decline in convergence order in this
scheme at~$\mathscr{I}^+$. Interestingly, this is not what we
observe. Figure~\ref{fig_conv_scri_stable} shows clean second order
convergence of~$\tilde{\sigma}^-$ at~$\mathscr{I}^+$ in the SBP-Stable
scheme, and we obtain similar results for~$\tilde{\psi}$
and~$\tilde{\sigma}^+$. The equivalent plot for the~SBP-TEM scheme
looks even better. We observe with that scheme a smaller amplitude of
the error at~$\mathscr{I}^+$ by about a factor of two.

\begin{figure*}
\centering
\includegraphics[width=0.325\textwidth]{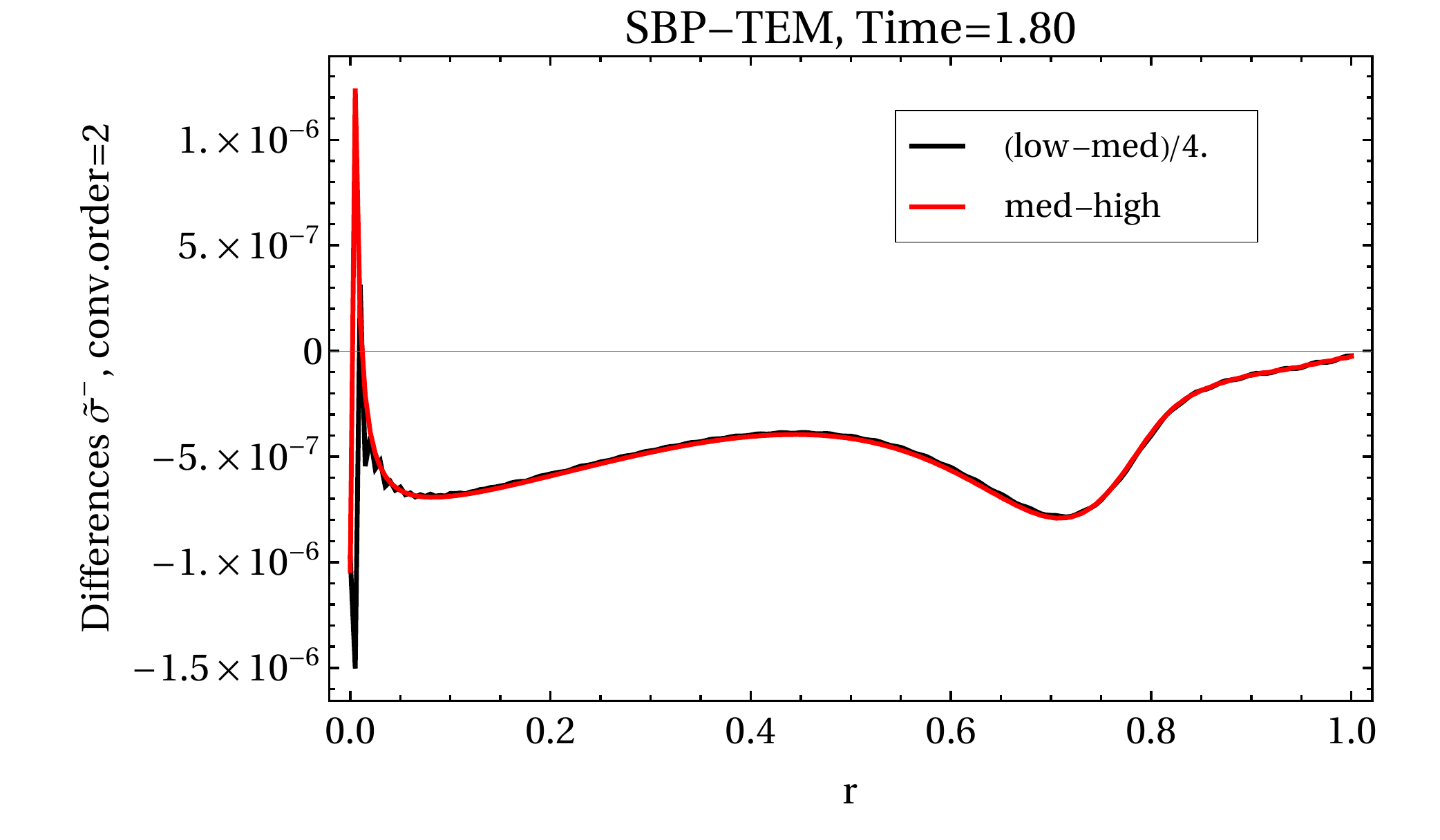}
\includegraphics[width=0.325\textwidth]{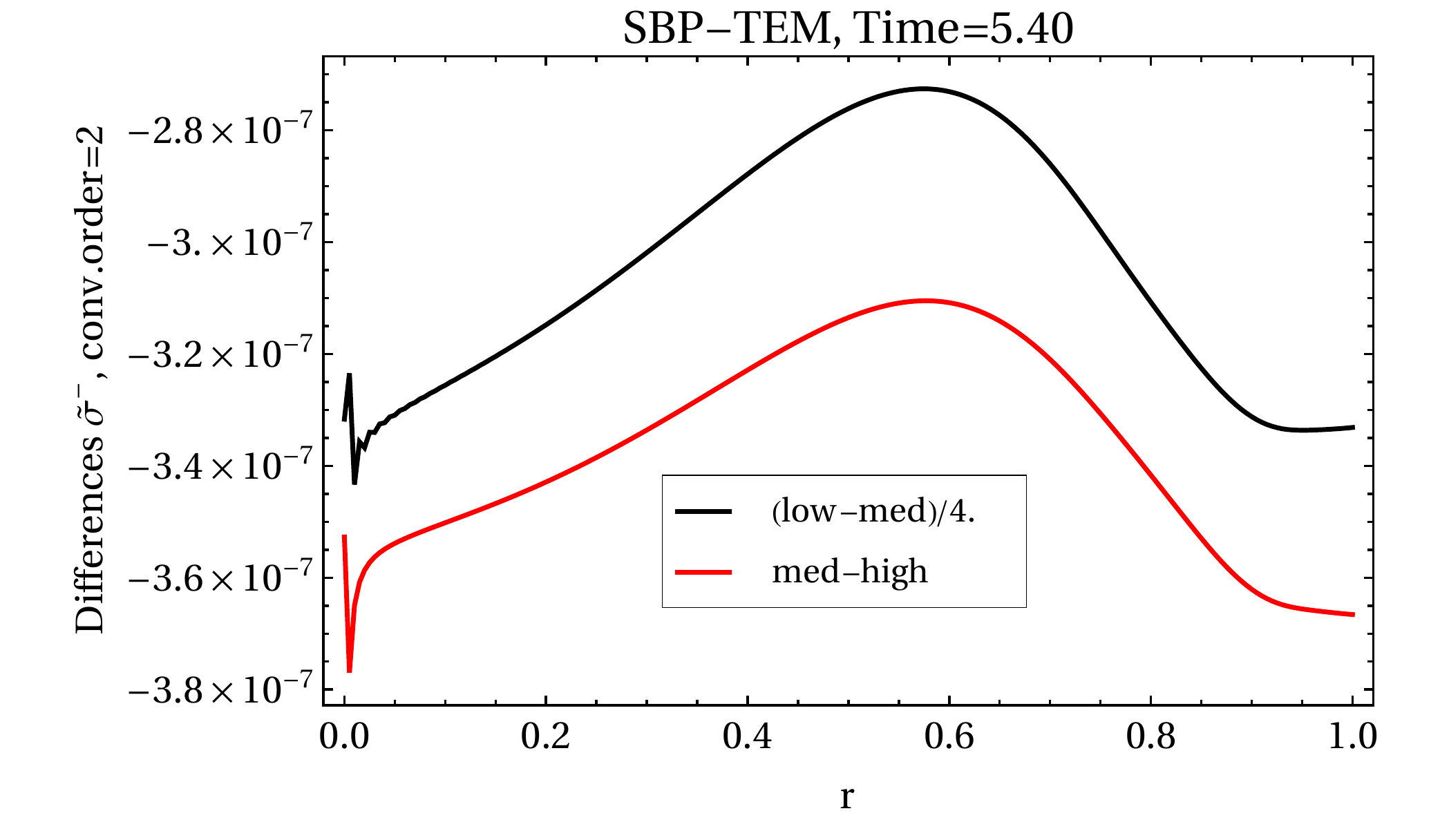}
\includegraphics[width=0.325\textwidth]{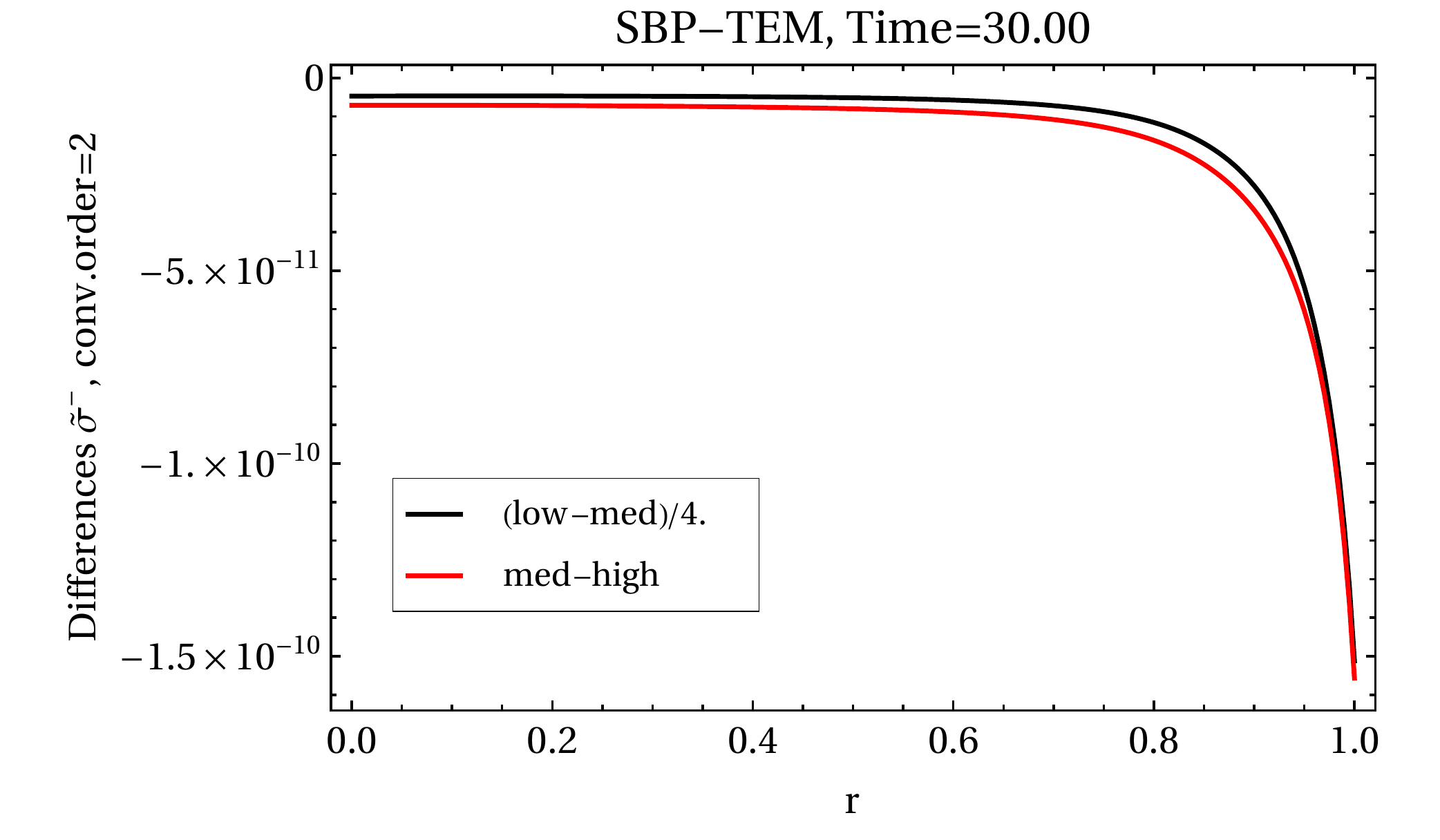}
\includegraphics[width=0.325\textwidth]{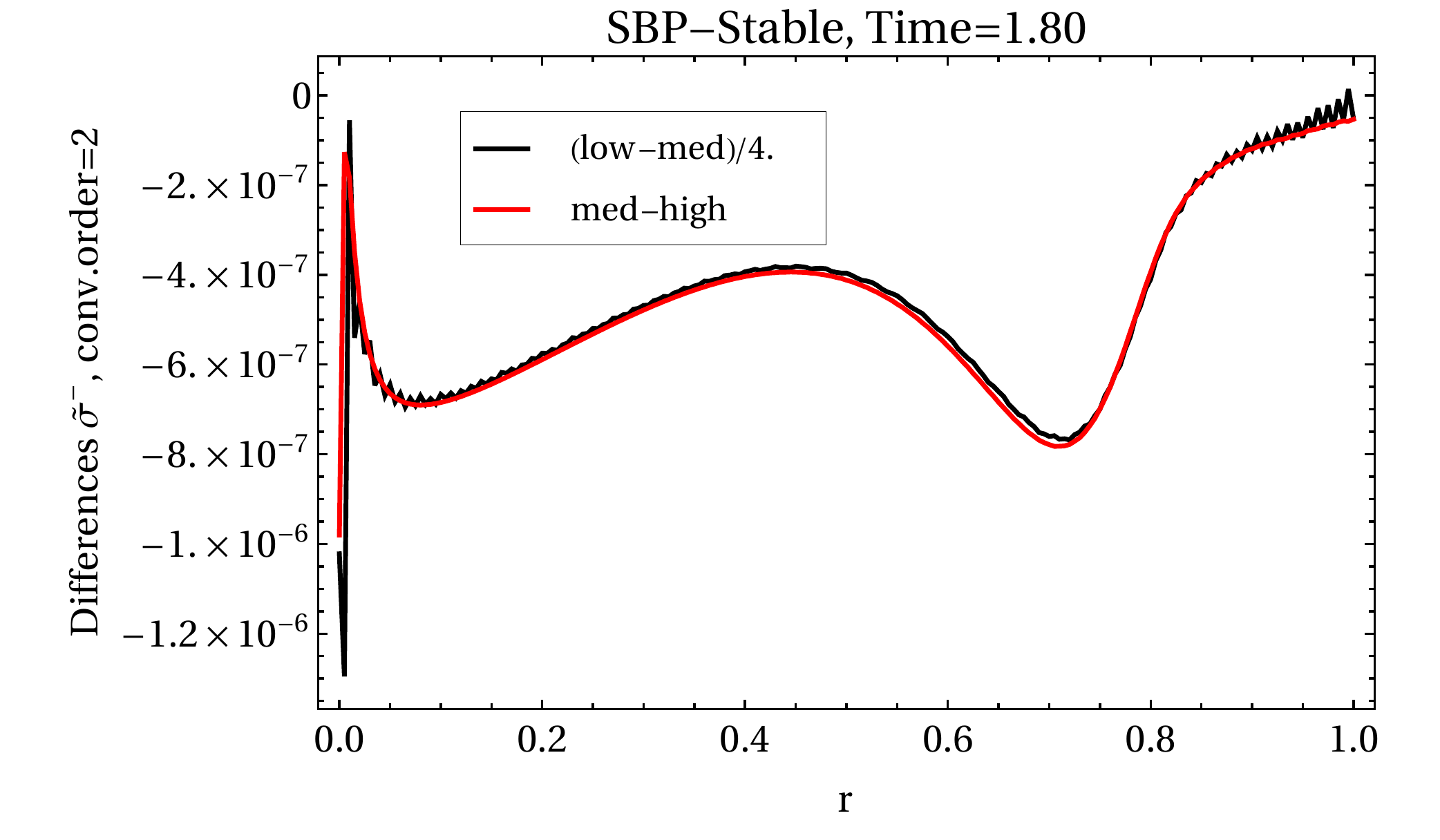}
\includegraphics[width=0.325\textwidth]{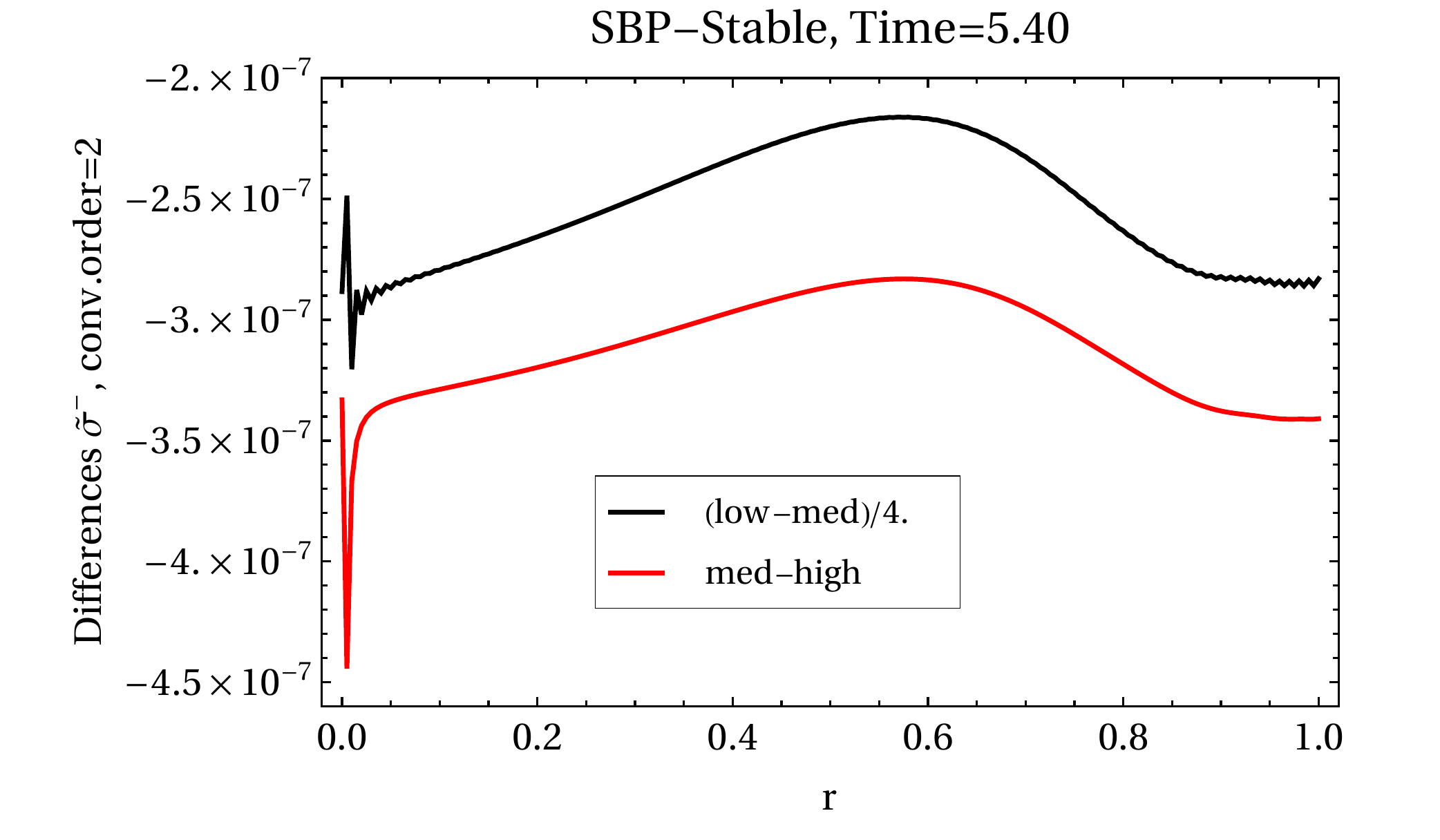}
\includegraphics[width=0.325\textwidth]{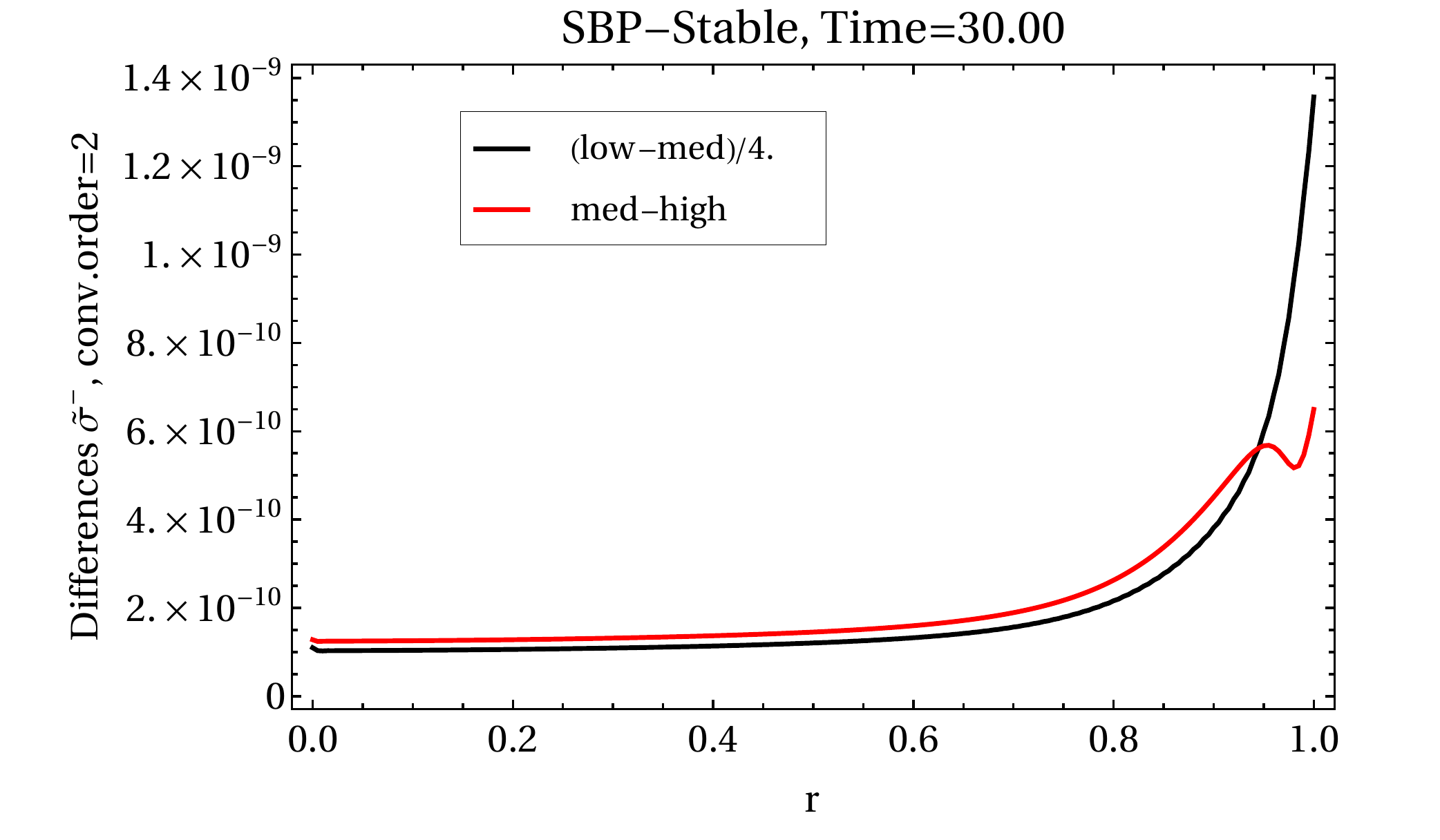}
\caption{Pointwise convergence of the scalar field obeying the
  LWE. The upper row shows the plots obtained from the~SBP-TEM
  discretization scheme at~$t=1.8$,~$5.4$ and~$30$ and the bottom
  lower row shows the equivalent plots obtained from the SBP-Stable
  scheme. We can observe how at late times the outer boundary starts
  affecting the pointwise convergence in the SBP-Stable scheme.}
\label{fig_LWE_point_conv}
\end{figure*}

In Fig.~\ref{fig_LWE_point_conv}, we compare pointwise convergence in
the two different schemes. The top row shows the pointwise convergence
curves in the~SBP-TEM scheme at three different instants, the bottom
the equivalent plots with the SBP-Stable scheme. The first column
shows how the noise at the origin dominates the errors generated on
the rest of the grid. We expect that this source of error could be
reduced by adjusting~$\chi$ to obtain smoothness at the origin. At
this instant, both sets of curves look essentially the same. In the
second column, we show equivalent plots at some intermediate time when
we observe a small wiggle on the norm convergence plot. As described
before, this wiggle is there due to the errors introduced by the
dissipation operator. At this instant, which corresponds to the small
wiggle in the convergence plot Fig.~\ref{fig_conv_order_en_norm}, we
can see that the plots for both schemes do not overlap. In the last
column, we see a typical pointwise convergence behavior at late
times. The bottom right plot explains the deviation in the norm
convergence in the SBP-Stable scheme in the physical energy. These
last panels clearly demonstrate the superiority of the~SBP-TEM scheme
over the~SBP-Stable one at late times on this initial data.

While working with the continuum equations, if we start with
constraint satisfying initial data, the equations of motion assure
that the constraint~\eqref{Constraint_cont} in the analytic solution
is satisfied for all times. However, in the discrete case, the
constraint~\eqref{Constraint_disc} is violated even for the initial
data. This violation is approximated in our scheme as
\begin{align}
  \hat{\mathcal{C}}_I = \frac{1}{2R'_I - 1}
  \left( \frac{h^2}{6} \Psi_I''' + \cdots \right) \, ,
\end{align}
for~$I = 0,\ldots,N$. Therefore, we expect the constraint violation to
converge at second order. In Sec.~\ref{Subsection:Constraints}, we
showed that, in the absence of dissipation~$\dot{\hat{\mathcal{C}}} =
0$, independent of the choice of the discretization scheme. Adding
dissipation terms to our equations however leads to a non-trivial form
of~$\dot{\hat{\mathcal{C}}}$. This is exactly what we observe in our
numerical results. For~$\epsilon > 0$, a near stationary constraint
violation appears on the grid, slowly evolving because of the
dissipation, but vanishing with increasing resolution.

\subsubsection{Linear Wave Equation with Potential,~$F= 1/\chi^2$}

The system~$F= 1/\chi^2$ and other models with potentials are
interesting for our methods for the following reason. In spherical
symmetry, Eq.~\eqref{LWEP} expressed in terms of the null
coordinates~$u = T - R$ and~$v = T + R$ shows that the rescaled
field~$\bar{\psi} = R \psi$ satisfies the equation
\begin{align}
\p_u \p_v \bar{\psi} = \p_v \p_u \bar{\psi} = - F \bar{\psi}\, .
\end{align}
Since, in spherical symmetry,~$\p_u \bar{\psi}$ and~$\p_v \bar{\psi}$
represent the characteristic variables, respectively, the above
equation simply means that all the `outgoing modes' of~$\bar{\psi}$,
and hence of~$\psi$, are coupled to all the incoming ones via the
potential, and vice versa. This coupling is dangerous in the
hyperboloidal setup, because if high frequency incoming modes are
generated near~$\mathscr{I}^+$ they will necessarily be poorly
resolved on the grid. We are now considering $F=
1/\chi^2$, where the coupling, which is completely absent in the
LWE, decreases with increasing radius like~$1/R^2$ and is hence
absent at the last grid point. In the next section we consider a much
more extreme example.

\begin{figure}
\includegraphics[scale=0.45]{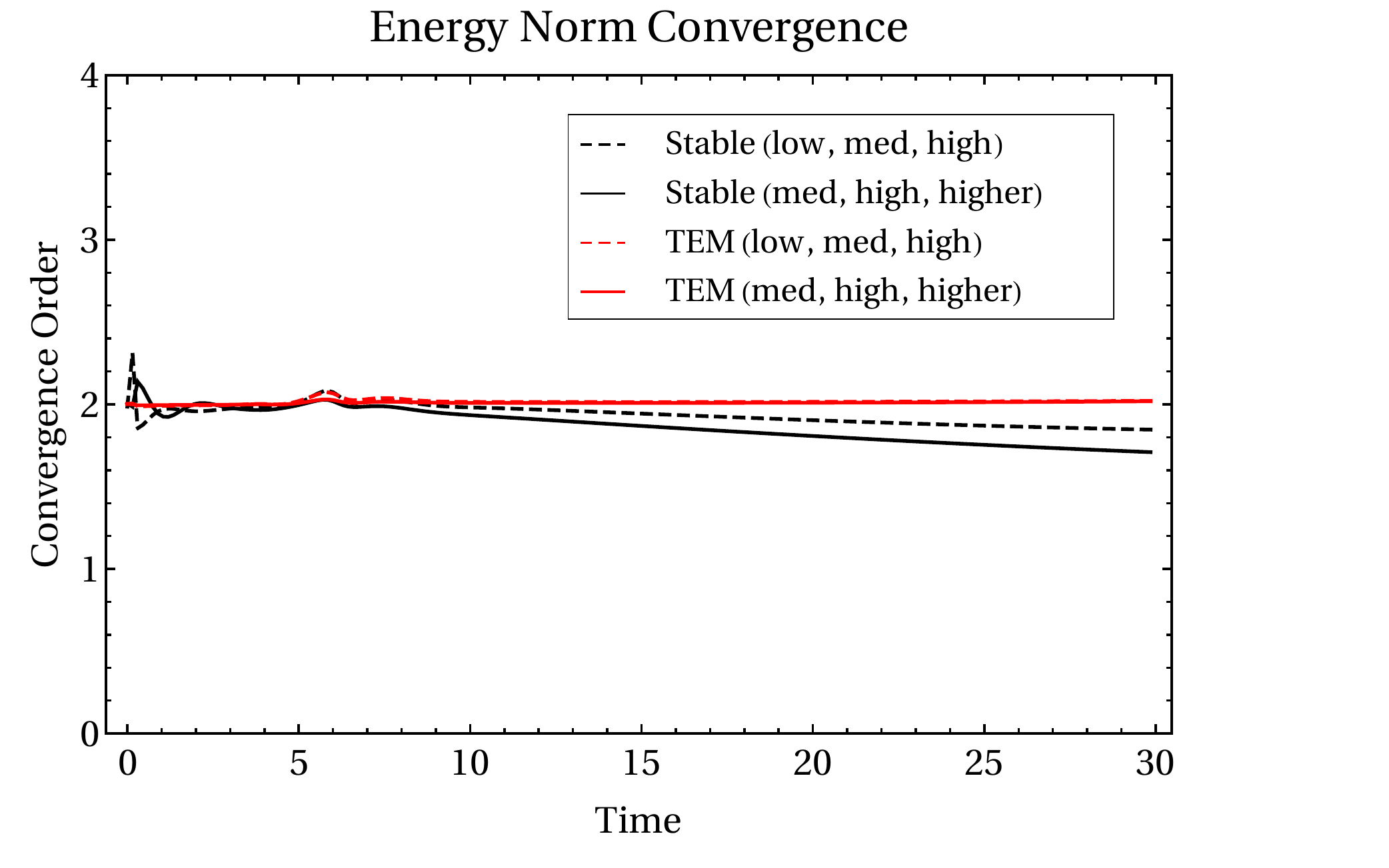}
\caption{Convergence order of the scalar field obeying the LWEP
  with~$F = 1/\chi^2$ in the energy norm given
  by~\eqref{Ehat_resc}. In this particular case we simply use the
  physical energy because it is not degenerate near infinity. The red
  curves correspond to the~SBP-TEM discretization and the black curves
  to the SBP-Stable one.}
\label{fig_LWEP_conv_order}
\end{figure}

Figure~\ref{fig_LWEP_conv_order} shows the convergence order in the
energy norm of the field obeying the~LWEP with~$F = 1/\chi^2$. In this
case, we observe an almost perfect convergence order at all times in
the~SBP-TEM scheme. However, the convergence order slowly decays in
the~SBP-Stable scheme once the data is very small and the error is
dominated by the lower order operators (in the derivatives and
dissipation) near the outer boundary. This plot also demonstrates the
superiority of the~SBP-TEM scheme over the~SBP-Stable one for this
family of initial data. However, we also observe very good second
order convergence at~$\mathscr{I}^+$ in both schemes, appearing very
similar to that shown in Fig.~\ref{fig_conv_scri_stable}.

\begin{figure}
\includegraphics[scale=0.5]{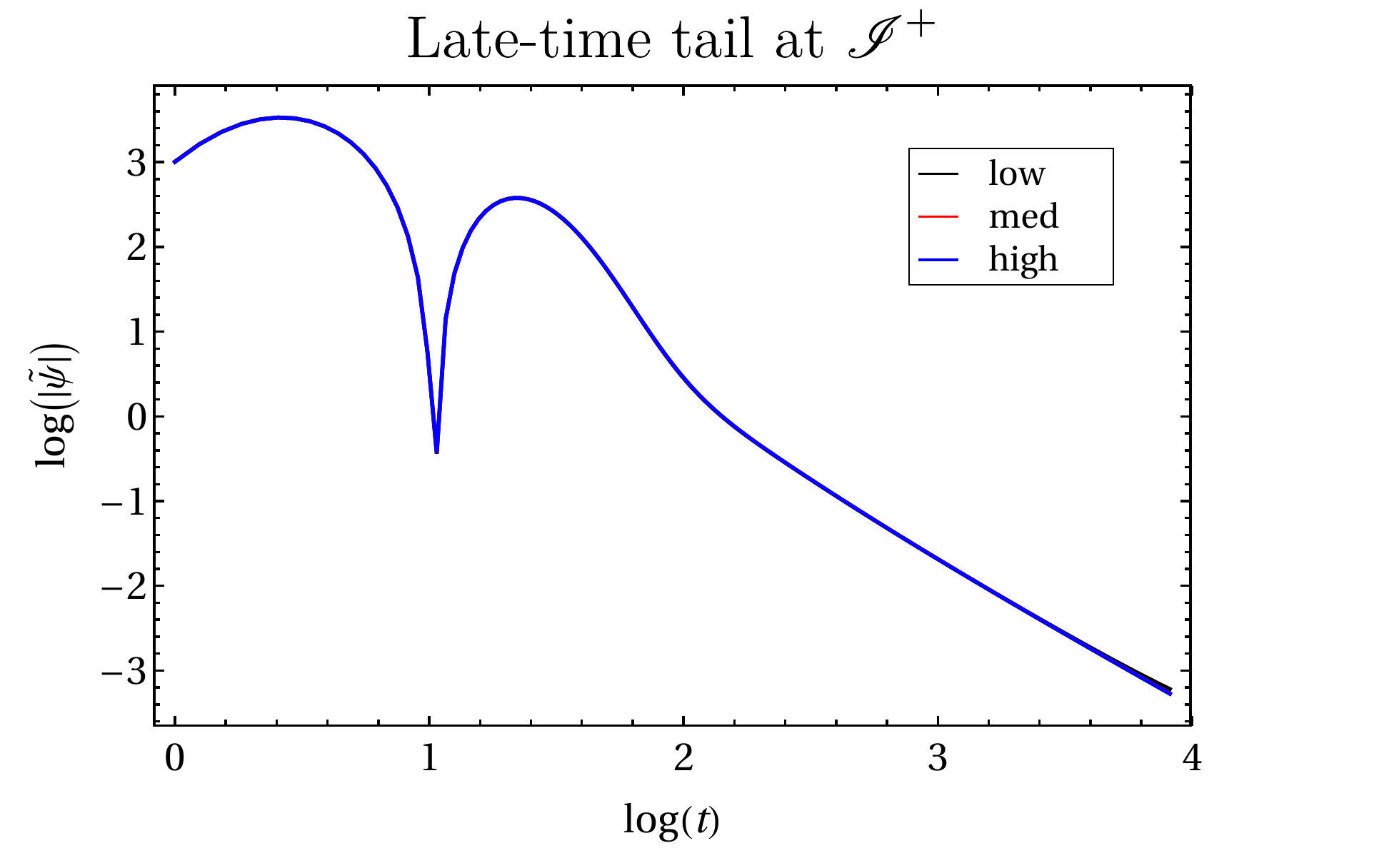}
\caption{Late time tail of a scalar field obeying the~LWEP with~$F =
  1/\chi^2$ at three different resolutions, with~$N = 200$, $400$
  and~$800$ respectively. This plot is generated for the initial data
  given by~\eqref{inidatagauss} with~$a = 100$ and~$\lambda = 1$. The
  slope of this tail is measured to be~$\approx - 1.74$.}
\label{fig_LWEP_Late_time_tail}
\end{figure}

Here, in contrast to the plain wave equation even given initial data
of compact support, part of the physical signal always remains on the
computational domain. The reason for the slower decay of the solution
is the coupling between the incoming and outgoing modes of the
solution as described above. Therefore, as in Price's law~\cite{Pri72}
we expect a late time tail at~$\mathscr{I}^+$ which decays like an
inverse power of time~$t$. This is what we observe in
Fig.~\ref{fig_LWEP_Late_time_tail}, which is constructed from
the~SBP-TEM scheme. We can see a perfect overlap of the curves
corresponding to three different resolutions for long times, up
until~$t = 50$ in the plot. As could be anticipated from the previous
figure however, this overlap is not as good in the~SBP-Stable
scheme. This result again demonstrates the superiority of the~SBP-TEM
scheme over the~SBP-Stable setup for this initial data.

\subsubsection{Linear Massive Klein-Gordon Equation,~$F=m^2$}
\label{subsubsection:LMKGE}

As an extreme example, we now consider the alternative
potential~$F=m^2$ with no decay near infinity. Despite the fact that
the hyperboloidal form of the equations of
motion~\eqref{pss_rescaled_chi_chi_sq_cont} have terms with divergent
coefficients of the form~$R'F\tilde{\psi}$ near infinity, the
continuum equations still make sense, at least within a large class of
initial data, because solutions decay faster than any inverse
polynomial in~$R$~\cite{Win88,Kla93}. A separate question is whether
or not we are able to find accurate approximate solutions in our
coordinates. Even given a usable setup with a conserved positive
energy at the semidiscrete level, such an energy would require a
restricted class of initial data that decay rapidly at infinity, and
so formal numerical stability~\cite{Tho98c} does not automatically
follow. Perhaps an alternative perspective is that the mass term is
effectively arbitrarily `stiff' near infinity, so that problems in
time integration could be foreseen.

We choose initial data for~$\tilde{\psi}$ that falls off fast enough
so that~$R' \tilde{\psi} \to 0$ as~$r \to r_\mathscr{I}$, which,
according to the continuum estimates mentioned above and as can be
deduced from~\eqref{en_cons_int_form}, should then hold true at later
times. The initial data given by~\eqref{inidatagauss} is one such
choice. Under this assumption all variables must vanish
at~$\mathscr{I}^+$. Unfortunately because of the singular coefficient
neither of our two schemes can be used without modification. We have
thus tried various different strategies to manage the singular
coefficients, including, for example, fixing all time derivatives
at~$\mathscr{I}^+$ to vanish. By so doing, we are able to perform
numerical evolutions and obtain very good energy conservation, even at
low resolutions. But unfortunately as soon as an outgoing pulse hits
the region near~$\mathscr{I}^+$ both norm and pointwise convergence
are completely lost, as high-frequency reflections propagate back into
the central region. Performing convergence tests at successively
higher resolutions does not help.

Presently it is not clear how, or even if, these difficulties can be
overcome. One possibility to obtain at least a consistent scheme with
a semidiscrete energy estimate would be to impose Dirichlet type
boundary conditions at a finite timelike boundary and to then take the
limit to~$\mathscr{I}^+$. But as mentioned above, even that would not
guarantee convergence. Another strategy might be to build a
discretization around the Bessel functions which naturally capture the
structure of solutions~\cite{Win88}. Final possibilities would be to
maintain a central, flat, slicing over the region of interest for the
massive field, or to simply admit defeat and modify the field
equations near~$\mathscr{I}^+$.

\section{Conclusions}\label{Conclusions}

In this series~\cite{HilHarBug16,GasHil18,GasGauHil19} of papers we
are developing a method to attach future null infinity to the
computational domain via hyperboloidal slices in numerical
relativity. There are several aspects to the problem. In the present
work we have focused on the properties of two approximation schemes
for first order reductions of linear wave equations. We call these
approximations SBP-Stable and SBP-TEM. The first of these is formally
stable, whilst the second is instead built so that troublesome
reflections from null infinity are minimized. Here we worked in
spherical symmetry with second-order accurate operators, but neither
of these simplifications was fundamental. We moreover expect that both
schemes can be straightforwardly lifted to treat nonlinear equations.

In our numerical experiments the two schemes behave comparably in many
tests. Although the SBP-TEM method is not formally stable even for the
flat-space wave equation, it seems unlikely that the user would
stumble across the expected class of `bad' initial data in
practice. If they did, the SBP-Stable method could be applied
instead. Concerning the SBP-Stable method, we seem to be forced to use
low-order operators near the outer boundary. In long evolutions the
errors associated with these operators are dominant. On the other
hand, we were positively surprised when using the SBP-Stable method
that pointwise convergence there is not too badly damaged for most of
the evolution.

The hyperboloidal coordinates that we employ are fundamentally adapted
to the clean resolution of outgoing waves. There is, therefore, a
limited class of models that can be accurately treated by their
use. We might anticipate, for example, that any model which generated
large amounts of incoming radiation near null infinity to be poorly
approximated by either of our schemes. To investigate this we studied
wave equations with different potentials. We found that when the
potential decays sufficiently fast near null infinity our methods
serve their purpose well but when this is not the case, as in the
massive Klein-Gordon equations, they cannot be directly applied and,
at least with naive adjustment, fail badly. Interestingly even if a
consistent method with a conserved norm could be found at the
semidiscrete level, it would not necessarily converge because the
equations of motion do not regularize. In the future it will be
desirable to unpick the relationship between the generation of
incoming radiation and the possibility to regularize a given model. It
would also be interesting to understand the slowest possible decay of
a potential that could be well treated by our (or any other) methods.

An important open question is whether or not {\it any} scheme could be
given that combines the advantages of both the SBP-Stable and SBP-TEM
setups, perhaps by using a careful upwinding discretization. For now,
however, our highest priority is to combine the methods we have
developed here with the regularization given in~\cite{GasGauHil19} for
nonlinear models to treat GR proper.

\acknowledgments

We are grateful to Abhay Ashtekar, Sanjeev Dhurandhar, Edgar Gasperin,
Jayant V. Narlikar and especially to Miguel Zilh\~ao for useful
discussions. This work was supported through the European Research
Council Consolidator Grant 647839, the FCT Programs~IF/00577/2015,
PTDC/MAT-APL/30043/2017, the PhD researcher Decree-Law no. 57/2016 of
August 29 (Portugal) and Project~No.~UIDB/00099/2020. All authors
would like to thank Navajbai Ratan Tata Trust (NRTT) grant for
supporting various visits of the authors to IUCAA and SG's visit to
CENTRA, T\'ecnico, Lisboa. This paper was assigned the LIGO preprint 
number LIGO-DCC-P2000514.

\bibliography{refs}

\end{document}